\documentclass[aps,prd,twocolumn,superscriptaddress,floatfix,nofootinbib,natbib,longbibliography]{revtex4-1}
\usepackage{graphicx,longtable,mathrsfs,color,array}
\usepackage[usenames,dvipsnames]{xcolor} 
\usepackage{amssymb,amsmath,mathtools,mathrsfs,slashed} 
\usepackage{epsfig,subfigure,placeins,float} 
\usepackage{booktabs,longtable,ctable,multirow} 
\usepackage{exscale,relsize} 
\usepackage[normalem]{ulem} 
\usepackage{fontawesome5}
\usepackage{enumerate}
\usepackage{enumitem}
\usepackage{threeparttable}
\usepackage{booktabs}
\usepackage[colorlinks,linkcolor=blue,citecolor=blue,urlcolor=blue ]{hyperref}

\makeatletter
\@fpsep\textheight
\makeatother

\usepackage[stable]{footmisc}
\usepackage{color}
\newcommand{\github}[1]{\href{#1}{\faGithubSquare}}

\newcommand{\pkdgrav}{\texttt{PkDGraV3}}

\newcommand{\healpix}{\texttt{HEALPix} }

\newcommand{\hillipop}{\texttt{HiLLiPoP}}

\DeclareUnicodeCharacter{2212}{\textminus}
\begin{document}
\author{Alexander Reeves}
\email{areeves@phys.ethz.ch}
\affiliation{Institute for Particle Physics and Astrophysics, ETH Zürich, Wolfgang-Pauli-Strasse 27, CH-8093 Zürich, Switzerland}
\author{Simone Ferraro}
\affiliation{Lawrence Berkeley National Laboratory, 1 Cyclotron Road, Berkeley, CA 94720, USA}
\affiliation{Department of Physics, University of California, Berkeley, CA 94720, USA}
\author{Andrina Nicola}
\affiliation{Jodrell Bank Centre for Astrophysics, Department of Physics and Astronomy, The University of Manchester, Manchester M13 9PL, UK}
\author{Alexandre Refregier}
\affiliation{Institute for Particle Physics and Astrophysics, ETH Zürich, Wolfgang-Pauli-Strasse 27, CH-8093 Zürich, Switzerland}

\title{Multiprobe constraints on early and late time dark energy}

\date{Received \today; published -- 00, 0000}

\begin{abstract}
We perform a multiprobe analysis combining cosmic microwave background (CMB) data from \textit{Planck} and the Atacama Cosmology Telescope (ACT), ACT CMB lensing, and large-scale structure (LSS) measurements from the Dark Energy Spectroscopic Instrument (DESI), including DESI Legacy Imaging Survey (LS) galaxies and baryon acoustic oscillations (BAOs). We present the first $5\times2$pt analysis of ACT DR6 lensing, DESI LS, and \textit{Planck} ISW. Within $\Lambda$CDM, this yields $S_8 = \sigma_8(\Omega_m/0.3)^{0.5} = 0.819 \pm 0.016$, in good agreement with primary CMB inferences and provides a sound-horizon-free Hubble constant constraint of $H_0 = 70.0 \pm 4.4$ km s$^{-1}$ Mpc$^{-1}$. Then, combining with CMB primary and BAO, we reconfirm a CMB–BAO discrepancy in the $\Omega_m$--$\frac{D_v}{r_d}$ plane, which is heightened when combining BAO with the $5\times2$pt data vector. We explore two dark-energy extensions that may reconcile this: an early-time modification, early dark energy (EDE), and late-time dynamical dark energy (DDE) parameterized by $w_0w_a$. For CMB primary+BAO+$5\times2$pt, we find a $3.3\sigma$ preference for DDE over $\Lambda$CDM, while EDE is modestly favoured at $2.3\sigma$. The models address different shortcomings of $\Lambda$CDM: DDE relaxes the neutrino mass bound ($M_\nu<0.17$eV vs. $<0.050$eV under $\Lambda$CDM), making it compatible with neutrino oscillation measurements, while EDE raises the Hubble constant to $H_0=70.5\pm1.2\,\mathrm{km\,s^{-1}\,Mpc^{-1}}$, easing the discrepancy with SH0ES. However, neither model resolves both issues simultaneously. Our analysis indicates that both DDE and EDE remain viable extensions of $\Lambda$CDM within current uncertainties and demonstrates the capacity of combined probes to place increasingly stringent constraints on cosmological parameters.
\end{abstract}
\maketitle

\section{Introduction}
\label{sec:Introduction}
Multiprobe techniques are at the forefront of modern cosmological analysis. Such analyses offer significant advantages over single-probe measurements. These include the ability to identify and mitigate systematic effects that do not correlate between datasets and the breaking of parameter degeneracies to provide tight constraints on cosmological models. In recent years, numerous studies in this direction have been performed, including the $3\times2$pt analysis of large-scale structure (LSS) galaxy surveys combining galaxy clustering and weak lensing measurements~\cite{DES:2021wwk, Heymans:2020gsg, Sugiyama:2023fzm}, a similar combination but exchanging cosmic microwave background (CMB) lensing for galaxy weak lensing~\cite{Farren:2024rla}, and comprehensive analyses combining CMB and several LSS tracers (see e.g.~\cite{Nicola:2016qrc, Nicola:2016eua, Reeves:2025axp, DES:2022urg, Xu:2023qmp}). 

Recent analysis of the second data release of the Baryon Acoustic Oscillation (BAO) data from the Dark Energy Spectroscopic Instrument (DESI DR2) shows a $\sim 3\sigma$ preference for an evolving dark energy equation of state when combined with CMB data from Planck~\cite{DESI:2025dr2, DESI:2025fii}. This is driven by an internal $\sim2\sigma$ tension between CMB and BAO data when analyzed under $\Lambda\mathrm{CDM}$~\cite{DESI:2025dr2}. While dynamical dark energy is a plausible model to explain these data, the evidence is not yet conclusive. In particular, several recent papers have suggested that altering pre-recombination physics, for example, via an early dark energy (EDE) component~\cite{Chaussidon:2025npr, Poulin:2025nfb} or by a modified recombination history~\cite{Mirpoorian:2025rfp, Lynch:2024hzh}, while keeping a cosmological constant, can also partially resolve this discrepancy.

In this work, we confront these recent hints with LSS and CMB data using a multiprobe approach. Building on the framework introduced in Refs.~\cite{Reeves:2025axp, Reeves:2023kjx} and incorporating new datasets, we analyze, for the first time, a \(5\times2\)pt LSS combination of ACT DR6 CMB lensing, Legacy Imaging Survey (LS) galaxy clustering, and \textit{Planck} PR3 ISW, measuring all auto- and cross-correlations. The \(2\times2\)pt subset of this data vector, comprising the LS auto-correlations and their cross-correlation with CMB lensing, has already been shown to place strong constraints on the lensing amplitude \(S_8\) within \(\Lambda\mathrm{CDM}\)~\cite{Sailer:2024coh, Qu:2024sfu, Kim:2024dmg}. Moreover, Ref.~\cite{Sabogal:2025jbo} demonstrated that this subset, when combined with Type Ia supernova (SNe) data, provides robust evidence for a DDE scenario even in the absence of primary CMB data. 

We begin by deriving \(\Lambda\mathrm{CDM}\) constraints, including deriving a sound horizon-free $H_0$ value and assessing goodness of fit under this model. We then combine with primary CMB data from \textit{Planck} PR4 and ACT DR6 and DESI DR2 BAO measurements, to explore two extensions in the dark energy sector that can reconcile the mild tensions between these data observed under $\Lambda$CDM. The extensions represent modifications to early- and late-time physics respectively: an early-Universe modification exemplified by an axion-like EDE model and a dynamical dark energy model with a time-evolving equation of state (denoted hereafter by DDE). In particular, we ask whether contemporary LSS and BAO data favour an evolving \(w(z)\) or an EDE component when analysed alongside the CMB; how these extensions affect derived parameters such as \(S_8\) and the neutrino mass sum; and what changes arise when we confront these models with external data from Pantheon+ SNe~\cite{Brout:2022vxf} and local distance ladder $H_0$ inferences from SH0ES~\cite{Riess:2021jrx}. We further aim to clarify the phenomenological drivers of any improvement in fit beyond \(\Lambda\)CDM.

The rest of this paper is organized as follows: In Section~\ref{sec:datasets} we introduce the datasets and measurements used in our analysis. In Section~\ref{sec:modelling}, we describe our analysis pipeline, including details of the likelihoods, modeling assumptions, covariance estimation and inference framework. Section~\ref{sec:theory_models} presents the theoretical models behind the extensions to $\Lambda$CDM under consideration. We report our results in Section~\ref{sec:Results}, first discussing the constraints on $\Lambda$CDM parameters from the $5\times2$pt data and goodness of fit. We then combined with CMB primary and BAO data and examine the EDE and DDE dark energy extensions and determine model preference from the relative goodness of fit. We also discuss these findings in the context of additional external datasets (supernovae data and external priors on $H_0$ or $S_8$). Finally, Section~\ref{sec:Conclusion} summarizes our conclusions and outlines future prospects for distinguishing early vs. late-time new physics with upcoming data.

\section{Datasets}\label{sec:datasets}

In this section, we briefly outline the datasets used in this analysis. 

\paragraph{DESI Legacy Imaging Survey}

We use the LS luminous red galaxy (LRG) catalogue~\cite{Zhou:2023gji}, drawn from the \textit{g\,r\,z} imaging of the LS DR9~\cite{Dey:2019} for the photometric clustering part of our analysis. The images were taken with DECam on the CTIO 4m Blanco telescope, the Bok 2.3m telescope, and the 4m Mayall telescope (MzLS), providing $\sim\!20\,000\;\mathrm{deg}^{2}$ of uniform photometry with well-characterized selection functions and photometric redshifts. We use the ``Main LRG sample'', which covers $16,700\;\mathrm{deg}^{2}$ on the sky, and comprises approximately $12$ million galaxies. The redshift distributions are characterized using 2.3 million spectroscopic redshifts from DESI's Survey Validation~\cite{DESI:2023dwi}. The data are distributed in four tomographic redshift bins in the range $0.2\lesssim z \lesssim1.4$. A detailed description of the photometric selection criteria is given in Ref.~\cite{Zhou:2023gji}\footnote{These data are publicly available here: \url{https://data.desi.lbl.gov/public/papers/c3/lrg_xcorr_2023/}.}.

\paragraph{ACT DR6 CMB Primary and Lensing}
For the ACT DR6 primary CMB, we use the \texttt{P-ACT} combination adopted by the ACT collaboration as their baseline dataset. This consists of the \textit{Planck} 2018 temperature and polarization spectra at low multipoles ($30<\ell<1000$ for TT and $30<\ell<650$ for TE/EE), combined with ACT DR6 data at high multipoles~\cite{ACTDR6:PowerSpectra}. Following the ACT collaboration, we also add the \texttt{sroll2} low-$\ell$ EE likelihood~\cite{Pagano:2019tci}. For the $2<\ell<30$ TT part of the likelihood, we use the compressed likelihood from Ref.~\cite{Prince:2021fdv} which is based on the \textit{Planck} PR3 \texttt{commander}-derived likelihood; the latter is used in the official ACT analysis, while the former was shown to produce identical constraints with reduced computational overhead~\cite{Prince:2021fdv}. We adopt the compressed ``ACT-lite'' CMB-only likelihood where foregrounds are pre-marginalized. This has been shown by the ACT collaboration to yield cosmological parameters consistent with the baseline \texttt{MFLike} likelihood (which explicitly varies foreground parameters) while significantly reducing the computational cost~\cite{ACTDR6:PowerSpectra}.

In addition, we use the gravitational-lensing convergence maps reconstructed from the same ACT DR6 data, covering a total sky area of approximately $9{,}400\,\mathrm{deg}^2$~\cite{ACT:2023dou}. The lensing pipeline applies both quadratic and iterative estimators to multi-frequency temperature and polarization maps to reconstruct the CMB lensing potential. The ACT DR6 public release includes the reconstructed lensing maps, their associated masks, and simulations of the noise properties\footnote{The ACT DR6 lensing data are available at this URL: \url{https://lambda.gsfc.nasa.gov/product/act/actadv_dr6_lensing_maps_info.html}}.

\paragraph{\textit{Planck} Primary CMB Anisotropies}
When deriving constraints, we compare results by exchanging between the CMB primary likelihood of ACT DR6 and \textit{Planck} PR4. The \textit{Planck} PR4 processing (``NPIPE’’) delivers improved calibration and foreground mitigation relative to earlier releases~\cite{Planck:2020olo}. We adopt the \hillipop\ version of the \textit{Planck} PR4 likelihood~\cite{Tristram:2023haj}. We use a compressed version of the likelihood described in Appendix~\ref{appendix:compressed_hillipop}, which has an order of magnitude fewer data points and allows for faster inference while producing identical cosmological constraints. We make the rewritten \texttt{JAX} version of this likelihood (including the corresponding compression matrix) publicly available\footnote{\url{https://github.com/alexander-reeves/jax-loglike}}. 

For our ISW analysis, we use the \textit{Planck} PR3 temperature maps\footnote{We do not expect any significant change to our results upon exchanging with the PR4 NPIPE temperature map, as the large-scale temperature signal is essentially unchanged in this reprocessing~\cite{Planck:2020olo}.}. We use the \texttt{Commander} maps, as these are the recommended maps for measurements on large angular scales\footnote{Available at the Planck Legacy Archive, \url{http:/pla.esac.esa.int}}. 

\paragraph{DESI DR2 Baryon Acoustic Oscillations}
The second public release of the Dark Energy Spectroscopic Instrument (DESI) survey provides more than $14$ million spectroscopic redshifts gathered with the 5000‑fiber DESI spectrograph on the 4m Mayall Telescope at Kitt Peak National Observatory~\cite{DESI:instrument}.  
DR2 covers $\sim\!7500\;\mathrm{deg}^{2}$ and is split into four tracer classes—0.3 million Bright Galaxy Survey galaxies ($0.1<z<0.4$), 2.14 million LRGs ($0.4<z<1.1$), 2.43 million ELGs ($0.8<z<1.6$) and 0.86 million quasars ($0.8<z<2.1$)—yielding BAO measurements across $0.1<z<2.1$~\cite{DESI:2025dr2}. In this work, we use the public DESI DR2 BAO likelihood, comprising 13 distance measurements at seven effective redshifts spanning $0.30\le z_{\rm eff}\le 2.33$ (BGS, LRG1, LRG2, LRG3+ELG1, ELG2, QSO, Ly$\alpha$). The likelihood consists of measurements of $\{D_M/r_d,\,H\,r_d\}$ for all tracer classes except BGS, where $D_V/r_d$ is used~\cite{DESI:2025dr2}.

\paragraph{Pantheon\,+ Type Ia Supernovae}
When adding SNe to our data combination, we employ the Pantheon\,+ compilation of 1550 spectroscopically confirmed SNe~Ia spanning \(0.001<z<2.26\) and comprising 1701 light curves~\cite{Brout:2022vxf}. To mitigate sensitivity to host–galaxy peculiar velocities and local bulk flows at very low redshift, we remove the 111 SNe with \(z<0.01\); our working sample therefore contains 1439 SNe (1590 light curves) over \(0.01<z<2.26\)~\cite{Brout:2022vxf}.

\section{Modeling, data vector, covariance, inference pipeline}\label{sec:modelling}  

\subsection{Modeling}\label{subsec:modelling}

Our modeling pipeline closely follows the framework developed in Refs.~\cite{Reeves:2023kjx,Reeves:2025axp}. For the LS LRGs, ISW cross-correlations and ACT DR6 CMB-lensing cross-correlations, we define our own modeling choices.  All other elements of the framework—primary CMB TTTEEE, DESI BAO, CMB lensing auto-correlation and Pantheon\,+ distances—are instead implemented using \texttt{JAX} translations of the official publicly available likelihoods from \texttt{Cobaya}~\cite{Torrado:2020dgo}.

\subsubsection{Theoretical predictions}\label{subsec:theory_pred}
To produce theoretical predictions for the LSS tracers (CMB lensing, galaxy clustering, ISW) in this pipeline, we employ the \texttt{FKEM}~\cite{Fang:2019xat} beyond Limber algorithm as implemented in \texttt{CCL} v3.1~\cite{LSSTDarkEnergyScience:2018yem} (see also Ref~\cite{Reymond:2025ixl} for a recent \texttt{JAX}-based beyond Limber implementation). This computes the full 3D angular power spectrum projection integral,
\begin{align}
C^{XY}_{\ell}
&= \int_{0}^{\infty}\!\mathrm{d}\chi
   \int_{0}^{\infty}\!\mathrm{d}\chi'\;
   W^{X}(\chi)\,W^{Y}(\chi') \nonumber\\[4pt]
&\quad \times
   \int_{0}^{\infty}\! \frac{2}{\pi}\,k^{2}\,\mathrm{d}k\;
   P\!\big(k;\chi,\chi'\big)\,
   j_{\ell}(k\chi)\,j_{\ell}(k\chi')\,.
\end{align}
where $P(k;\chi,\chi')$ is the unequal time matter power spectrum, $j_{\ell}$ is the spherical bessel function, $\chi$ is the comoving distance, and $W^{X/Y}$ are window functions specific to the probes in the pipeline. We use this full computation for multipoles $\ell < 60$. For $\ell \geq 60$, we set-up \texttt{CCL} to transition transition to the standard Limber approximation~\cite{limber_approx}
\begin{equation}
C^{XY}_{\ell}
 \;=\;
\int \! \frac{\mathrm{d}\chi}{\chi^{2}}\,
P\!\left(k=\frac{\ell+\tfrac12}{\chi}, \chi\right)\,
W^{X}\!(\chi)\,W^{Y}\!(\chi).
\end{equation}
We chose the transition at $\ell=60$ based on Fig. 4 of Ref.~\cite{Sailer:2024coh} where the Limber approximation is shown to be inaccurate at $>1\%$ for $\ell \lesssim60$ for LS LRG auto-correlations. 

For the LS projected galaxy overdensity, we adopt a linear galaxy bias $b_i$ fit independently in each of the four tomographic bins, include magnification using the standard magnification kernel with a free number-count slope parameter $s_i$ per bin, and add a constant projected shot-noise term with one free parameter per bin, following Ref.~\cite{Sailer:2024coh}. We adapt the \texttt{CCL} code to enable the use of the \texttt{AxiClass}~\cite{Poulin:2018dzj} backend Boltzmann solver with the appropriate parameter settings in order to make theoretical predictions for the EDE model.

For the CMB primary, we compute TT, TE, and EE spectra with \texttt{CLASS}~\cite{Blas:2011rf} (or \texttt{AxiClass}~\cite{Poulin:2018dzj} in the case of EDE) using the precision settings detailed in Appendix~A of Ref.~\cite{ACT:2025tim}. The precision settings are required to ensure unbiased constraints when using the high-$\ell$ ACT DR6 CMB primary likelihood. DESI BAO and Pantheon\,+ SN distances and sound horizon predictions are produced from the same \texttt{CLASS} evaluation. 

\subsubsection{Scale cuts}

The scale cuts used in this analysis are summarized in Table~\ref{tab:scale_cuts}. For galaxy clustering—both in auto- and cross-correlations—we restrict to scales where a linear-bias description is adequate. Concretely, in each redshift bin we set \(\ell_{\max}\) by mapping angular to three-dimensional modes with the Limber approximation. We then require \(k<0.1\,h\,\mathrm{Mpc}^{-1}\), ensuring the analysis remains safely within the linear regime, where the difference between the linear and non-linear matter power spectrum is $\lesssim 1.5\%$ (we estimate this using \texttt{CLASS}~\cite{Blas:2011rf} in the best-fit \textit{Planck} $\Lambda\mathrm{CDM}$ cosmology). The minimum multipole is conservatively fixed to \(\ell_{\min}=30\) for galaxy auto-correlations to mitigate the impact of large-scale redshift-space distortions~\cite{SDSS:2006egz} and to suppress box-size effects in the simulation-based covariance matrix. 

For ACT DR6 CMB lensing we analyze the auto-spectrum over the extended range \(40 \le L \le 1300\), which the collaboration deemed reliable after unblinding, following a simulation-based re-evaluation of potential foreground-induced biases~\cite{ACT:2023kun,ACT:2023ubw}. Finally, for each cross-correlation we adopt the more conservative of the two scale cuts defined by the probes entering the cross-spectrum, except for ISW cross-correlations where we use $\ell_{\text{min}}=20$ to enhance the signal-to-noise ratio (S/N), which peaks on larger scales. We checked that the recovered ISW covariance matrix matches an analytic estimate within $5\%$ on the diagonals, showing that we are not significantly affected by box-size effects. 

\begin{table}[h]
\centering
\begin{tabular}{ccc}
    \hline
    Spectrum & $\ell$-range & \# bins \\
    \hline
    \multicolumn{3}{c}{LSS Auto-Correlations} \\
    \hline
    $\delta_{g1}\,\delta_{g1}$ & 30--185  & 5 \\
    $\delta_{g2}\,\delta_{g2}$ & 30--215  & 5 \\
    $\delta_{g3}\,\delta_{g3}$ & 30--235  & 5 \\
    $\delta_{g4}\,\delta_{g4}$ & 30--245  & 5 \\
    $\kappa_{\rm CMB}\,\kappa_{\rm CMB}$           & 40--1300 & 18 \\
    \hline
    \multicolumn{3}{c}{LSS Cross-Correlations} \\
    \hline
    $\delta_{g1}\,\kappa_{\rm CMB}$ & 40--185 & 5 \\
    $\delta_{g2}\,\kappa_{\rm CMB}$ & 40--215 & 5 \\
    $\delta_{g3}\,\kappa_{\rm CMB}$ & 40--235 & 5 \\
    $\delta_{g4}\,\kappa_{\rm CMB}$ & 40--245 & 5 \\
    $\delta_{g1}\,T$      & 20--90  & 4 \\
    $\delta_{g2}\,T$      & 20--90  & 4 \\
    $\delta_{g3}\,T$      & 20--90  & 4 \\
    $\delta_{g4}\,T$      & 20--90  & 4 \\
    $\kappa_{\rm CMB}\,T$           & 40--90  & 4 \\
    \hline
\end{tabular}
\caption{\textbf{Multipole ranges and binning for LSS auto- and cross-correlations.}\label{tab:scale_cuts}}

\end{table}

\subsubsection{Neural-network emulators}

As in our previous work~\cite{Reeves:2025axp,Reeves:2023kjx}, we employ \texttt{JAX}-based neural-network emulators to produce rapid theoretical predictions at each sampled point in parameter space. We draw $10^6$ points in a Latin hypercube in the cosmological parameter space and train dense feed-forward networks to emulate the outputs of our theoretical predictions module (see Sec.~\ref{subsec:theory_pred}). The pipeline mirrors Ref.~\cite{Reeves:2025axp}, with only a single change for the EDE runs where we replace baseline \texttt{CLASS} with \texttt{AxiCLASS}. 

Each neural network emulator is a multilayer perceptron with four hidden layers and 512 units per layer, using the custom activation function
\begin{equation}
\label{eqt:activation}
    f(\vec{x}) \;=\; \bigg(\vec{\gamma} \;+\;\big(1 + \exp(- \vec{\beta}\cdot\vec{x})\big)^{-1}\,(1-\vec{\gamma})\bigg)\,\cdot\,\vec{x},
\end{equation}
where $f(\vec{x})$ is the activation function that describes how the input vector $\vec{x}$ is mapped to the output vector in each layer and  \(\vec{\beta}\) and \(\vec{\gamma}\) are trainable parameters following the approach in Ref.~\cite{SpurioMancini:2021ppk}. We produce separate emulators for $C_\ell^{TT}$, $C_\ell^{TE}$, $C_\ell^{EE}$, $C_\ell^{\kappa_{\rm CMB}\kappa_{\rm CMB}}$, BAO sound horizon-distance ratios, and SNe distances. For the remaining LSS probes, we train a single joint emulator whose target is the concatenated vector of all binned LSS spectra. The networks are trained using \texttt{tensorflow}~\cite{tensorflow2015-whitepaper}. We subsequently translate the weights and biases to the \texttt{JAX} ecosystem using \texttt{flax}~\cite{flax2020github}. 

\begin{figure*}[htb]
    \centering
    \includegraphics[width=0.9\linewidth]{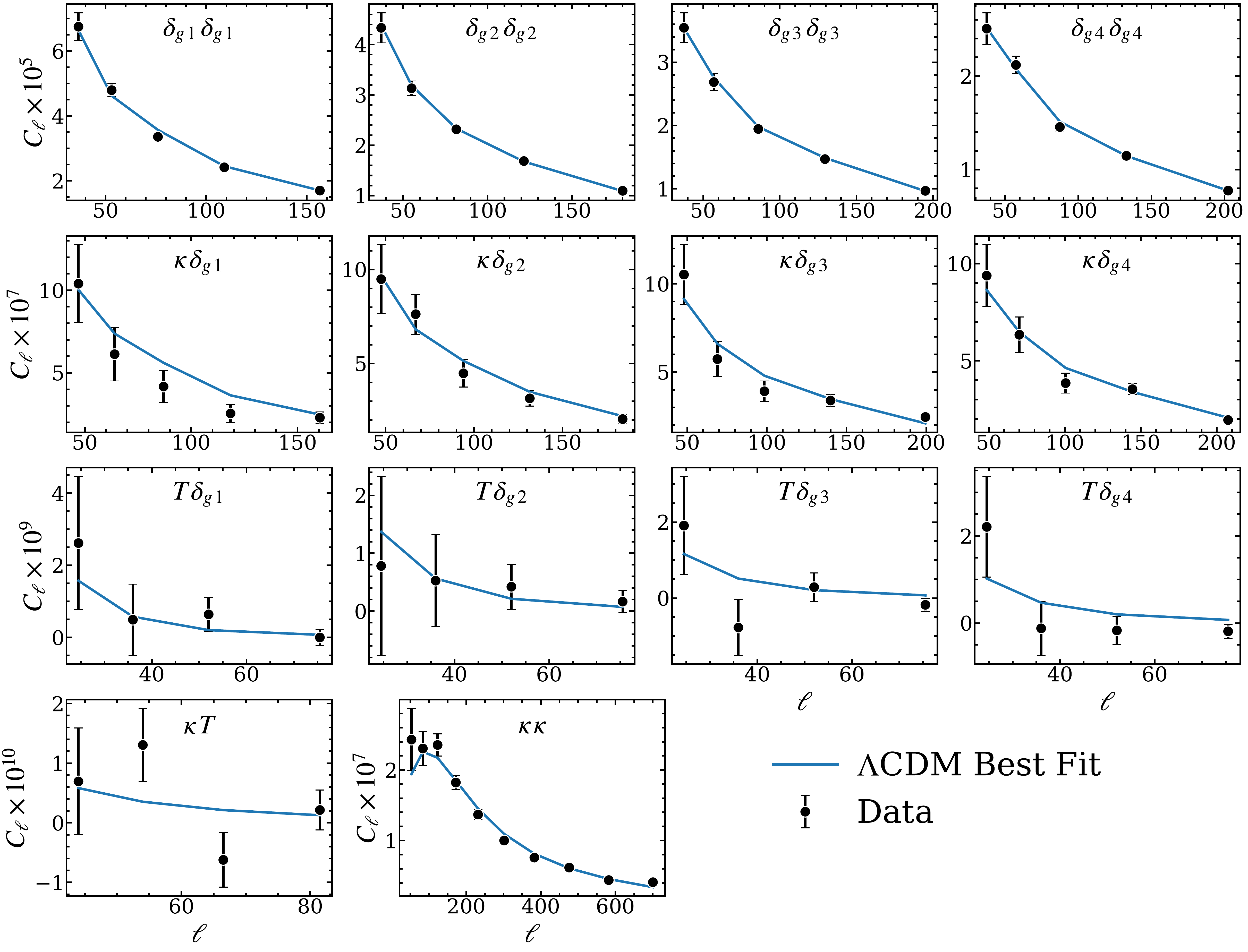}
    \caption{\textbf{The measured $5\times2$pt data vector with corresponding simulation-based error bars}. We show also the best-fit prediction for these spectra under $\Lambda$CDM.}
    \label{fig:datavector}
\end{figure*}

\subsection{Data vector}
\subsubsection{Mask deconvolution \label{subsection:mask_deconvolution}}
In this pipeline, we deal with projected maps of both simulations and observational data, which have a survey mask that generally couples modes together. The survey mask introduces mode coupling between multipoles. We can relate the ensemble-averaged cut-sky power spectrum $\langle\tilde{C}^{XY}_\ell\rangle$ to the full-sky spectrum $C_\ell$ through a mode coupling matrix $M_{\ell \ell'}$:
\begin{equation}
\langle\tilde{C}_{\ell}^{XY}\rangle = \sum_{\ell '} M_{\ell \ell'} C_{\ell '}.
\end{equation}
Recovery of the full-sky $C_\ell$ from cut-sky data by inverting the coupling matrix is generally not possible due to information loss from partial sky coverage. To make progress, we employ the MASTER algorithm~\cite{Hivon:2001jp}, implemented in \texttt{NaMaster}~\cite{Alonso:2018jzx}, which approximates the true angular power spectrum as piecewise constant over multipole bins. This approximation enables inversion of the binned mode coupling matrix, providing unbiased estimates of the full-sky bandpowers. To address pixelization effects, we use the window function obtained from \texttt{healpy}'s \texttt{pixwin} function as an effective ``beam'' in the \texttt{NaMaster} \texttt{NmtField} object for each map \footnote{This pixelization correction approach is applicable at high $\ell/\mathrm{NSIDE}$ only when the source density is sufficiently high (see appendix A of Ref.~\cite{Nicola:2020lhi}). Alternative methods that circumvent pixelization entirely and thereby avoid these issues are discussed in Refs.~\cite{Wolz:2024dro, BaleatoLizancos:2023jbr}.}.

\subsubsection{Measuring the data vector}
To produce the $5\times2$pt data vector, we follow Ref.~\cite{Reeves:2025axp}, starting from galaxy maps for the LS LRGs, the ACT DR6 lensing convergence map, and the \textit{Planck} PR3 \texttt{Commander}-derived temperature map. Each of the maps are produced at \texttt{HEALPix} $\mathrm{NSIDE}=2048$ with the respective survey masks applied. We then use \texttt{NaMaster} to compute mask-deconvolved bandpowers following the procedure outlined in Sec.~\ref{subsection:mask_deconvolution}. For CMB lensing auto-correlations, we instead make use of the official ACT DR6 lensing bandpowers instead of deriving our own data vector directly from the convergence map. For the ACT DR6 lensing cross-correlations we apply a Monte Carlo normalization correction following the procedure outlined in Appendix~\ref{appendix:mc_norm}. Our measured data vector can be seen in Fig~\ref{fig:datavector}. 

\subsection{Covariance matrix}
\begin{table}[h]
\centering
\begin{tabular}{cc}
    \hline
    Parameter & Value \\
    \hline
    $\Omega_{m}$ & 0.315 \\
    $A_s$ & $2.1\times10^{-9}$ \\
    $n_{s}$ & 0.9649 \\
    $\Omega_{b}$ & 0.0493 \\
    $H_{0}$(Km/s/Mpc) & 67.3 \\
    $M_\nu$(eV) & 0.06 \\
    \hline
\end{tabular}
\caption{\textbf{Fiducial Cosmological parameters used in simulations for the covariance matrix}. This matches the best-fit parameters from \textit{Planck} 2018~\cite{Planck:2018vyg}.}
\label{tab:fidcosmo}
\end{table}

We build the low-$z$ covariance from $2\,000$ mock map–level realizations based on 200 independent $N$-body light-cones. Each light-cone is generated with \pkdgrav\ \cite{Potter:2016ttn} in the \textit{Planck} 2018 fiducial cosmology (Table~\ref{tab:fidcosmo}) with a boxsize of \texttt{LBOX}=2250\,Mpc$/h$ and $2080^{3}$ particles. The resulting lightcones are post-processed with the publicly available \texttt{UFalcon} software\footnote{\href{https://cosmology.ethz.ch/research/software-lab/UFalcon.html}{https://cosmology.ethz.ch/research/software-lab/UFalcon.html}} \cite{Sgier:2020das, Reeves:2023kjx} to generate correlated galaxy-density, ISW and CMB lensing signal maps at $\mathrm{NSIDE}=2048$.

We add noise realizations separately for each probe. For DESI galaxies, the noise is added as a Poisson draw of random positions inside the DESI mask. For the ISW maps, we add FFP10 end-to-end simulations containing both CMB signal and noise components to the ISW signal maps following Ref.~\cite{Reeves:2023kjx}. Whilst this adds an additional ISW signal component to the simulated maps, this is entirely subdominant to the other contributions, so it has a negligible impact. This method was shown to produce covariances that agree well (within $5\%$ on the diagonals) with analytically-derived Gaussian ISW covariance matrices in Ref.~\cite{Reeves:2023kjx}. For CMB lensing, we create noise-only maps from the public ACT DR6 simulations\footnote{\url{https://lambda.gsfc.nasa.gov/product/act/actadv_dr6_lensing_maps_info.html}} by subtracting the input signal $a_{\ell m}$s from the full simulated $a_{\ell m}$s and converting the residuals to \healpix\ maps at $\mathrm{NSIDE}=2048$ with the \texttt{alm2map} routine from \texttt{healpy}. Treating the noise as additive in this way is accurate to better than $3.5\%$ in the noise power spectrum \cite{Nicola:2016qrc}, which is adequate for covariance estimation.

We apply the respective survey masks to each simulation before we measure pseudo-$C_{\ell}$s following the same procedure outlined in Sec.~\ref{subsection:mask_deconvolution}. The covariance matrix is then estimated from the ensemble of the $2\,000$ pseudo-$C_{\ell}$ mock realizations of the data vector. When inverting this matrix for parameter inference, we correct for finite-sample bias with the Hartlap factor~\cite{Hartlap:2006kj}. Finally, we replace the simulated ACT lensing auto-covariance block with the covariance matrix supplied by the ACT collaboration while retaining the simulation-derived cross-covariance blocks (i.e., the $<C_\ell^{\kappa_{\rm CMB} \kappa_{\rm CMB}}C_\ell^{XY}>$ parts where $X$, $Y$ represent two other probes in the pipeline).

We checked that the resulting covariance matrix is well converged by varying the number of simulations used and computing the mean relative difference of the diagonals of the matrix. We find that after $\sim 1500$ realizations, the relative differences are $\lesssim1\%$ implying sufficient convergence for cosmological analysis. We also checked that the simulation-based covariance agrees within 10\% along the diagonal with an independent analytical calculation for the same matrix.

\subsection{Likelihood}
In our joint analyses, we form the sum the likelihood contributions from the CMB primary, BAO, SNe, and the combined \(5\times 2\)pt LSS data vector, under the assumption that inter-probe cross-covariances are negligible. For the CMB primary likelihoods, this approximation is justified because the cross-correlation with LSS tracers is small compared to the dominant primary CMB power spectra~\cite{Schmittfull:2013uea} (this statement may require re-examination for future high-precision surveys~\cite{Peloton:2016kbw, Kou:2025hvg}). For the BAO--LSS cross-covariance, we test this approximation explicitly in Appendix~\ref{appendix:bao_information}, and find that the BAO information content in the LSS data vector is negligible compared to the information content of the reconstructed BAO likelihood.

When combining CMB lensing with primary CMB power spectra, we follow Ref.~\cite{Farren:2024rla} and forward-model the linear response of the lensing reconstruction to the measured primary spectra, updating to first order the estimator normalization and reconstruction biases (e.g., \(N^{(0)}\), \(N^{(1)}\)). This is required for a consistent analysis, as these quantities necessarily depend on the given CMB primary realization. When the primary CMB power spectra are not included, we adopt a \emph{CMB-marginalized} covariance matrix for the CMB lensing auto-spectrum. In this setup, uncertainties in the CMB two-point functions are propagated as an additional contribution to the covariance.

\subsection{Inference and minimization}
The priors used in this analysis are presented in Table~\ref{tab:priors}. For cosmological parameters, we generally employ wide flat priors. For the galaxy nuisance parameters, we follow the prescription in Ref.~\cite{Sailer:2024coh} in placing uniform priors on the galaxy bias and Gaussian priors based on the measured values from Ref.~\cite{Zhou:2023gji} on the shot noise and magnification-slope parameters. Following the recent DESI DR2 EDE analysis of Ref.~\cite{Chaussidon:2025npr}, we adopt a Gaussian prior on \(\log_{10} a_c\) centered at the scale factor near matter-radiation equality. This choice reduces prior volume in low-likelihood regions corresponding to EDE transitions that occur much later or much earlier than equality~\cite{DAmico:2020ods,Smith:2020rxx}. Our baseline results without placing the Gaussian prior on \(\log_{10} a_c\) are shown in Appendix~\ref{appendix:ede_prior}, where we find that using a flat prior makes a negligible difference to our conclusions. 

\begin{table*}[t]
\centering
\begin{tabular}{lll}
\hline
\textbf{Parameter} & \textbf{Description} & \textbf{Prior / fixed value}\\
\hline
\multicolumn{3}{c}{\textbf{Base $\Lambda$CDM}}\\
\hline
$\omega_{\rm c}$ & Physical cold dark matter density & $\mathcal{U}(0.08,\,0.20)$ \\
$\omega_{\rm b}$ & Physical baryon density          & $\mathcal{U}(0.01875,\,0.02625)$ \\
$\ln(10^{10}A_{\rm s})$                  & Primordial amplitude            & $\mathcal{U}(2.5,\,3.5)$ \\
$n_{\rm s}$                              & Scalar spectral index           & $\mathcal{U}(0.84,\,1.10)$ \\
$h$                                      & Hubble parameter                & $\mathcal{U}(0.4,\,1.0)$ \\
\hline
\multicolumn{3}{c}{\textbf{LSS nuisance (galaxy / baryonic)}}\\
\hline
$b_{g,i}$ $(i=1..4)$                     & Linear galaxy bias per bin      & $\mathcal{U}(0.1,\,5.0)$ \\
\multirow{4}{*}{$\mathrm{SN}^{2\mathrm{D}}(z_i)$} & \multirow{4}{*}{Projected shot noise (per bin)} 
  & bin 1: $\mathcal{N}(4.07,\,1.22)$ \\ 
& & bin 2: $\mathcal{N}(2.25,\,0.675)$ \\
& & bin 3: $\mathcal{N}(2.05,\,0.615)$ \\
& & bin 4: $\mathcal{N}(2.25,\,0.675)$ \\
\hline
\multirow{4}{*}{$s_\mu(z_i)$} & \multirow{4}{*}{Number-count (magnification) slope (per bin)} 
  & bin 1: $\mathcal{N}(0.972,\,0.0972)$ \\
& & bin 2: $\mathcal{N}(1.044,\,0.1044)$ \\
& & bin 3: $\mathcal{N}(0.974,\,0.0974)$ \\
& & bin 4: $\mathcal{N}(0.988,\,0.0988)$ \\
$\log_{10} T_{\rm AGN}$                  & Baryonic feedback amplitude     & $\mathcal{U}(7.1,\,8.5)$ \\
\hline
\multicolumn{3}{c}{\textbf{Extended cosmology}}\\
\hline
$\log_{10} a_c$                          & EDE critical scale factor       & $\mathcal{N}(-3.531,\,0.10)$\,$^\ddagger$ \\
$f_{\rm EDE}(a_c)$                       & EDE fraction at $a_c$           & $\mathcal{U}(0,\,0.8)$ \\
$\theta_i$                               & Initial field displacement      & $\mathcal{U}(0.1,\,\pi)$ \\
$w_0$                                    & DE equation-of-state (today)    & $\mathcal{U}(-3.0,\,0.5)$ \\
$w_a$                                    & Time variation of DE EoS        & $\mathcal{U}(-3.0,\,2.0)$ \\
$M_\nu=\sum m_\nu$            & Sum of neutrino masses          & $\mathcal{U}(0,\,1.5)\,\mathrm{eV}$\,$^\dagger$ \\
\hline
\end{tabular}

\vspace{2mm}
\par\small \textit{Notes:}   
$^\dagger$\,When not varied the sum of neutrino masses is fixed to $M_\nu=0.06$eV.\\
$^\ddagger$\,Gaussian prior on $\log_{10}a_c$ as described in the text (see also Ref.~\cite{Chaussidon:2025npr}).

\caption{\textbf{Parameters varied in the analysis and associated priors.}
Unless stated otherwise, priors are flat and uninformative.
CMB-foreground nuisance parameters in the ACT DR6 and \hillipop\ likelihoods are \emph{not} listed; for those we follow the respective collaboration prescriptions.
Shot-noise and magnification-slope priors follow \cite{Sailer:2024coh} as described in the text.}
\label{tab:priors}
\end{table*}

We use the \texttt{emcee} package~\cite{Foreman-Mackey:2012any} for performing cosmological inference. We leverage \texttt{JAX}'s natural vectorization via \texttt{vmap} and GPU acceleration, to greatly accelerate our inference. Running on a single NVIDIA A100 Tensor Core GPU allows for converged chains in under four hours for every analysis set-up explored in this work with most taking under 30 minutes (see also~\cite{Reeves:2025bxn} for a similar approach). We deem our MCMC chains to be converged by verifying that the autocorrelation time $\tau$~\cite{Foreman-Mackey:2012any} stabilized and that each walker satisfied $N_{\text{steps}}>100\,\tau$ for all parameters. 

To find the best-fit parameters and $\chi_{\text{min}}^2$ values, we first use a simulated annealing approach~\cite{Herold:2024enb, Schoneberg:2021qvd, Hannestad:2000wx}. That is, we first sample the full posterior with \texttt{emcee}, then \emph{whiten} the parameter space by rescaling each direction by its posterior standard deviation.  Twelve starting points are drawn from a small Gaussian ball ($0.1\sigma$ in every dimension) around the posterior mode.  From each starting point, we launch a simulated-annealing chain whose acceptance probability at an effective temperature~$T$ is
\begin{equation}
P_{\mathrm{acc}}(\theta\!\to\!\theta')=\min\!\left[1,
\exp\!\bigl(-\tfrac{\Delta E}{T}\bigr)\right],
\end{equation}
where $\Delta E = E(\theta')-E(\theta)$ and $E=-\ln\mathcal{L}$.  The temperature is lowered in four linear steps from $T=0.75$ to $T=0.01$ and run a maximum of $2.5\times10^4$ steps at each rung, transitioning to the next rung when this limit is reached or earlier if the best-fit does not improve in a window of $200$ steps. After the annealing ladder is finished, the best candidate from each chain is further optimized using \texttt{ADAM}~\cite{adamxyz}, a first-order gradient-based stochastic optimizer, which is implemented in the \texttt{optax} package in \texttt{JAX}~\cite{deepmind2020jax}. We checked that the typical spread between the minimum $\chi^2$ of the twelve different starting points is $\sigma ( \chi_{\rm min}^2) \lesssim0.2$ in each case. 

\subsection{Assessing model preference using Wilks' theorem}

Following the methodology of recent DESI analyses (see e.g.~\cite{DESI:2025dr2}), we assess model preference between $\Lambda$CDM and extensions using Wilks' theorem. Two models are said to be \emph{nested} when the simpler model is obtained as a special case of the extended one for specific parameter values. In our case, taking $f_{\mathrm{EDE}}\to0$ and $(w_0, w_a)\to(-1,0)$ for the EDE and DDE models respectively recovers the $\Lambda$CDM predictions. Wilks' theorem~\cite{wilks_theorem} states that, under the assumptions of large sample size, regularity of posteriors, and that the simpler (nested) model is the true underlying model, the test statistic
\begin{equation}
    \Delta \chi^2 = -2 \ln \left( \frac{\mathcal{L}_{\Lambda\mathrm{CDM}}}{\mathcal{L}_{\mathrm{ext}}} \right)
\end{equation}
asymptotically follows a chi-squared distribution with degrees of freedom equal to the difference in the number of free parameters between the two models. We can then assess model preference for the extended model by quantifying how unlikely it would be to observe the measured decrease in $\chi^2$ if $\Lambda$CDM were the true underlying model.

\section{Theory models}\label{sec:theory_models}
In this work, we explore two extensions of the dark energy sector that can help to reconcile CMB and BAO data. 

\subsection{EDE}
The simplest EDE model comprises an ultra-light scalar field displaced from the minimum of its corresponding potential. The field is initially frozen by Hubble friction, leading to an enhanced expansion rate in the pre-recombination Universe. Once the Hubble rate drops below it effective mass, the field rolls down the potential and oscillates around its minimum. The canonical EDE model features a pseudoscalar (axion-like) field with the following potential:
\begin{eqnarray}
V(\phi) = m^2f^2 \left [ 1-\cos \left ( \frac{\phi}{f} \right ) \right ] ^n\,,
\label{eq:potential}
\end{eqnarray}
where $m$ and $f$ are the mass and decay constant of the EDE field~\footnote{For examples of other EDE(-like) models, see~\cite{McDonough:2021pdg, Niedermann:2021ijp, Zumalacarregui:2020cjh, Oikonomou:2020qah, Ye:2021iwa}.}. With this choice of potential, once the field becomes dynamical, it decays as a fluid with an effective equation of state $\langle w_{\phi} \rangle = (n-1)/(n+1)$. Throughout this work we fix $n=3$, which has been shown to produce a cosmologically viable model~\cite{Smith:2019ihp, Poulin:2018dzj}. This choice means the energy density will dilute at a rate, $\rho_{EDE}\propto(1+z)^{4.5}$, that is faster than matter, meaning the effect of EDE is localized in time to the period just prior to recombination. 

The particle physics parametrization $m$ and $f$ can be traded for the phenomenological parameters $f_{\rm EDE}$ and $a_c$: at scale factor $a_c$, EDE's fractional contribution to the energy density, $f_{\rm EDE}$, is maximal. The physics of our EDE model is then governed by four parameters: $f_{\rm EDE}$, $a_c$, $n$, and the initial misalignment angle $\theta_i = \phi_i/f$, with $\phi_i$ the initial field value. Increasing \(f_{\rm EDE}\) raises the early-Universe expansion rate \(H(z)\) and thereby reduces the characteristic sound horizon scales (both the BAO ruler at the drag epoch, \(r_{\rm drag}\), and the sound horizon at recombination, \(r_s(z_*)\)). The CMB very precisely measures the acoustic angular scale,
\begin{equation}
    \theta_* \equiv \frac{r_s(z_*)}{D_A(z_*)}\,,
\end{equation}
where \(D_A(z_*)\) is the angular diameter distance to the surface of last scattering. Consequently, if \(r_s(z_*)\) decreases, maintaining the observed \(\theta_*\) requires a smaller \(D_A(z_*)\), which in turn drives a higher inferred \(H_0\) (for fixed physical densities). We adopt the EDE model implementation in the modified \texttt{CLASS}~\cite{Blas:2011rf} code \texttt{AxiCLASS}\footnote{https://github.com/PoulinV/AxiCLASS}~\cite{Smith:2019ihp, Poulin:2018dzj} for deriving the numerical results reported here.

\subsection{Dynamical dark energy} 

We consider a dynamical dark energy model with the Chevallier--Polarski--Linder (CPL) form~\cite{Chevallier:2000qy,Linder:2002et} 
\begin{equation}
    w(a)=w_0+w_a(1-a),
\end{equation}
where \(w_0\) is the present-day equation of state and \(w_a\) describes its linear redshift evolution. Theoretically, CPL can be thought of as a first-order expansion about \(a=1\) that provides a controlled, accurate approximation to broad classes of smooth quintessence dynamics near the present epoch: for canonical scalar fields with either thawing or freezing evolution, the late-time distance and linear-growth histories over \(0\lesssim z\lesssim 2\) are well captured by the CPL parametrization~\cite{Linder:2002et}. This is further motivated by the limits of current data; a recent DESI DR2 extended analysis, which analyzed non-parametric reconstructions (such as binning and Gaussian processes) for the evolution of $w(z)$, found consistent behavior between methods and concluded that the two-parameter \(w_0w_a\) description captures the dominant late-time evolution constrained by these data~\cite{DESI:2025fii}. 

\section{Results}\label{sec:Results}

\subsection{$\Lambda \mathrm{CDM}$}

\begin{figure}
    \centering
    \includegraphics[width=0.9\linewidth]{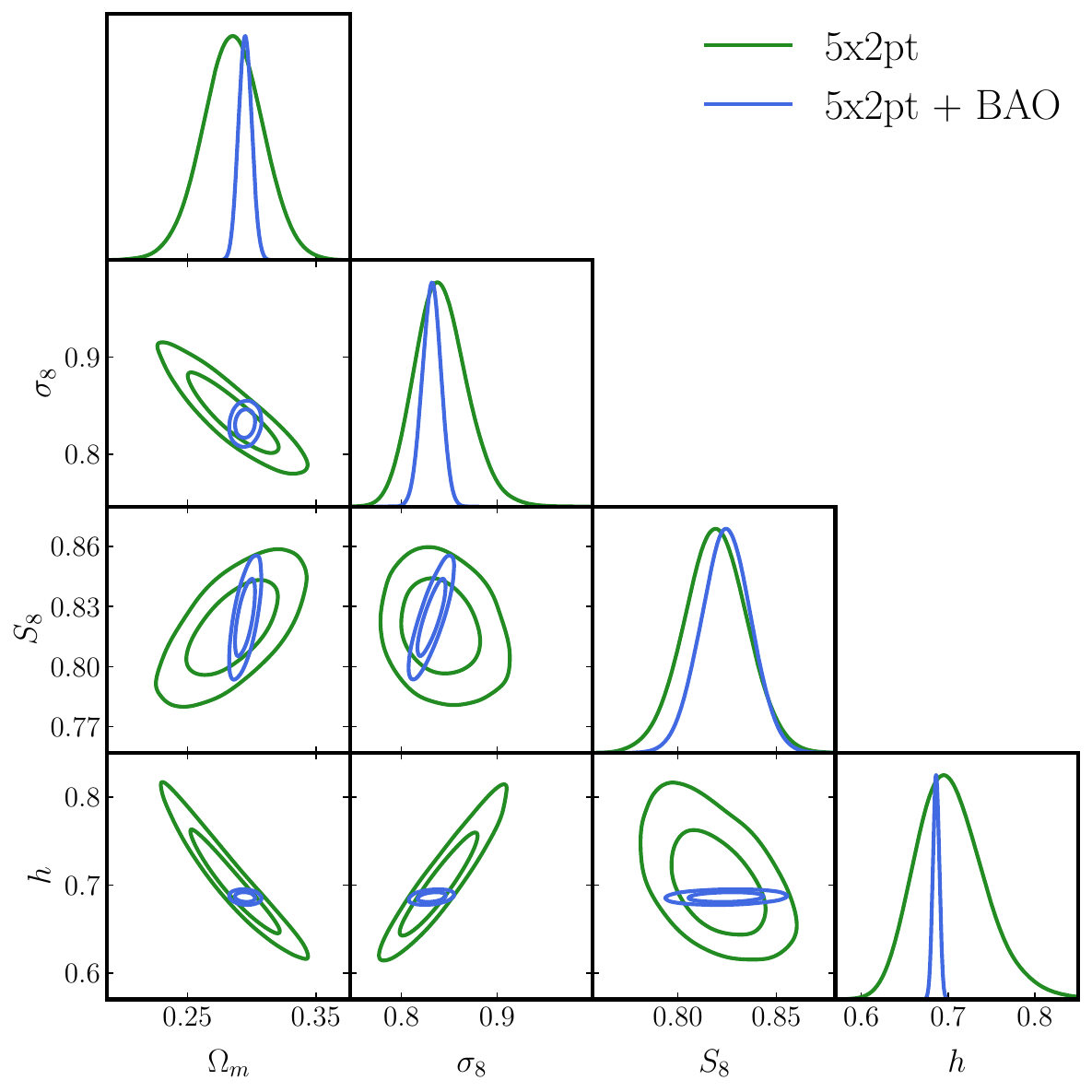}
    \caption{\textbf{Contours for $\Lambda \mathrm{CDM}$ model with LSS data.} \emph{Green contours}: $5\times2$pt refers to the combination of ACT DR6 lensing, LS LRGs clustering and \textit{Planck} ISW. \emph{Blue contours}: same $5\times2$pt combination as before with added DESI DR2 BAO data. Note for the constraints shown here we fix $n_s=0.9649$, and $\omega_b = 0.02236$.}
    \label{fig:lcdm_contours}
\end{figure}

We display the $\Lambda$CDM constraints of our $5\times2$pt LSS data combination in Figure~\ref{fig:lcdm_contours}. From the $5\times 2$-point data vector alone we obtain
\begin{equation}
S_8 \equiv \sigma_8\sqrt{\Omega_m/0.3} \;=\; 0.819 \pm 0.016,
\end{equation}
which is in excellent agreement with both the \textit{Planck} PR4 \hillipop\ value, $S_8 = 0.819 \pm 0.014$~\cite{Tristram:2023haj}, and the ACT~DR6 \texttt{P-ACT} value,
$S_8=0.830 \pm 0.014$~\cite{ACTDR6:PowerSpectra}. Our measurement also agrees with the latest KiDS-Legacy weak-lensing result, $S_8=0.815\pm0.012$~\cite{Wright:2025xka}, as well as with a recent multi-probe ($3\times2$pt) analysis combining \textit{Planck} and ACT~DR6 lensing with unWISE galaxies, which found $S_8=0.816\pm0.015$~\cite{Farren:2024rla}. Our result is also consistent (within $1\sigma$) with a recent DESI-DR1 joint 3D clustering $+$ LS galaxy$\times$CMB-lensing analysis which finds $S_8=0.808\pm0.017$~\cite{Maus:2025rvz}.

We further compare our results to several recent analyses using the 2$\times$2pt subset of our data vector (CMB-lensing$\times$galaxy and galaxy auto), which excludes the ACT
auto-spectrum. Ref.~\cite{Sailer:2024coh} reports $S_8 = 0.790^{+0.024}_{-0.027}$ from the \emph{ACT-only} cross and $S_8 = 0.775^{+0.019}_{-0.022}$ when combining ACT with
\textit{Planck} lensing (see also the companion paper Ref.~\cite{Kim:2024dmg} who also find a lower amplitude of fluctuations compared to CMB primary inferences). Ref.~\cite{Qu:2024sfu}, using a DESI LS catalogue that was \emph{not} calibrated using DESI Spectra from~\cite{Hang:2020gwn}, finds $S_8=0.772\pm0.040$ for ACT-only lensing and $0.765\pm0.032$ for ACT+\textit{Planck} lensing. In Appendix~\ref{appendix:comapre_to_noah} we show that the \emph{2$\times$2pt} subset of our pipeline yields fully consistent constraints with the publicly available chains from Ref.~\cite{Sailer:2024coh}. The multiprobe constraints therefore yield a slightly higher $S_8$ inference at the $\sim 1-1.5\sigma$ level compared to the $2\times2$pt analysis, especially when the latter are derived with the ACT+\textit{Planck} lensing map. This is consistent with the general finding that the ACT DR6 lensing auto- power-spectrum analysis favors an $S_8$ close to the primary-CMB prediction (the ACT collaboration finds $S_8=0.840\pm0.028$ when combining ACT DR6 lensing with BAO measurements from SDSS and 6dF~\cite{Madhavacheril:2023}), so including it alongside galaxy clustering lifts the central value of the joint $S_8$ relative to the $2\times2$pt constraint.

Note for the above results and Fig.~\ref{fig:lcdm_contours}, we fix the poorly constrained baryon density and spectral tilt to their respective \textit{Planck} best-fit values~\cite{Planck:2018vyg} ($n_s=0.9649$ and $\omega_b = 0.02236$). If we instead allow $\omega_b$ and $n_s$ to vary with a flat prior from Table~\ref{tab:priors}, we find a slightly broader constraint of $S_8=0.827\pm0.020$. Adding the DESI BAO likelihood, keeping $n_s$ and $\omega_b$ fixed as before, sharpens the constraints by providing a strong constraint on $\Omega_m$ and hence breaking degeneracy, yielding
\begin{align}
\sigma_8 &= 0.832 \pm 0.010, \\
\Omega_m &= 0.295 \pm 0.005.
\end{align}
Overall, these results demonstrate the power of combining CMB lensing with LSS tracers to produce competitive cosmological constraints without recourse to primary-CMB anisotropies. The results suggest that the (mostly linear) scales and redshifts probed by our analysis do not exhibit the `lensing is low' trend seen in some previous galaxy weak lensing cross-correlation studies~\cite{Leauthaud:2016jdb, DES:2021wwk, Heymans:2020gsg, Sugiyama:2023fzm}.

\subsubsection{Sound horizon free $H_0$}
\begin{figure}
    \centering
    \includegraphics[width=\linewidth]{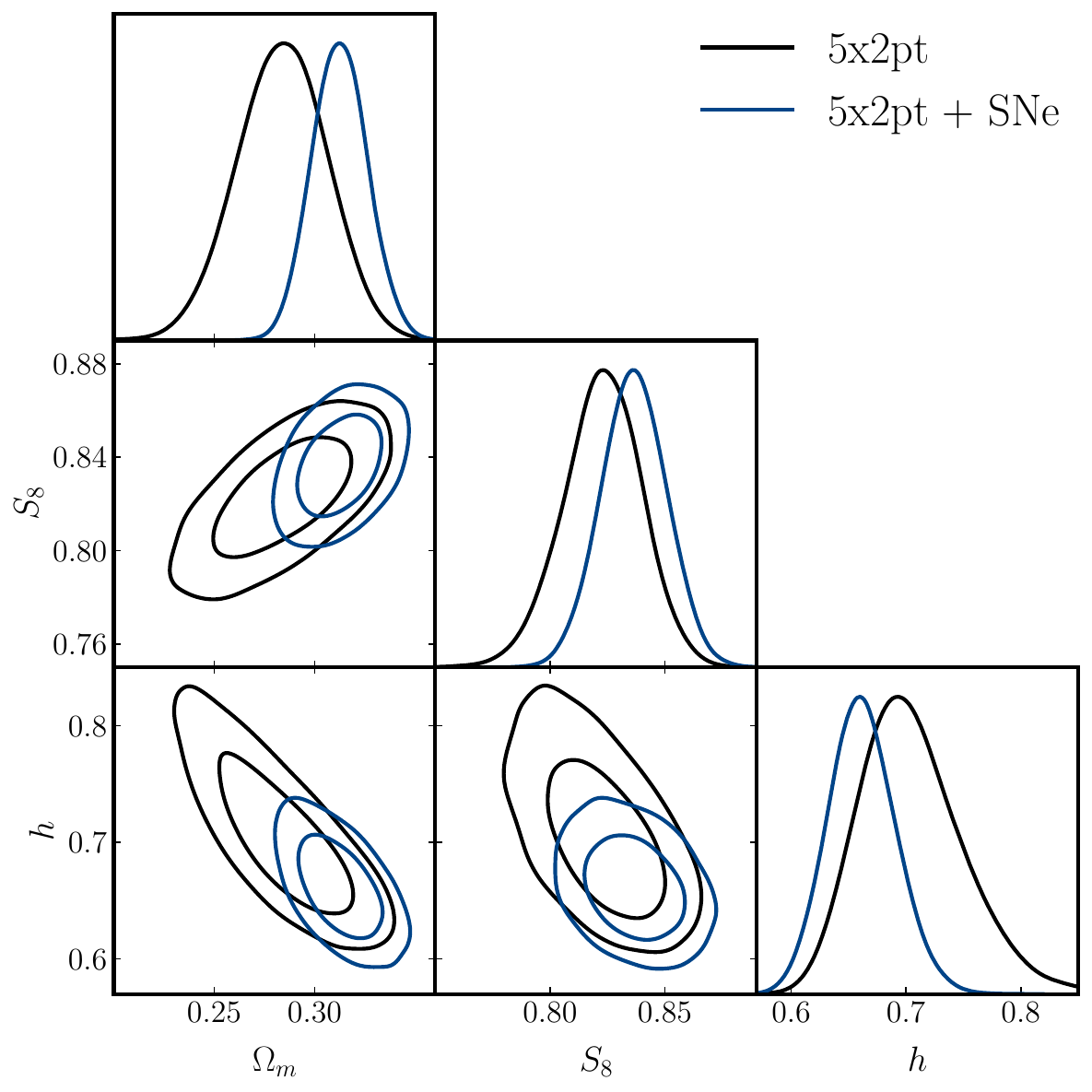}
    \caption{\textbf{Sound horizon free $H_0$ measurements.} The $5\times2$pt+SNe contour combines the Pantheon+ SNe dataset with our LSS data combination. A Gaussian prior is placed on the spectral tilt and $\omega_b$ is varied in a flat uniform prior as described in the text.}
    \label{fig:sounds_horizon_free_h}
\end{figure}
With our $5\times2$pt dataset alone, we are able to weakly constrain the Hubble parameter:
\begin{equation}
H_0 \;=\; 70.0 \pm 4.0 {\rm km\,s^{-1}\,Mpc^{-1}},
\end{equation}
this broad constraint is consistent within $1\sigma$ with both the value of $H_0$ preferred by CMB primary inferences and local measurements of $H_0$ from SH0ES~\cite{Riess:2021jrx}. We note that this constraint is found assuming a fixed $n_s=0.9649$ and $\omega_b=0.02236$, when instead allowing $\omega_b$ to vary within a wide flat prior from Table~\ref{tab:priors} and imposing a conservative $n_s$ prior centered on, but ten times broader than, the \textit{Planck} constraint~\cite{Planck:2018vyg} ($n_s \sim \mathcal{N}(0.9649, 0.04)$), we find a slightly broader constraint of
\begin{equation}
H_0 \;=\; 70.0 \pm 4.4 {\rm km\,s^{-1}\,Mpc^{-1}}.
\end{equation}
This determination is free of sound horizon information as BAO features are washed out when integrating over the projected kernels,\footnote{See Appendix~\ref{appendix:bao_information} for a detailed investigation into the (negligible) residual BAO information in the $5\times2$pt combination.} the constraint comes instead from projecting the matter-radiation equality scale $k_{eq}$ at different redshifts~\cite{Zaborowski:2024car}. Consistent with this picture, we find our $H_0$ constraint to be insensitive to the prior placed on $\omega_b$: fixing $\omega_b=0.02236$ leaves the $H_0$ results unchanged. This indicates the information is coming from $k_{eq}$ and distances and not $r_s$, which depends strongly on $\omega_b$. We can further combine with Pantheon+ SNe data to break some of the $\Omega_m-h$ residual degeneracy, yielding
\begin{equation}
H_0 \;=\; 66.2 \pm 3.1 {\rm km\,s^{-1}\,Mpc^{-1}}.
\end{equation}
These results serve as a cross-check of $H_0$ constraints from CMB and BAO data, which rely on the $r_s$\footnote{Throughout this discussion we do not carefully distinguish between the sound–horizon scales used in BAO and CMB analyses, denoted \(r_d\) and \(r_s\), respectively. Strictly speaking, \(r_s\) refers to the sound horizon at photon decoupling, while \(r_d\) is defined as the sound horizon at the end of the baryon drag epoch, which occurs slightly after photon decoupling.} calibration scale. This also has implications for models that aim to resolve the Hubble tension by reducing the sound horizon \(r_s\) as this constraint will be unaffected by such mechanisms. However, with our current data, the precision is not yet sufficient to make decisive claims, as our constraints are compatible with both the CMB-based determinations of $H_0$ and the SH0ES results within $\sim 2\sigma$. Forecasts and other contemporary measurements suggest that upcoming galaxy-survey and CMB-lensing analyses will substantially tighten such \(r_s\)-independent constraints, and hence will be able to decisively weigh in on the Hubble tension~\cite{Farren:2021grl,Philcox:2022sgj, GarciaEscudero:2025lef}.

\subsection{Goodness of fit}
We analyze the goodness of fit of the LSS dataset under the assumption of  $\Lambda \mathrm{CDM}$. We compute the Bayesian probability to exceed (PTE) in $\Lambda \mathrm{CDM}$ using the posterior predictive distribution (PPD). This quantifies how extreme the observed data are compared to synthetic datasets generated from the posterior~\cite{Gelman1996, Meng1994}, providing a test of how well the $\Lambda \mathrm{CDM}$ model fits our datasets. For Gaussian likelihoods, this is computed as ~\cite{Sailer:2024coh}: 
\begin{equation}\label{eqt:pte}
    \mathrm{PTE} \;=\;
    \int \mathrm{d}\theta \,
    \left[
       1 \;-\;
       \frac{\gamma\!\bigl( \tfrac{N_d}{2},\, \tfrac{\chi^2(\theta)}{2}\bigr)}
            {\Gamma\!\bigl(\tfrac{N_d}{2}\bigr)}
    \right]
    P(\theta),
\end{equation}
where $N_d$ is the number of data points, $P(\theta)$ is the posterior, and $\gamma$ and $\Gamma$ are the incomplete and complete Gamma functions, respectively. This formula is only valid for Gaussian likelihoods so when assessing the goodness of fit of combinations with CMB data, we will always remove the low-$\ell$ \textit{Planck} based $TT/EE$ likelihoods. We show the results for the $5\times2$pt combination in Table~\ref{tab:goodness_of_fit}, finding reasonable values for both the $5\times2$pt analysis alone and with the addition of BAO data. When adding the ACT DR6 \texttt{P-ACT} likelihood, we do not find much change in the Bayesian PTE, suggesting that the combined data can be well fit in the $\Lambda$CDM model.\footnote{Note that we do not show the goodness of fit with \textit{Planck} PR4 \hillipop\. This is very low when fit to \textit{Planck} PR4 \hillipop\ likelihood alone, as described in detail in Refs.~\cite{Tristram:2023haj, Reeves:2025axp}. Hence, in combination with BAO and $5\times2$pt this remains very small, and as such it is difficult to interpret if combining the datasets leads to a significantly worse fit.}

\begin{table}[t]
\centering

\begin{tabular}{lccc}
\hline
\textbf{Dataset} & \boldmath$n_{\mathrm{data}}$ & \boldmath$\chi^2_{\mathrm{red}}$ & \textbf{PTE} \\
\hline
$5\times2$pt                & 78 & 0.98 & 0.36 \\
$5\times2$pt + DESI DR2 BAO & 91 & 0.96 & 0.44
\\
\texttt{P-ACT}$^\ddagger$ + $5\times2$pt + DESI DR2 BAO & 481 & 1.0112 & 0.36 \\
\hline
\end{tabular}
\par\small \textit{Notes:}   
$^\ddagger$\, \texttt{P-ACT} here is without additional $\ell<30$ $TT/EE$ \textit{Planck} likelihoods as these are non-Gaussian as described in the text.
\caption{\textbf{Goodness-of-fit for $\Lambda$CDM.} 
We show the reduced $\chi^2$ and corresponding probability-to-exceed (PTE) values for the baseline $5\times2$pt data vector and for the combination with DESI DR2 BAO. In both cases the fit is statistically acceptable, with PTE values in the expected range.}
\label{tab:goodness_of_fit}
\end{table}

\subsection{Extensions meet LSS}
\begin{figure*}[t]
    \centering
    \includegraphics[width=0.65\textwidth]{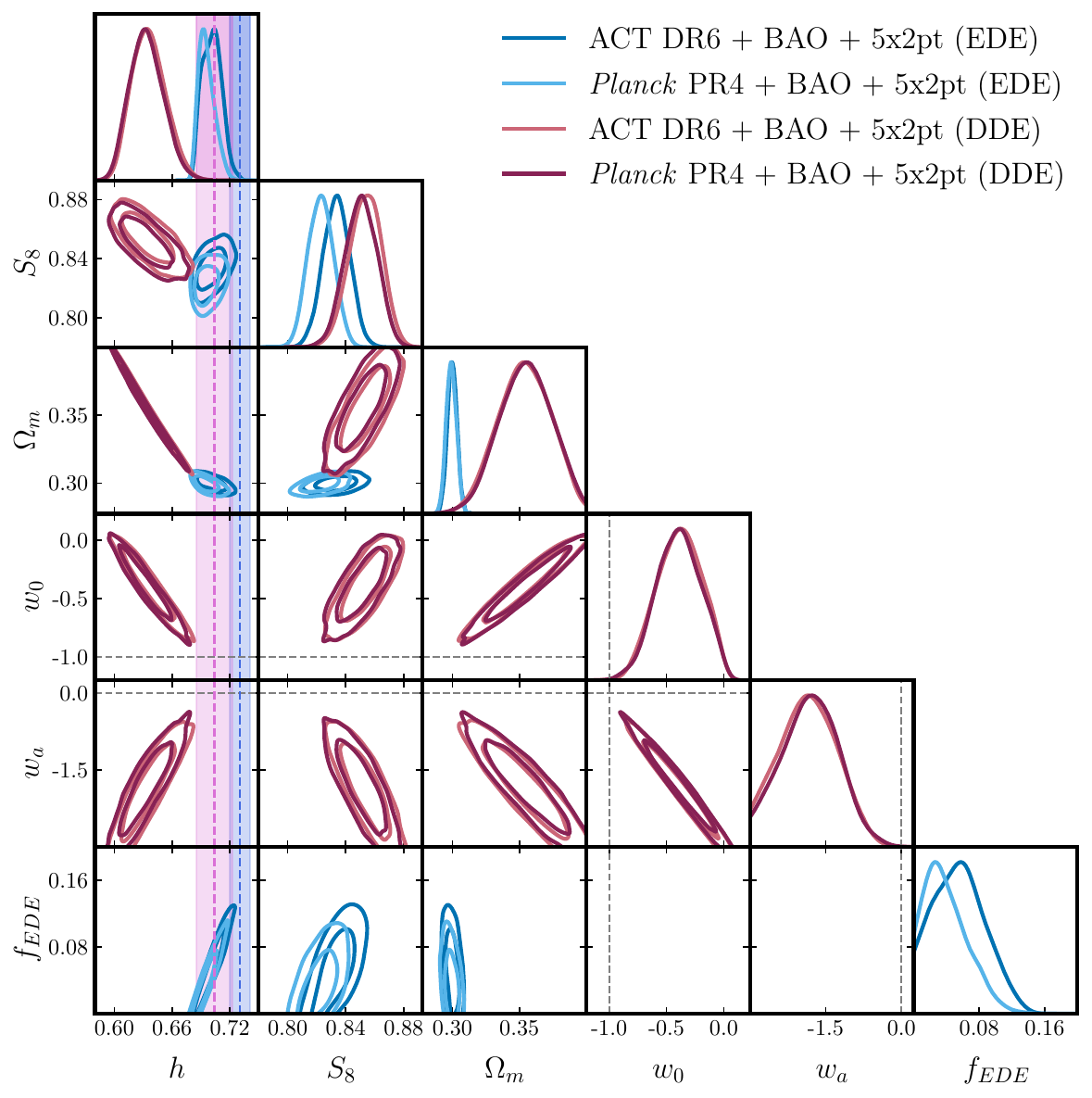}
    \caption{\textbf{Posterior constraints in extended dark–energy models.}
    Shown are posteriors for DDE and EDE from CMB, BAO, and LSS combinations.
    Vertical shaded bands show local $H_0$ constraints, the latest from CCHP ($H_0 = 70.4\pm1.9\,{\rm km\,s^{-1}\,Mpc^{-1}}$)~\cite{Freedman:2024eph} and SH0ES ($H_0 = 73.04\pm1.04\,{\rm km\,s^{-1}\,Mpc^{-1}}$)~\cite{Riess:2021jrx}. Dashed lines at $(w_0,w_a)=(-1,0)$ indicate the $\Lambda$CDM cosmological constant limit.}
    \label{fig:main_contours}
\end{figure*}

\begin{figure*}
    \centering
    \includegraphics[width=0.7\linewidth]{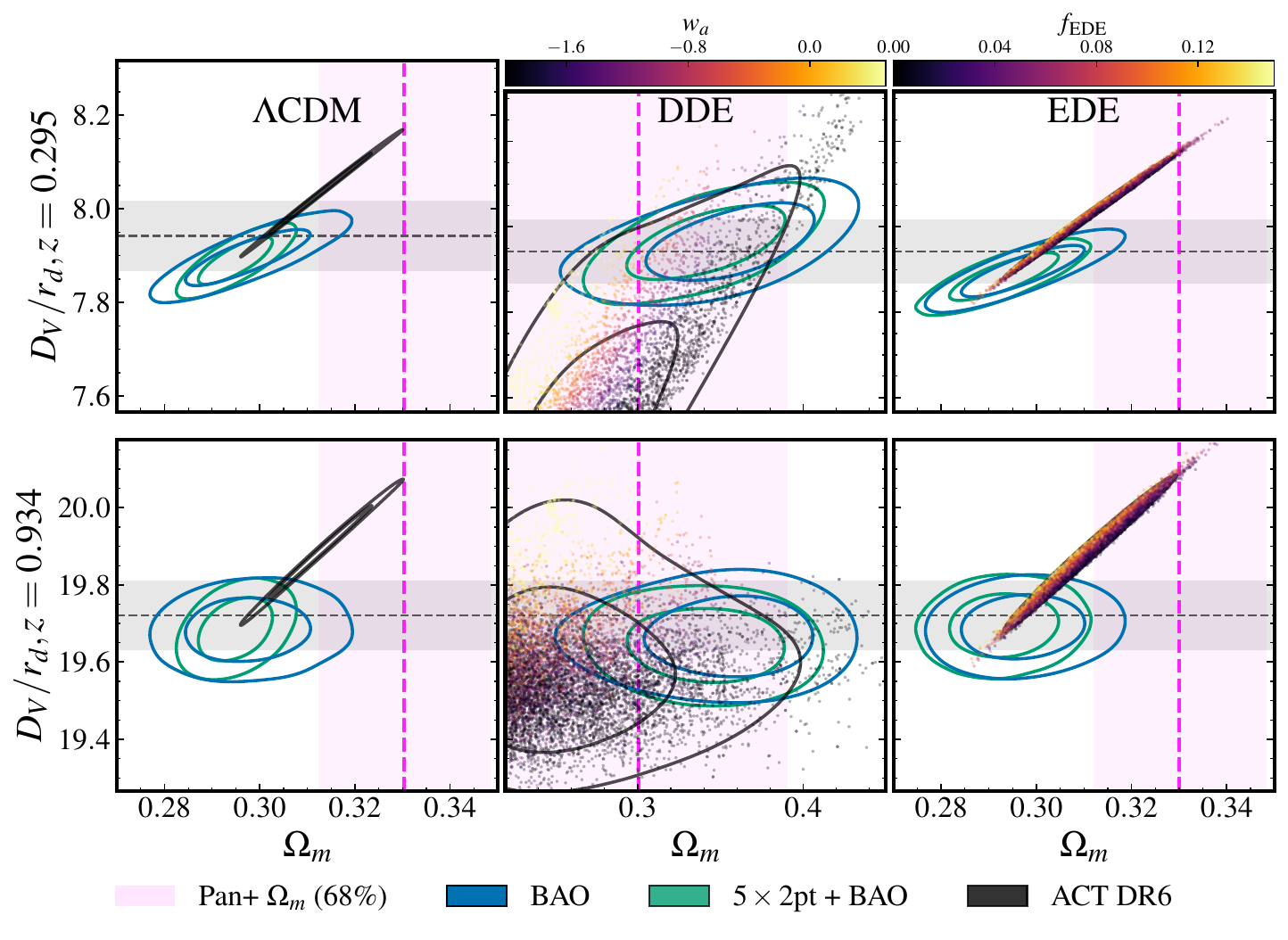}
    \caption{\textbf{Constraints in the $\Omega_m$--$\frac{D_v}{r_d}$ plane for different individual datasets in each of the three models analysed in this work}. The ACT DR6 constraints here are without lensing auto-correlation data which is instead included in the $5\times2$ data combination.}
    \label{fig:tension_plot}
\end{figure*}

\begin{table*}[t]
\centering
\setlength{\tabcolsep}{4pt}
\renewcommand{\arraystretch}{1.1}
\resizebox{\textwidth}{!}{%
\begin{tabular}{llccccccccc}
\hline
\textbf{Model} & \textbf{Dataset} &
\boldmath$H_0$~[km\,s$^{-1}$\,Mpc$^{-1}$] &
\boldmath$\Omega_m$ &
\boldmath$S_8$ &
\boldmath$w_0$ &
\boldmath$w_a$ &
\boldmath$f_{\rm EDE}$ &
\boldmath$\Delta\chi^2$ &
\boldmath$n_\sigma$ \\
\hline
\multirow{4}{*}{$\Lambda$CDM}
 & ACT DR6 $+$ BAO                     & $68.5\pm0.3$ & $0.300\pm0.004$ & $0.808\pm0.008$ & --- & --- & --- & --- & --- \\
 & ACT DR6 $+$ BAO $+$ $5\times2$pt    & $68.3\pm0.3$ & $0.303\pm0.003$ & $0.820\pm0.007$ & --- & --- & --- & --- & --- \\
 & \textit{Planck} PR4 $+$ BAO         & $68.3\pm0.3$ & $0.300\pm0.004$ & $0.805\pm0.009$ & --- & --- & --- & --- & --- \\
 & \textit{Planck} PR4 $+$ BAO $+$ $5\times2$pt  & $68.2\pm0.3$ & $0.303\pm0.004$ & $0.817\pm0.007$ & --- & --- & --- & --- & --- \\
\hline
\multirow{4}{*}{DDE}
 & ACT DR6 $+$ BAO                     & $63.9\pm1.9$ & $0.350\pm0.021$   & $0.849\pm0.023$ & $-0.44\pm0.22$ & $-1.70\pm0.60$ & ---             & $8.9$  & $2.3$ \\
 & ACT DR6 $+$ BAO $+$ $5\times2$pt    & $63.6\pm1.8$ & $0.351\pm0.021$   & $0.854\pm0.011$ & $-0.40\pm0.20$ & $-1.84\pm0.56$ & ---             & $14.6$ & $3.3$ \\
 & \textit{Planck} PR4 $+$ BAO         & $64.0\pm2.0$ & $0.346\pm0.022$   & $0.840\pm0.015$ & $-0.49\pm0.22$ & $-1.50\pm0.63$ & ---             & $6.8$  & $1.8$ \\
 & \textit{Planck} PR4 $+$ BAO $+$ $5\times2$pt  & $63.4\pm1.8$ & $0.360\pm0.021$   & $0.851\pm0.011$ & $-0.40\pm0.20$ & $-1.80\pm0.55$ & ---             & $13.5$ & $3.0$ \\
\hline
\multirow{4}{*}{EDE}
 & ACT DR6 $+$ BAO                     & $70.5\pm1.2$ & $0.299\pm0.004$ & $0.827\pm0.013$ & ---            & ---            & $<0.14$ (95\%CL) ($0.069\pm0.041$) & $8.7$  & $1.8$ \\
 & ACT DR6 $+$ BAO $+$ $5\times2$pt    & $70.1\pm1.0$ & $0.300\pm0.004$ & $0.830\pm0.010$ & ---            & ---            &  $<0.12$ (95\%CL) ($0.057\pm0.033$) & $11.4$ & $2.3$ \\
 & \textit{Planck} PR4 $+$ BAO         & $69.5\pm0.8$ & $0.299\pm0.004$ & $0.813\pm0.011$ & ---            & ---            & $<0.09$ (95\%CL) ($0.037\pm0.027$) & $4.1$  & $0.7$ \\
 & \textit{Planck} PR4 $+$ BAO $+$ $5\times2$pt  & $69.6\pm0.8$ & $0.300\pm0.004$ & $0.826\pm0.009$ & ---            & ---            & $<0.09$ (95\%CL) ($0.041\pm0.026$) & $6.7$  & $1.4$ \\
\hline
\end{tabular}}
\caption{\textbf{Summary of cosmological constraints and model preferences.}
We report marginalized $68\%$\,CL constraints on $H_0$, $\Omega_m$, $S_8$, and, where applicable, $(w_0,w_a)$ for DDE and $f_{\rm EDE}$ for EDE. 
$\Delta\chi^2$ and $n_\sigma$ quantify the improvement relative to the corresponding $\Lambda$CDM fit using Wilks’ theorem. 
Dashes indicate parameters not defined for a given model.
CMB\,$+$\,BAO entries exclude CMB lensing; the $5\times2$pt rows add LRG$\times\kappa_{\rm CMB}$, $\kappa_{\rm CMB}\kappa_{\rm CMB}$, and ISW. We report both the $95\%$ confidence intervals and the mean $\pm\sigma$ for the EDE constraints.}
\label{tab:main_results}
\end{table*}

The constraints for the two extended dark energy models are collected in Table~\ref{tab:main_results} and displayed in Fig.~\ref{fig:main_contours}.  
The novelty of our analysis lies in the inclusion of the $5\times2$pt LSS data vector; relative to the already well-studied CMB\,+\,BAO baseline, this extra information strengthens the preference for extensions beyond $\Lambda$CDM. For DDE, adding $5\times2$pt to \texttt{P-ACT}$+$BAO raises the improvement in the global fit with respect to $\Lambda \mathrm{CDM}$ from $\Delta\chi^2=8.9$ ($n_\sigma=2.3$) to $\Delta\chi^2=14.3$ ($n_\sigma=3.2$), while for the equivalent combinations for \hillipop\ the gain is from $\Delta\chi^2=6.8$ ($1.8\sigma$) to $\Delta\chi^2=13.47$ ($3.0\sigma$). For the combination with \texttt{P-ACT} this represents the highest SNe-independent preference for DDE found to date, slightly higher than the $3.1\sigma$ found by Ref.~\cite{DESI:2025dr2} for the combination of \textit{Planck} PR4 (\texttt{CamSpec} version) and DESI DR2 BAO. The inclusion of the  $5\times2$pt likelihood also \emph{increases} the preference for EDE relative to CMB\,+\,BAO alone, albeit more modestly than for DDE. For \texttt{P-ACT}$+$BAO the EDE improvement rises from $\Delta\chi^2=8.7$ ($1.8\sigma$) to $11.4$ ($2.3\sigma$) when $5\times2$pt is added; for \hillipop$+$BAO it grows from $4.1$ ($0.7\sigma$) to $6.7$ ($1.4\sigma$). 

The mechanisms underlying these statistical improvements can be understood through examination of Fig.~\ref{fig:tension_plot}, which projects posteriors for various datasets onto the $\Omega_m$--$D_v/r_d$ plane at two redshifts where DESI makes BAO measurements ($z = 0.295$ and $z = 0.934$). Within the $\Lambda$CDM framework, the figure demonstrates an offset between CMB-derived constraints (Planck PR4, ACT DR6) and BAO+LSS measurements, manifesting as a $\sim2\sigma$ discrepancy in the inferred matter density and a smaller shift in the preferred value of $\frac{D_v}{r_s}$~\footnote{See also the recent SPT-3G D1 analysis, which finds a $2.8\sigma$ discrepancy in the $\Omega_m$-$hr_d$-plane.} when combining SPT-3G, \textit{Planck}, and ACT data~\cite{SPT-3G:2025bzu}). The $5\times2$pt LSS data vector (green contours in Fig.~\ref{fig:tension_plot}) functions as an independent probe that tightens constraints around the BAO-preferred region, thereby amplifying the statistical preference for extended models that can reconcile this discrepancy. The DDE model (middle panel) operates by freeing the dark energy equation of state, which primarily modifies $D_v$ while leaving $r_d$ unchanged, enabling the model to reconcile CMB and BAO constraints through altered expansion history. The colored points show samples from the posterior and demonstrate that models further from a cosmological constant (with $w_0>-1$, $w_a<0$) bring the CMB inference close to the preferred values from BAO+LSS data. 
The EDE model (right panel) can partially resolve this discrepancy through a geometric mechanism. EDE reduces the sound horizon scale $r_s$ at recombination; to maintain a good fit to the observed CMB acoustic peaks necessitates an increase in the Hubble parameter $H_0$. This elevated $H_0$ produces a fractional decrease in the angular diameter distance $D_v$ to BAO positions that exceeds the fractional decrease in $r_d$ itself. Consequently, models with higher EDE fractions (as shown in the brighter scatter points) systematically shift the CMB-preferred region in the $\Omega_m$--$D_v/r_d$ plane laterally toward the BAO measurements, as visible in the rightmost panel where EDE contours extend toward the BAO-preferred parameter space.
Both extended models also modify the CMB-inferred $\Omega_m$ allowing for lower inferred values compared to $\Lambda$CDM which further lowers the discrepancy between BAO and CMB constraints. The purple shaded regions in Fig.~\ref{fig:tension_plot} represent the $\Omega_m$ constraint from Pantheon+ SNe when fit in the respective models. In $\Lambda$CDM and EDE these data tend to prefer a higher $\Omega_m$ than the $5\times2$pt + BAO contours, whilst the constraint is significantly broadened and shifted to lower values in DDE. We will revisit the implications of this for model preference in the next section. 

Our constraint for the combination with ACT DR6 data of $f_{EDE}<0.12$ exactly matches the corresponding constraint derived by the ACT collaboration for the combination with DESI Y1 BAO and ACT DR6 lensing~\cite{ACT:2025tim}. However, as in the DDE scenario, the inclusion of the $5\times2$pt dataset strengthens the preference for the EDE extension over $\Lambda\mathrm{CDM}$ to a higher significance than previously found in analyses of combinations of data with ACT DR6 primary data \footnote{We note that our results display a lower preference for the EDE scenario compared to combinations of data with the older ACT DR4 likelihood. This can be traced to a fluctuation in the EE power spectrum of ACT DR4 around $\ell\sim500$ and a broad feature in the TE spectrum of \textit{Planck} and ACT DR4 both of which could be slightly better fit in the EDE scenario compared to $\Lambda$CDM~\cite{Hill:2021yec}.} when not including additional SH0ES $H_0$ priors (which we will explore later). Specifically, while previous combinations of ACT DR6 primary data with CMB lensing and BAO measurements found $\Delta\chi^2 \sim 6$--$7$~\cite{Poulin:2025nfb, ACT:2025tim} in favor of EDE, our analysis incorporating the $5\times2$pt LSS data yields $\Delta\chi^2 = 11.4$. We note that a recent SPT-3G analysis also found a mild preference for EDE when combining SPT-3G\footnote{See also Ref.~\cite{LaPosta:2021pgm} who found a mild preference for EDE when not including the full \textit{Planck} data in an earlier release of SPT data.}, \textit{Planck}, and ACT-DR6 primary data with DESI BAO data~\cite{SPT-3G:2025vyw}, finding $\Delta \chi^2=10.8$ with respect to $\Lambda$CDM fit to the same combination of data, which is marginally lower than the $\Delta \chi^2=11.4$ we find for our baseline combination with \texttt{P-ACT}. 

The EDE fits consistently push the Hubble constant to $H_0\simeq70\,{\rm km\,s^{-1}\,Mpc^{-1}}$, moving in the direction of the local distance ladder determination by SH0ES ($H_0=73.04 \pm 1.04\rm km\,s^{-1}\,Mpc^{-1}$~\cite{Riess:2021jrx}) and thereby reducing the tension relative to $\Lambda$CDM to $\sim 2 \sigma$, and the results are consistent with the CCHP value ($H_0 = 70.4\pm1.9\,{\rm km\,s^{-1}\,Mpc^{-1}}$)~\cite{Freedman:2024eph} of the Hubble constant. By contrast, for DDE we see a shift in the central value of $H_0$ to lower values, although the size of the contour is significantly broadened so the 1D Gaussian tension with SH0ES remains around the $\Lambda\mathrm{CDM}$ value of $\sim 4.5\sigma$. We will revisit this point of difference in reconciling the late-time distance ladder $H_0$ measurements in the next section. 

For the DDE model, the results are remarkably robust to the choice of CMB primary likelihood used with \texttt{P-ACT} preferring only a slightly larger deviation from $\Lambda\mathrm{CDM}$ when analyzed under Wilks' theorem, and the contours are almost indistinguishable in Fig.~\ref{fig:main_contours}. However, the preference for EDE is clearly larger in the ACT DR6 combinations than in those based on \textit{Planck} PR4, and this is also reflected in the contours. which display a larger preference for non-zero $f_{EDE}$. These results mirror those in  Ref.~\cite{Poulin:2025nfb}, who analyzed the CMB + DESI DR2 combination and found that ACT DR6 allows a broader range of nonzero $f_{\rm EDE}$ than \textit{Planck} PR4\footnote{Note that this analysis used the \texttt{CamSpec} version of the \textit{Planck} PR4 likelihood~\cite{Rosenberg:2022sdy}.}. At the level of this shift in model preference ($1.4\sigma$ for \hillipop\ combinations vs. $2.3\sigma$ for \texttt{P-ACT}) this may be simply due to statistical fluctuations, or a hint of a small systematic offset between the two CMB experiments; we do not further investigate this in this work. We also note that Ref.~\cite{McDonough:2023qcu} showed that the updated \textit{Planck} PR4 likelihoods place tighter limits on $f_{EDE}$ compared to the older \textit{Planck} PR3 likelihoods. 

The results in this section demonstrate the impact of including our novel $5\times2$pt LSS data vector when analyzing dark energy model extensions. As we have understood through examining Fig.~\ref{fig:tension_plot}, this generally bolsters the preference for models that can help to resolve CMB-BAO discrepancies present in $\Lambda$CDM. In the following sections, we will further examine these extended models, first in the context of a free neutrino mass and then with several external datasets.

\subsection{Neutrino masses}
\begin{figure}
    \centering
    \includegraphics[width=0.9\linewidth]{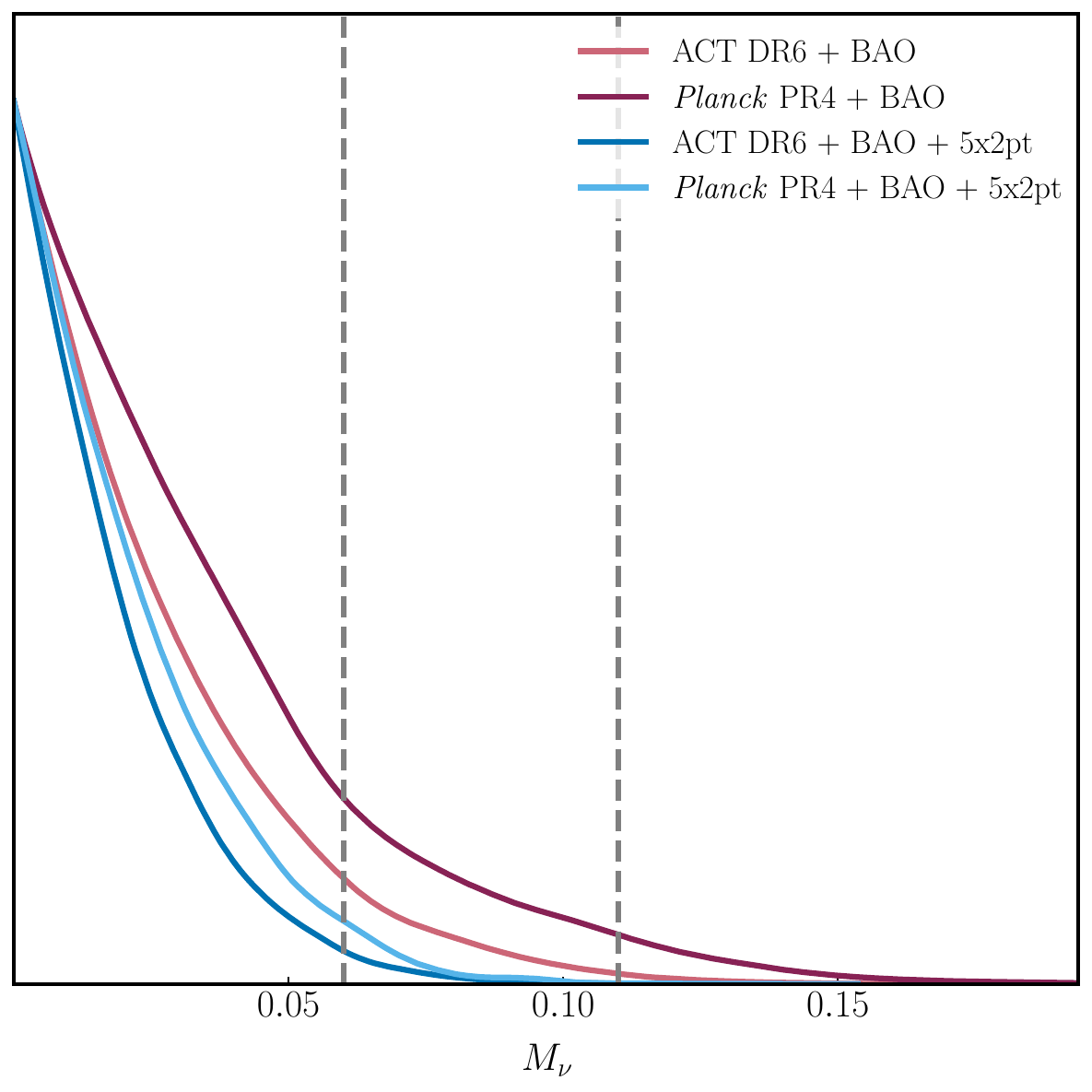}
    \caption{\textbf{Neutrino mass constraints in $\Lambda$CDM for CMB+BAO comapred with CMB+BAO+$5\times2$pt}. The lower limits implied by the normal ordering (NO) and inverted ordering (IO) are shown in the grey vertical bands in order from left to right.}
    \label{fig:mnu_contours_lcdm}
\end{figure}

\subsubsection{Neutrino masses in $\Lambda$CDM}

\begin{table*}[t]
\centering
\setlength{\tabcolsep}{6pt}
\renewcommand{\arraystretch}{1.2}
\begin{tabular}{lc}
\hline
\textbf{Dataset + Model} & \textbf{$M_\nu$ [eV]} \\
\hline
ACT DR6 $+$ BAO $+$ $5\times2$pt ($\Lambda$CDM)     & $<0.050$ \\
\textit{Planck} PR4 $+$ BAO $+$ $5\times2$pt ($\Lambda$CDM) & $<0.056$ \\
ACT DR6 $+$ BAO $+$ $5\times2$pt (EDE)              & $<0.088$ \\
\textit{Planck} PR4 $+$ BAO $+$ $5\times2$pt (EDE)  & $<0.094$ \\
ACT DR6 $+$ BAO $+$ $5\times2$pt (DDE)         & $0.081 \pm 0.052$ \,(95\% CL: $<0.171$) \\
\textit{Planck} PR4 $+$ BAO $+$ $5\times2$pt (DDE) & $0.074 \pm 0.049$ \,(95\% CL: $<0.164$) \\
\hline
\end{tabular}
\caption{\textbf{Constraints on $M_\nu$ for combinations of CMB, BAO and $5\times2$pt.}
For most models we quote the 95\% CL upper limit (``$<$''). For the DDE we report the posterior mean and 1$\sigma$ error and also show the corresponding 95\% upper limit in parentheses.}
\label{tab:mnu_constraints}
\end{table*}

Massive neutrinos free-stream while relativistic and, upon becoming non-relativistic, impede the growth of cosmic structure, imprinting a percent-level scale-dependent suppression in the matter power spectrum and a corresponding reduction in the CMB lensing amplitude~\citep{Lesgourgues:2006nd,Wong:2011ip,Dvorkin:2019jgs}. Neutrinos also importantly affect the expansion rate at late times in a mass-dependent manner by contributing to the energy density during matter domination once they have become non-relativistic. These physical effects allow low-redshift structure probes and CMB lensing to tighten bounds on $M_\nu$ beyond what primary CMB anisotropies alone can provide. In Fig.~\ref{fig:mnu_contours_lcdm} we show constraints when allowing for a free neutrino mass in the $\Lambda$CDM scenario. This plot demonstrates the significant additional constraining power on neutrino mass coming from the addition of our new $5\times2$pt LSS data vector. We find an upper limit of $M_\nu<0.050$eV for the combination of CMB + BAO + $5\times$2pt with \texttt{P-ACT} and $M_\nu<0.056$eV for the equivalent combination with \hillipop\ instead. These results are markedly more stringent than the upper limit found by the ACT collaboration of $M_\nu<0.089$eV found with the combination of \texttt{P-ACT} + CMB lensing + DESI Y1 BAO, and the baseline results from DESI, who find $M_\nu < 0.0774$eV when combining DESI DR2 BAO with \hillipop~\cite{DESI:2025dr2, ACT:2025tim}. Our result is, however, slightly less stringent than the recent SPT-3G D1 $M_\nu<0.048$eV found when combining \textit{Planck} PR4, ACT DR6 and SPT-3G data with DESI DR2 BAO.

There is ongoing debate in the literature regarding the compatibility of cosmological neutrino mass constraints with limits from neutrino oscillations, which require $M_\nu \ge 0.06\,\mathrm{eV}$ under normal ordering (or $M_\nu \ge 0.11\,\mathrm{eV}$ in the inverted ordering scenario). Under the baseline $\Lambda\mathrm{CDM}$ framework, several recent analyses combining CMB and BAO have found posteriors that peak at negative values of $M_\nu$ when analytically extending the neutrino mass range to allow for (unphysical) negative values~\cite{Craig:2024tky,Green:2024xbb, DESI:2025ejh}. Given the more stringent upper limits we find with our baseline combination of data under $\Lambda$CDM, this tension is increased relative to the CMB+BAO analyses; we find the normal and inverted ordering scenarios (NO and IO) are ruled out at and $97.64\%$ and $99.99\%$ respectively for the baseline combination with \texttt{P-ACT} and $96.07\%$ and $99.93\%$ for the combination with \hillipop. These results, therefore, strongly motivate further exploring this apparent tension with neutrino oscillation measurements.   

One possible origin is data/likelihood systematics that bias parameter estimates. For example, Refs.~\cite{Sailer:2025lxj, Jhaveri:2025neg} showed that a systematically low reionization optical depth, $\tau$, inference would bias neutrino masses to low values. Furthermore, using \textit{Planck}~PR3 and, to a lesser degree, the \texttt{CamSpec} implementation for \textit{Planck}~PR4, there is a preference for excess acoustic-peak smoothing, often captured by an effective $A_L>1$, which can artificially strengthen neutrino-mass limits~\cite{Reeves:2023kjx,Elbers:2024sha}. By contrast, the \hillipop\ PR4 likelihood used here shows no such preference for $A_L>1$, and in our \texttt{P-ACT} combination the preference for $A_L>1$ is mitigated by cutting the \textit{Planck} PR3 temperature power spectrum at $\ell=1000$. Furthermore, it was demonstrated in Ref.~\cite{Reeves:2023kjx} that the consistent addition of multiprobe LSS data to CMB data enabled systematics-robust neutrino mass constraints even in the presence of a peak-smearing systematic effect. Therefore, we expect these results are not significantly impacted by anomalously high peak smearing. Alternatively, the issue may be model-related. Several recent studies demonstrate that extending beyond $\Lambda\mathrm{CDM}$, for example by allowing for a dynamical dark energy component, can relax neutrino-mass upper limits and help to reconcile the cosmological posteriors with oscillation priors~\cite{Reeves:2025axp,DESI:2025ffm,Shao:2024mag,RoyChoudhury:2024wri}. We explore this possibility for the two extensions discussed in this work in the forthcoming section. 

\subsubsection{Extensions meet neutrino masses}

\begin{figure}
    \centering
    \includegraphics[width=0.9\linewidth]{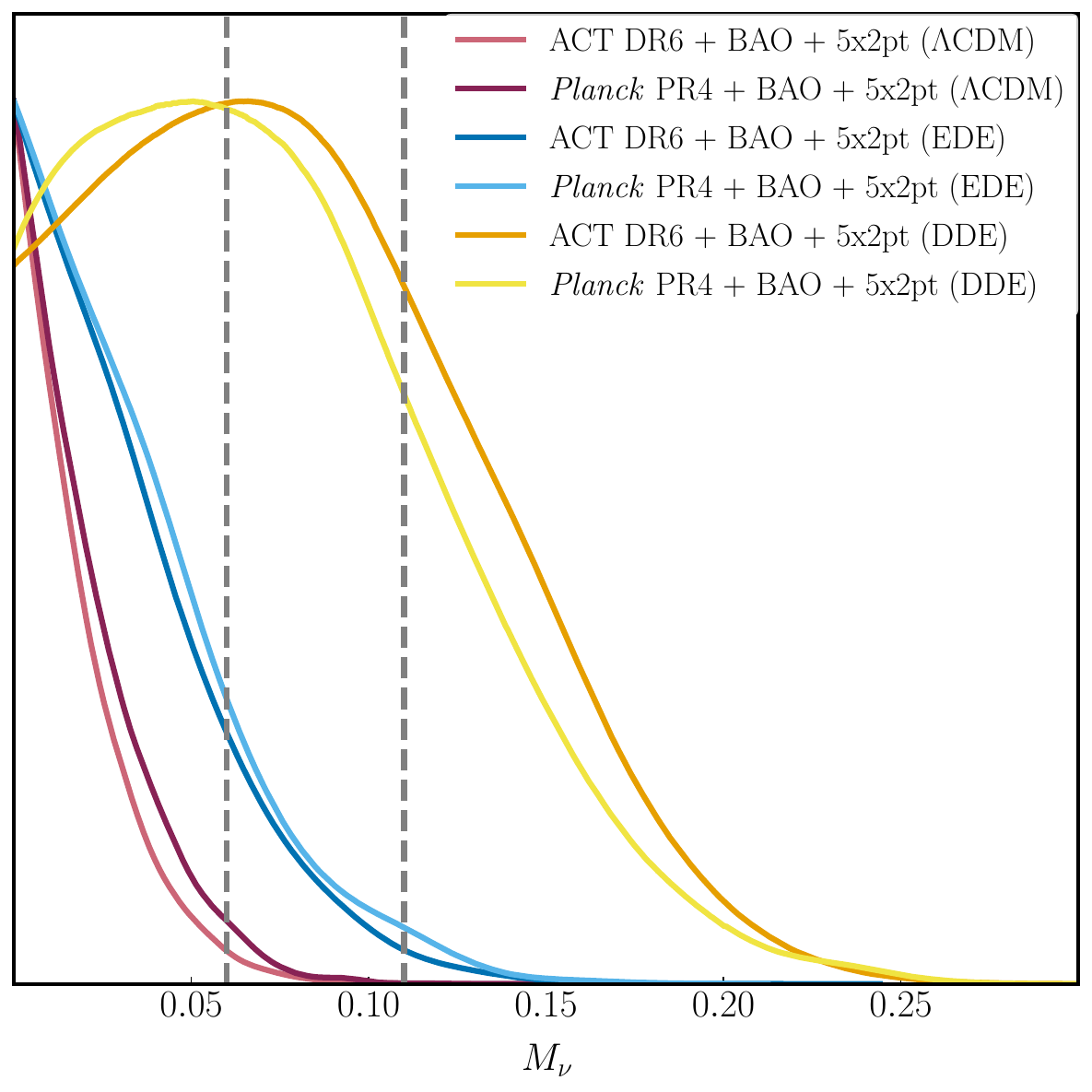}
    \caption{\textbf{Neutrino mass constraints for CMB + $5\times2$pt + BAO for the three models explored in this work}. The lower limits implied by the normal ordering (NO) and inverted ordering (IO) are shown in the grey vertical bands in order from left to right.}
    \label{fig:mnu_contours}
\end{figure}

\begin{figure}
    \centering
    \includegraphics[width=0.9\linewidth]{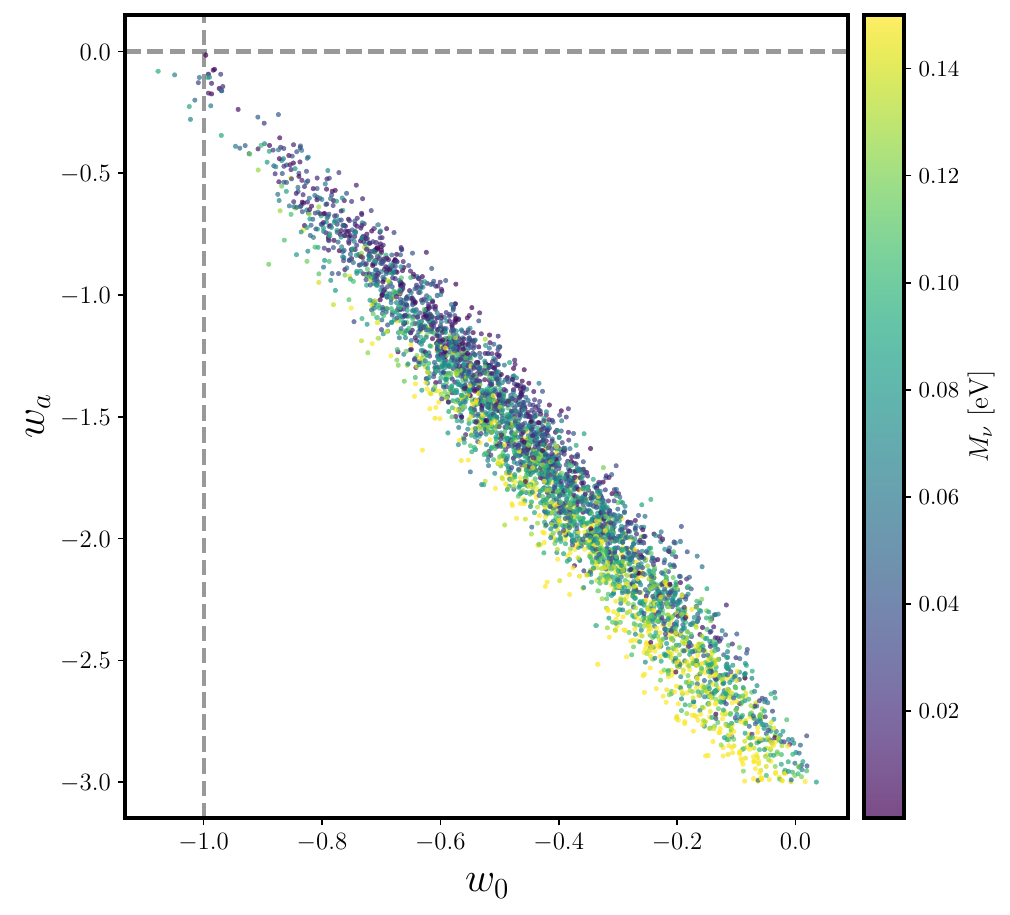}
    \caption{\textbf{Scatter plot of neutrino mass values in the $w_0w_a$ plane for the \texttt{P-ACT} + $5\times2$pt + BAO chain.} The $\Lambda\mathrm{CDM}$ limit is denoted at the crossing of the two grey dashed lines.}
    \label{fig:act_dr6_scatter_mnu}
\end{figure}

In Fig.~\ref{fig:mnu_contours} and Table~\ref{tab:mnu_constraints} we compare how extending the baseline $\nu\Lambda$CDM model to either DDE or EDE modifies the posterior of the summed neutrino mass, $M_\nu$, when fit to our full data combination. Allowing a dynamical dark energy sector relaxes the upper limit on $M_\nu$ and shifts the posterior peak to $M_\nu>0$, yielding $M_\nu<0.171~\mathrm{eV}$ for combinations with \texttt{P-ACT} and $M_\nu<0.164~\mathrm{eV}$ for the corresponding constraints with \hillipop. This result agrees well with previous studies of neutrino masses in this model; for example, DESI finds $M_\nu<0.177~\mathrm{eV}$ when combining DR2 BAO with \textit{Planck} PR4 (\texttt{CamSpec)} representing a significant broadening of the constraint in $\Lambda\mathrm{CDM}$. By contrast, introducing EDE in our analysis produces only a small shift of the $M_\nu$ posterior toward larger values: the posterior still peaks at $M_\nu=0$\,eV and the 95\% upper limits are only marginally broader than the $\Lambda$CDM counterparts, ruling out NO/IO at $83.3\%$/$98.4\%$ for the baseline combination with \texttt{P-ACT}. This limited impact of EDE on $M_\nu$ is in line with the few studies that vary EDE and $M_\nu$ jointly: Ref.~\cite{Qu:2024lpx} find an upper limit $M_\nu<0.096~\mathrm{eV}$ (95\%\,CL) when fitting EDE with Planck~PR4 CMB, ACT~DR6 and Planck PR4 CMB lensing, and DESI~Y1 BAO, while Ref.~\cite{Reeves:2022aoi} showed earlier that freeing $M_\nu$ does not appreciably increase the allowed EDE fraction once BAO distances are included.

The different behavior with respect to neutrino mass constraints can be traced to parameter degeneracies when fitting the extended models to the full combination of data. Increasing EDE fraction ($f_{\rm EDE}$) demands a compensating rise in the cold-dark-matter density ($\omega_{\rm cdm}$) to maintain the CMB fit~\cite{Poulin:2018dzj}. As shown in Ref.~\cite{Reeves:2022aoi}, this strongly limits the extent to which the value of $M_\nu$ can be raised in models with $f_{EDE}>0$, as the effect of the increased dark matter density on the BAO scale is similar to that of raising $M_\nu$, which behaves like dark matter once it becomes non-relativistic, meaning there is not more room to raise $M_\nu$ whilst maintaining a good fit to BAO data. In the DDE case, however, departures from a cosmological-constant equation of state affect the BAO distances in a manner that is opposite in sign to raising $M_\nu$, so the combined posterior allows for larger values of $M_\nu$. This is demonstrated in Fig.~\ref{fig:act_dr6_scatter_mnu}, where for model parameters further from the $\Lambda$CDM limit, a higher inferred $M_\nu$ is preferred by the data to maintain a good fit to BAO distance ratios. Correspondingly, the constraints in the $w_0w_a$ plane are degraded when allowing for a free neutrino mass. 

These results suggest that even in light of the new $5\times2$pt data vector, dynamical dark energy can offer a viable solution to resolve the apparent tension between oscillation measurements and cosmological neutrino mass constraints under $\Lambda$CDM, while EDE only marginally improves the situation. This is a clear point of difference between the models and adds some supporting evidence for DDE whilst at the same time not completely ruling out EDE. In the next section, we confront the extended models with several external data sets to assess their compatibility.

\subsection{Adding external datasets}
\begin{figure}
    \centering
    \includegraphics[width=\linewidth]{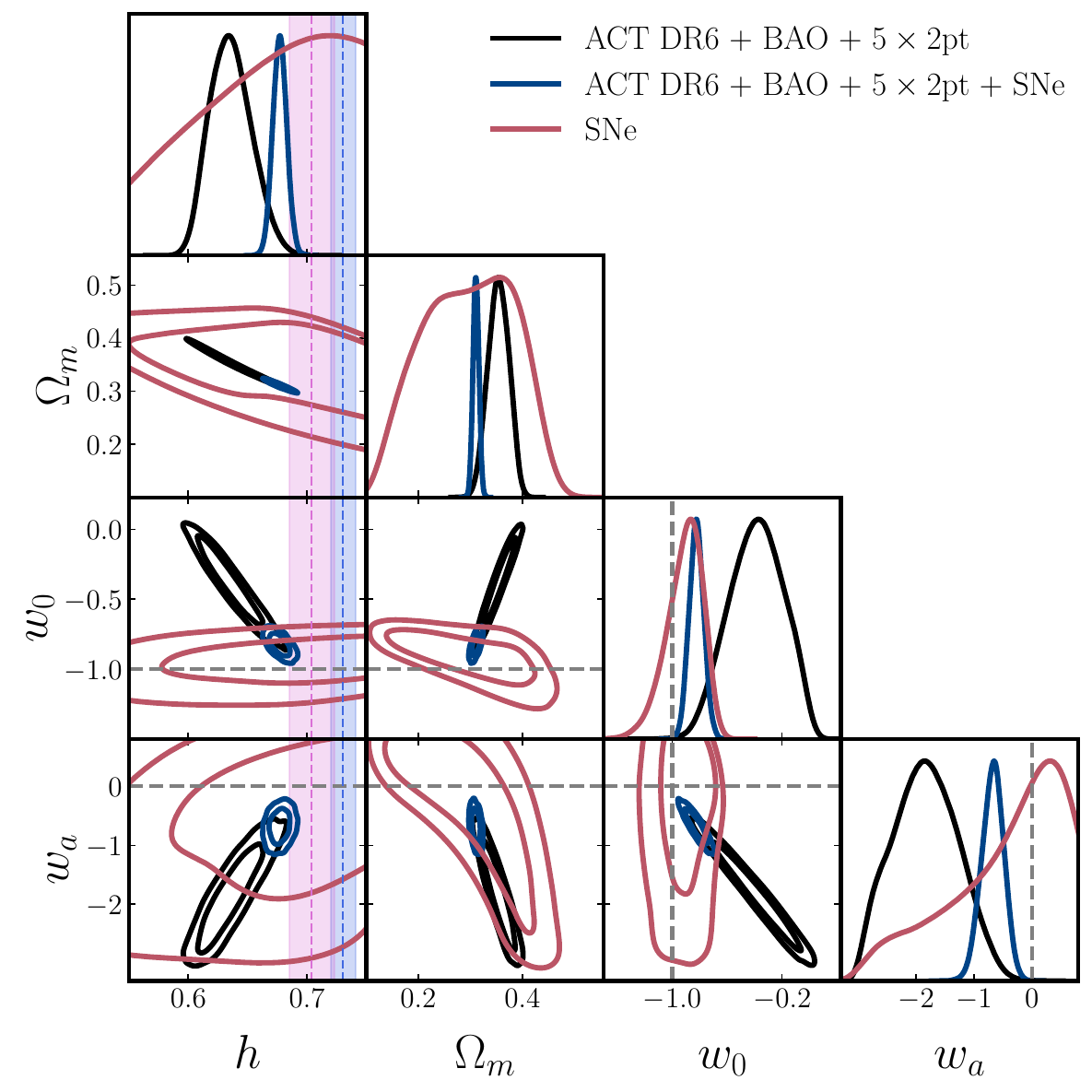}
    \caption{\textbf{Contours for the DDE model using the baseline and selected external data.}
    \emph{Black}: baseline; \emph{blue}: baseline + Pantheon+ (uncalibrated SNe); \emph{red}: Pantheon+ supernovae alone.}
    \label{fig:w0wa_with_external}
\end{figure}

\begin{figure}
    \centering
    \includegraphics[width=\linewidth]{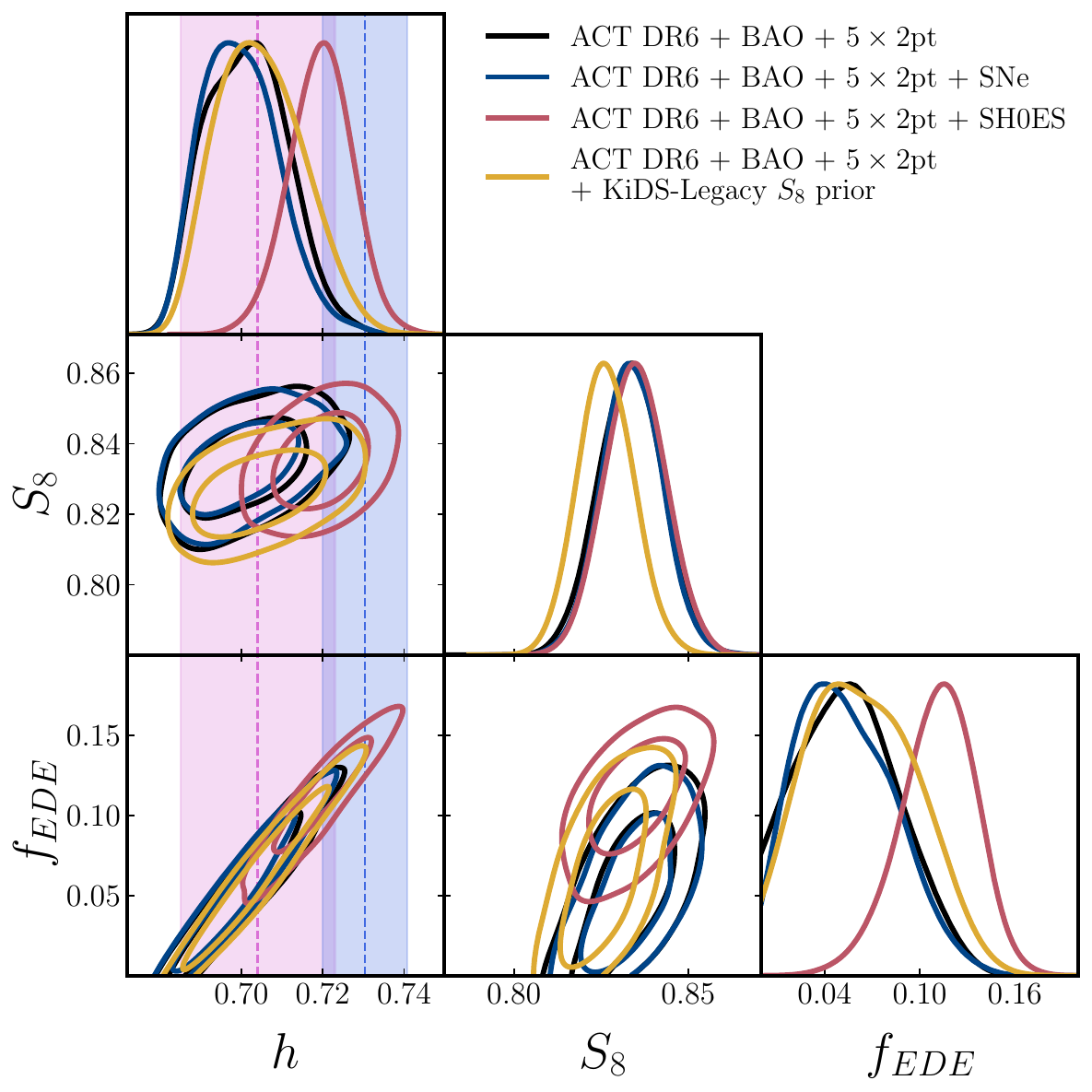}
    \caption{\textbf{Contours for the EDE model using the baseline and selected external data.}
    \emph{Black}: baseline; \emph{blue}: baseline + Pantheon+ (uncalibrated SNe); \emph{red}: baseline + SH0ES $H_0$ prior; \emph{yellow} baseline + KiDS-Legacy $S_8$ prior.}
    \label{fig:ede_with_external}
\end{figure}

\begin{table}[t]
\centering
\footnotesize
\setlength{\tabcolsep}{2.5pt}
\renewcommand{\arraystretch}{1.08}
\resizebox{\columnwidth}{!}{%
\begin{tabular}{llrrrr}
\hline
\textbf{Model} & \textbf{Dataset} &
\multicolumn{2}{c}{\textbf{\textit{Planck} PR4}} &
\multicolumn{2}{c}{\textbf{ACT DR6}} \\
 &  & $\Delta\chi^{2}$ & $n_\sigma$ & $\Delta\chi^{2}$ & $n_\sigma$ \\
\hline
\multirow{2}{*}{DDE}
 & Baseline  & 13.5 & 3.0 & 14.6 & 3.3 \\
 & $+\,$SNe                      & 8.3  & 2.2 & 9.5  & 2.5 \\
\hline
\multirow{4}{*}{EDE}
 & Baseline  & 6.7  & 1.4 & 11.4 & 2.3 \\
 & $+\,$SNe                      & 6.6  & 1.4 & 10.3 & 2.1 \\
& $+\,$KiDS Legacy prior & 6.3 & 1.3 & 10.0 & 2.0 \\
 & $+\,$SH0ES prior              & 23.6 & 4.0 & 28.3 & 4.5 \\
\hline
\end{tabular}}
\caption{\textbf{Preferences for DDE and EDE with external datasets.}
Entries show best-fit $\Delta\chi^{2}\equiv\chi^{2}_{\Lambda\mathrm{CDM}}-\chi^{2}_{\mathrm{model}}$ and the Gaussian-equivalent significance $n_\sigma$ for the baseline CMB\,+\,BAO\,+\,$5\times2$pt combination (first row in each block) and after additionally including each external dataset (subsequent rows).}
\label{tab:model_ext_prefs_compact}
\end{table}

In this section, we explore the early and late-time dark energy modifications in the context of other external datasets. Here we ask if the fit of these models to the CMB, BAO and $5\times2$pt data is consistent with external data, namely Supernovae data from Pantheon+ and an external $H_0$ prior from SH0ES~\cite{Riess:2021jrx}. 
\begin{table}[t]
\centering
\footnotesize
\setlength{\tabcolsep}{2.5pt}
\renewcommand{\arraystretch}{1.08}
\begin{tabular}{ll}
\hline
\textbf{Dataset} & \boldmath$f_{\rm EDE}$ \\
\hline
Baseline (\texttt{P-ACT} + BAO + $5\times2$pt)      & $f_{\mathrm{EDE}} < 0.12$ (95\% CL) \\
+ SNe                               & $f_{\mathrm{EDE}} < 0.10$ (95\% CL) \\
+ KiDS-Legacy prior                 & $f_{\mathrm{EDE}} < 0.12$ (95\% CL) \\
+ SH0ES prior                       & $f_{\mathrm{EDE}} = 0.11 \pm 0.02$ \\
\hline
\end{tabular}
\caption{\textbf{EDE constraints from ACT DR6 with external datasets.} Entries follow the rule: if the detection significance is below $2\sigma$, report a one-tailed 95\% upper limit; otherwise report the mean $\pm 1\sigma$. Baseline is ACT DR6 + BAO + $5\times2$pt.}
\label{tab:fede_act_dr6}
\end{table}

\paragraph{Adding Pantheon+ SNe (uncalibrated).}
Supernova distances tighten the late-time expansion constraint by providing a measure of the shape of $H(z)$ in the low-redshift Universe (until $z=2.26$ for Pantheon+). This significantly improves the constraining power in the $w_0w_a$ plane by adding additional expansion rate constraints at late times, essentially giving a relatively strong constraint on $w_0$ whilst $w_a$ is weakly constrained (see the red contour in Fig.~\ref{fig:w0wa_with_external}). When adding these data to the baseline combination, the posterior becomes significantly tighter and is pulled toward $w_0=-1$. We measure, via Wilks’ theorem, a reduced preference over $\Lambda\mathrm{CDM}$ of $2.5\sigma$ for combinations with ACT DR6 and $2.1\sigma$ for \hillipop. Inspecting Fig.~\ref{fig:w0wa_with_external}, we can understand this as there is a mild ($\sim 1 \sigma$) tension between the SNe alone constraints in the $w_0w_a$ plane where the SNe data generally restrict the dark energy equation of state today to be closer to $w_0=-1$ than the central value of the baseline constraint. We note that the DESI collaboration also find a small reduction in the preference for DDE when analyzing the combination of CMB, BAO, and Pantheon+ data compared to the CMB+BAO preference\footnote{Note the DESI team explored two other supernovae datasets (Union3~\cite{Rubin:2023jdq} and DES Y5 SNe~\cite{DES:2024jxu}), both of which show a slightly stronger preference for a DDE scenario in combination with CMB and BAO data; we expect to see a similar trend if we substituted the Pantheon+ dataset for either one of these.}. This also has implications for the neutrino mass constraint in DDE when adding SNe to the data combination; we find a tighter neutrino mass upper limit of $M_\nu<0.11$eV, and the posterior now peaks at $M_\nu=0$eV when allowing a free neutrino mass. This is due to the fact that the addition of SNe limits the departure from a cosmological constant in the $w_0w_a$ plane, and as per Fig.~\ref{fig:act_dr6_scatter_mnu} this means there is less scope for higher neutrino mass inferences. Nonetheless, this constraint still remains more compatible with oscillation measurements than the $\Lambda$CDM and EDE baseline constraints. 

For the EDE model, adding Pantheon+ very slightly reduces the preference for the EDE scenario and tightens the upper limit to $f_{EDE}<0.10$ (95\% CL); this can be understood by the fact that Pantheon+ prefers a higher $\Omega_m$ compared to the $5\times2$pt and BAO combination. Examining Fig.~\ref{fig:tension_plot}, we can see that the EDE model, when fit to CMB data alone, prefers a lower $\Omega_m$ as $f_{EDE}$ is increased; hence, including data that prefer a larger $\Omega_m$ limits the extent to which $f_{EDE}$ can be raised while maintaining a good global fit to the datasets. The EDE model does not affect late-time dynamics, and hence the $\Omega_m$ constraint from Pantheon+ alone is essentially unchanged in the EDE model (the same is not true for DDE).

\paragraph{Compatibility with SH0ES.}
In this section, we discuss the compatibility of the two extended dark energy models with the SH0ES local distance ladder measurement ($H_0=73.04 \pm 1.04\rm km\,s^{-1}\,Mpc^{-1}$~\cite{Riess:2021jrx}). For DDE, our baseline constraints yield $H_0=63.6\pm1.8~\mathrm{km\,s^{-1}\,Mpc^{-1}}$ (\texttt{P\text{-}ACT}) and $H_0=63.4\pm1.8~\mathrm{km\,s^{-1}\,Mpc^{-1}}$ (\hillipop), corresponding to $4.54\sigma$ and $4.64\sigma$ 1D Gaussian tensions. As seen in Fig.~\ref{fig:w0wa_with_external}, the degeneracy directions imply that moving farther into the DESI\,BAO\,+\,CMB–preferred region with $w_0>-1$ and $w_a<0$ drives the inferred $H_0$ even lower, worsening the disagreement with SH0ES. It should be noted that the Gaussian tension in the baseline $\Lambda$CDM model is around the same level as seen for DDE, but this is because the error bar of the $H_0$ inference is greatly enlarged compared to $\Lambda\mathrm{CDM}$ which compensates for the lower central value. This model, hence, does not necessarily worsen this problem from a Bayesian constraint point of view, yet the best-fit model has a more discrepant value. Given the level of discrepancy, we do not additionally combine with a SH0ES prior on $H_0$ when fitting to the DDE model. Overall, the DDE model cannot effectively reconcile late-time local distance ladder measurements with the inferred $H_0$ from CMB, BAO, and LSS data.

The story is different for EDE, which was originally designed to help resolve this tension. Because EDE's fit to our \emph{baseline} combination is only mildly inconsistent with SH0ES (Gaussian tensions of $2.0\sigma$ and $2.6\sigma$ for \texttt{P\text{-}ACT} and \hillipop, respectively), we do analyze the situation when adding a SH0ES prior to our baseline dataset. Incorporating this prior shifts the posterior toward higher $H_0$ (closer to SH0ES) and yields a well-constrained $f_{\rm EDE}=0.11\pm0.02$. Relative to the corresponding $\Lambda$CDM fit, EDE is then favored at $4.5\sigma$ for the \texttt{P-ACT} combination and at $4.0\sigma$ for the \hillipop\ combination. Thus, EDE can reconcile the local distance-ladder determination of $H_0$ with the global dataset without compromising the overall quality of fit; we find a Bayesian PTE=$0.472$ when adding the SH0ES prior in the EDE scenario compared to $0.476$ for the baseline combination. 

\paragraph{Adding a \textit{KiDS}-Legacy prior for EDE.}
\begin{figure}
    \centering
    \includegraphics[width=0.9\linewidth]{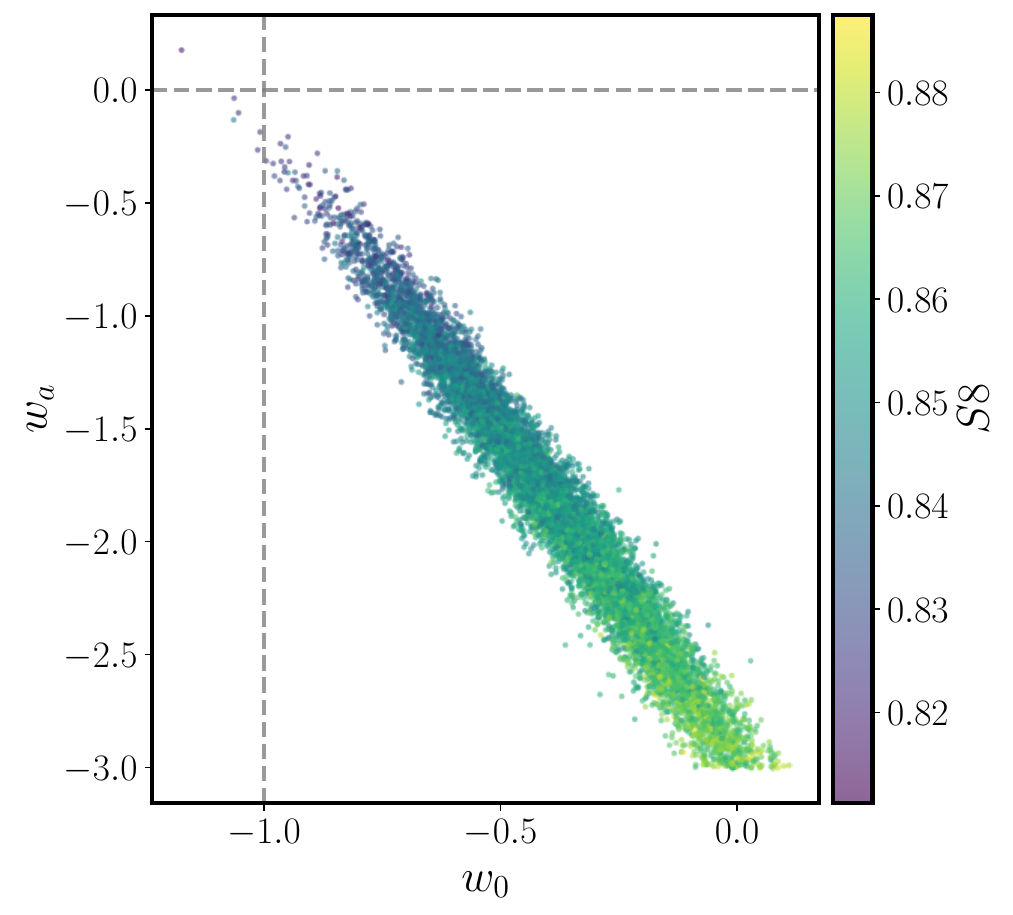}
    \caption{\textbf{Scatter plot of $S_8$ values in the $w_0w_a$ plane for the \texttt{P-ACT} + $5\times2$pt + BAO chain.} Models far from a cosmological constant (crossing point of grey lines) prefer higher $S_8$ values. \label{fig:S_8_w0wa}}
\end{figure}

To understand the effect of adding weak lensing data to the full data combination when analyzing the EDE model, we include a Gaussian $S_8=0.815\pm0.012$ prior based on the latest KiDS-Legacy results\footnote{We note Ref.~\cite{Hill:2020osr} validated that placing a Gaussian prior on $S_8$ is a very good approximation to including the full weak lensing dataset when analyzing EDE with CMB, BAO, and lensing data (see their Appendix B).}. As this updated likelihood was not made publicly available at the time of completing this work, this provides a useful estimate of the level of parameter shift expected when including this dataset. In EDE scenarios, increasing $f_{\rm EDE}$ generally raises the inferred $S_8$ when fitting CMB data~\cite{Hill:2020osr,Vagnozzi:2023nrq}. Ref.~\cite{Hill:2020osr} showed that this exacerbates the ``$S_8$ tension,'' disfavouring EDE when analyzing this model with a joint weak-lensing prior from KiDS-1000, DES-Y3, and HSC-Y1. Re-examining this with a prior based on the latest KiDS-Legacy result, we find that the EDE preference remains stable—ACT: $\Delta\chi^2=10.3$ ($n_\sigma=2.5$), \textit{Planck}: $\Delta\chi^2=6.9$ ($n_\sigma=1.9$)—and the $95\%$\,CL upper limit on $f_{\rm EDE}$ is unchanged. This result would suggest that the latest weak lensing data, which show a significantly lower $S_8$ tension, do not provide a sufficiently strong limit on $f_{EDE}$ to decisively rule out this model. 

For DDE, we cannot simply place a Gaussian prior on $S_8$, as the $S_8$ constraint from the KiDS-Legacy data alone will be significantly changed in the DDE scenario. In future work, it would be interesting to consistently combine our analysis with weak lensing data, including all cross-correlations. It was demonstrated in Ref.~\cite{Reeves:2025axp} that the addition of KiDS-1000 data to the combination of CMB and BAO data reduced the preference for the dynamical dark energy scenario. This is because models far from a cosmological constant, when fit to CMB and BAO data, prefer a higher $S_8$ value than in $\Lambda\mathrm{CDM}$ (see Fig.~\ref{fig:S_8_w0wa}) and hence tend to worsen the discrepancy with the lower $S_8=0.777\pm0.017$ value measured in KiDS-1000~\cite{KiDS:2020suj}. We expect the most recent KiDS-Legacy data to be more compatible with the dynamical dark energy scenario, given its preference for an increased $S_8$. 

\section{Discussion and conclusions}\label{sec:Conclusion}

In this work, we analyzed a multiprobe $5\times2$pt combination of ACT DR6 CMB lensing, LS LRGs, and \textit{Planck} ISW for the first time. The $5\times$2pt data vector is well described by a $\Lambda\mathrm{CDM}$ model with the computed Bayesian PTE$=0.36$. In $\Lambda$CDM, we derived a constraint on the lensing amplitude of $S_8=0.819\pm0.016$. We also found a sound-horizon free value of $H_0=70.0\pm4.4 \rm km\,s^{-1}\,Mpc^{-1}$ from the $5\times2$pt data vector alone and  $H_0=66.2\pm3.1 \rm km\,s^{-1}\,Mpc^{-1}$ when additionally including Pantheon+ SNe data. These values agree very well with CMB determinations of these parameters from \textit{Planck} PR4 and ACT DR6. In particular, we do not observe the `lensing is low` trend for the combination of LSS data used here, suggesting that if the hints for low-$S_8$ seen in some previous galaxy weak lensing analyses are confirmed to be physical in nature, the new physics required to explain these data resides in scales and redshifts not probed by our $5\times2$pt combination. 
 
Despite the fact that $\Lambda\mathrm{CDM}$ fits the $5\times2$pt LSS, BAO and CMB data individually very well (see e.g., ~\cite{Tristram:2023haj, DESI:2025dr2, ACTDR6:PowerSpectra} and our PTE values in Tab.~\ref{tab:goodness_of_fit}), when put together, we see a mild preference for extended models. This can be traced to a ($\sim2\sigma$) discrepancy between the posteriors under $\Lambda\mathrm{CDM}$ for the CMB and $5\times2$pt+BAO combination, which we exemplify in the $\Omega_m$--$\frac{D_v}{r_d}$ plane (Fig.~\ref{fig:tension_plot}). We investigated two dark energy extensions that can help reconcile this discrepancy: an early-Universe component prior to recombination (EDE) and a late-time dynamical equation of state (DDE). To assess the robustness of our results, we employed two complementary CMB primary likelihoods: \textit{Planck} PR4 (\hillipop) and ACT DR6 (\texttt{P-ACT}). In our \emph{baseline} CMB+$5\times2$pt+BAO combinations, the dynamical model is preferred over $\Lambda$CDM at $3.3\sigma$ (combination with \texttt{P-ACT}) and $3.0\sigma$ (combination with \hillipop), whereas EDE yielded a milder preference of $2.3\sigma$ (combination with \texttt{P-ACT}) and $1.4\sigma$ (combination with \hillipop). We demonstrated that adding our new $5\times2$pt LSS data vector beyond CMB+BAO provides a stronger preference for extended dark energy in all cases; by $\simeq1\sigma$ for DDE and by $\simeq0.5\sigma$ for EDE, which we traced back to an enhanced tension in the $\Omega_m$--$\frac{D_v}{r_d}$ plane. 

To further investigate the dark energy extensions, we examined their impact on neutrino mass constraints, which exhibit tension with oscillation measurements within $\Lambda$CDM~\cite{Green:2024xbb, Craig:2024tky}. The inclusion of the $5\times2$pt data vector to CMB+BAO tightened the upper limit on $M_\nu$ in $\Lambda$CDM, worsening the apparent tension with oscillation data. We found the normal (inverted) hierarchy scenario to be ruled out at $97.64\%$ ($99.99\%$) when combining with \texttt{P-ACT} with an upper limit of $M_\nu<0.050$eV (95\% CL). The EDE scenario provides only marginal relief, with the $95\%$ upper limit relaxed to $\lesssim0.09$eV and the posterior still peaking at $M_\nu=0$eV. By contrast, the DDE model significantly relaxes the neutrino mass bound to $M_\nu\lesssim0.17$eV, with the posterior peaking at $M_\nu>0$eV for both CMB likelihoods. This relaxation occurred due to a degeneracy between dynamical dark energy and neutrino mass effects on BAO distances at late times, as illustrated in Fig.~\ref{fig:act_dr6_scatter_mnu}, where models with $w_0<-1, w_a<0$ allow for higher neutrino masses. At face value, these findings strengthen the case for a DDE scenario over EDE and $\Lambda$CDM, as in this model we can effectively reconcile neutrino constraints with oscillation data. The degeneracy between dynamical dark energy and neutrino mass, however, remains a fundamental limitation of achieving precise neutrino mass measurements in the DDE scenario with current data. 

We then considered adding several external datasets and priors to our baseline combination of data. Firstly, we combined with SNe data from Pantheon+, finding that in both cases the preference for departures from $\Lambda\mathrm{CDM}$ is slightly decreased. In the case of EDE, this can be understood as the Pantheon+ supernovae prefer a relatively high value of $\Omega_m=0.334\pm0.018$~\cite{Brout:2022vxf}, which is unchanged when analyzed under an EDE model, and this is incompatible with the high-$f_{EDE}$ scenario, which prefers a slightly lower $\Omega_m$ when fit to CMB data alone. The shift is, however, not significant enough to rule out the EDE scenario; for the combination with ACT DR6 \texttt{P-ACT}, we find an upper limit of $f_{EDE}<0.10$, compared to $f_{EDE}<0.12$ for the baseline data combination. The addition of Pantheon+ has a relatively bigger impact on the DDE model, where the contours in the $w_0w_a$ plane are significantly tightened compared to our baseline combination and the model preferences are decreased from $3.3\sigma$ ($3.0\sigma$) to $2.5\sigma$ ($2.2\sigma$) for the combination of data with ACT DR6 (\textit{Planck} PR4). This is due to the relatively strong constraint on $w_0$ provided by the Pantheon+ SNe data, which prefers a value close to $w_0=-1$ and hence limits the allowed departure from a cosmological constant in the combined constraints. This also has implications for the neutrino mass constraint in this model; this is tightened to $M_\nu<0.11$eV. While this remains more compatible with oscillation measurements than the baseline $\Lambda$CDM and EDE constraints, the posterior in this scenario peaks again at $M_\nu=0$eV. It would be interesting, in future work, to compare these findings when including other Supernovae datasets such as Union3~\cite{Rubin:2023jdq} and DES-Y5~\cite{DES:2024jxu}. 

Next, we assessed the compatibility of the two extended dark energy models with the SH0ES local distance ladder ($H_0=73.04 \pm 1.04\rm km\,s^{-1}\,Mpc^{-1}$~\cite{Riess:2021jrx}) measurement. Here, we see a more stark distinction between the models; when fit to the baseline combination of data, EDE brings the $H_0$ tension with SH0ES to $\lesssim2.5\sigma$. Including the SH0ES measurement as a prior on $H_0$ when fitting to the baseline dataset, we find a constraint of $f_{EDE}=0.11\pm0.02$ which peaks significantly away from $f_{EDE}=0$. The EDE scenario is favored at $\sim 4\sigma$ over the $\Lambda$CDM fit to the same data (also including the SH0ES prior), and the Bayesian PTE of the fit including the SH0ES prior is not changed compared to the baseline. These results demonstrate that EDE can do what it was originally designed for: help to resolve the Hubble tension while maintaining a good global fit to data. We also showed that including an $S_8$-prior based on KiDS-Legacy measurements~\cite{Wright:2025xka} does not alter the model preference for EDE, in contrast to previous studies using an $S_8$ prior based on the older KiDS-1000 dataset, DES-Y3, and HSC-Y1, which find this significantly reduces model preference (see e.g. Ref.~\cite{Hill:2020osr}). For DDE, our baseline Hubble constant constraints yield $H_0=63.6\pm1.8~\mathrm{km\,s^{-1}\,Mpc^{-1}}$ (\texttt{P\text{-}ACT}) and $H_0=63.4\pm1.8~\mathrm{km\,s^{-1}\,Mpc^{-1}}$ (\hillipop), corresponding to $4.54\sigma$ and $4.64\sigma$ 1D Gaussian tensions. The degeneracy directions imply that moving farther into the DESI\,BAO\,+\,CMB–preferred region with $w_0>-1$ and $w_a<0$ drives the inferred $H_0$ even lower, worsening the disagreement with SH0ES. In short, the DDE model cannot effectively reconcile late-time local distance ladder measurements with the inferred $H_0$ from CMB, BAO and LSS data. 

In summary, both the DDE and EDE extensions remain potentially viable explanations for the small discrepancies in current cosmological data, in the sense that neither model can be definitely ruled out with current data. However, neither model provides a compelling solution to all contemporary issues found when confronting $\Lambda$CDM with data. The DDE model generally yields an improved global fit to our baseline data and can reconcile the lower limit on $M_\nu$ from neutrino oscillation data, but it cannot address the high $H_0$ measurements from the late Universe. The EDE model offers a compelling (if partial) resolution to the Hubble tension, but its statistical preference over $\Lambda\mathrm{CDM}$ is not as strong, and neutrino mass constraints remain in some tension with oscillation data. A plethora of other model extensions exist that have the potential to resolve one or more of the mild discrepancies observed in the $\Lambda$CDM framework (see Ref.~\cite{CosmoVerseNetwork:2025alb} for a recent review); it would be interesting to explore a subset of these in a future study.

Looking ahead, upcoming observations will be pivotal in discriminating between $\Lambda$CDM and its contenders. On the CMB side, new high-resolution measurements from the next generation of experiments, such as the Simons Observatory, will provide independent cross-checks of the mild preference for EDE seen in ACT data. In parallel, forthcoming large-scale structure surveys (e.g., DESI Year~5, \textit{Euclid}~\cite{EUCLID:2011zbd}, and the Rubin Observatory’s LSST~\cite{LSSTScience:2009jmu}) will dramatically sharpen measurements of the low-$z$ expansion history and structure growth. In addition, improved supernova samples and gravitational lensing time-delay measurements will further pin down the late-Universe expansion rate. Together, these advances will either bolster the case for new physics—if the moderate hints reported here grow with more data—or else they will tighten the constraints enough to definitively reaffirm $\Lambda$CDM by ruling out its proposed alternatives. In either event, multiprobe techniques, capitalizing on the synergy of CMB, galaxy, and lensing surveys, will continue to be a powerful approach for testing the foundations of cosmology.

\newpage
\bibliography{ref}

\begin{thebibliography}{122}%
\makeatletter
\providecommand \@ifxundefined [1]{%
 \@ifx{#1\undefined}
}%
\providecommand \@ifnum [1]{%
 \ifnum #1\expandafter \@firstoftwo
 \else \expandafter \@secondoftwo
 \fi
}%
\providecommand \@ifx [1]{%
 \ifx #1\expandafter \@firstoftwo
 \else \expandafter \@secondoftwo
 \fi
}%
\providecommand \natexlab [1]{#1}%
\providecommand \enquote  [1]{``#1''}%
\providecommand \bibnamefont  [1]{#1}%
\providecommand \bibfnamefont [1]{#1}%
\providecommand \citenamefont [1]{#1}%
\providecommand \href@noop [0]{\@secondoftwo}%
\providecommand \href [0]{\begingroup \@sanitize@url \@href}%
\providecommand \@href[1]{\@@startlink{#1}\@@href}%
\providecommand \@@href[1]{\endgroup#1\@@endlink}%
\providecommand \@sanitize@url [0]{\catcode `\\12\catcode `\$12\catcode `\&12\catcode `\#12\catcode `\^12\catcode `\_12\catcode `\%12\relax}%
\providecommand \@@startlink[1]{}%
\providecommand \@@endlink[0]{}%
\providecommand \url  [0]{\begingroup\@sanitize@url \@url }%
\providecommand \@url [1]{\endgroup\@href {#1}{\urlprefix }}%
\providecommand \urlprefix  [0]{URL }%
\providecommand \Eprint [0]{\href }%
\providecommand \doibase [0]{http://dx.doi.org/}%
\providecommand \selectlanguage [0]{\@gobble}%
\providecommand \bibinfo  [0]{\@secondoftwo}%
\providecommand \bibfield  [0]{\@secondoftwo}%
\providecommand \translation [1]{[#1]}%
\providecommand \BibitemOpen [0]{}%
\providecommand \bibitemStop [0]{}%
\providecommand \bibitemNoStop [0]{.\EOS\space}%
\providecommand \EOS [0]{\spacefactor3000\relax}%
\providecommand \BibitemShut  [1]{\csname bibitem#1\endcsname}%
\let\auto@bib@innerbib\@empty
\bibitem [{\citenamefont {Abbott}\ \emph {et~al.}(2022)\citenamefont {Abbott} \emph {et~al.}}]{DES:2021wwk}%
  \BibitemOpen
  \bibfield  {author} {\bibinfo {author} {\bibfnamefont {T.~M.~C.}\ \bibnamefont {Abbott}} \emph {et~al.} (\bibinfo {collaboration} {DES}),\ }\bibfield  {title} {\enquote {\bibinfo {title} {{Dark Energy Survey Year 3 results: Cosmological constraints from galaxy clustering and weak lensing}},}\ }\href {\doibase 10.1103/PhysRevD.105.023520} {\bibfield  {journal} {\bibinfo  {journal} {Phys. Rev. D}\ }\textbf {\bibinfo {volume} {105}},\ \bibinfo {pages} {023520} (\bibinfo {year} {2022})},\ \Eprint {http://arxiv.org/abs/2105.13549} {arXiv:2105.13549 [astro-ph.CO]} \BibitemShut {NoStop}%
\bibitem [{\citenamefont {Heymans}\ \emph {et~al.}(2021)\citenamefont {Heymans} \emph {et~al.}}]{Heymans:2020gsg}%
  \BibitemOpen
  \bibfield  {author} {\bibinfo {author} {\bibfnamefont {Catherine}\ \bibnamefont {Heymans}} \emph {et~al.},\ }\bibfield  {title} {\enquote {\bibinfo {title} {{KiDS-1000 Cosmology: Multi-probe weak gravitational lensing and spectroscopic galaxy clustering constraints}},}\ }\href {\doibase 10.1051/0004-6361/202039063} {\bibfield  {journal} {\bibinfo  {journal} {Astron. Astrophys.}\ }\textbf {\bibinfo {volume} {646}},\ \bibinfo {pages} {A140} (\bibinfo {year} {2021})},\ \Eprint {http://arxiv.org/abs/2007.15632} {arXiv:2007.15632 [astro-ph.CO]} \BibitemShut {NoStop}%
\bibitem [{\citenamefont {Sugiyama}\ \emph {et~al.}(2023)\citenamefont {Sugiyama} \emph {et~al.}}]{Sugiyama:2023fzm}%
  \BibitemOpen
  \bibfield  {author} {\bibinfo {author} {\bibfnamefont {Sunao}\ \bibnamefont {Sugiyama}} \emph {et~al.},\ }\bibfield  {title} {\enquote {\bibinfo {title} {{Hyper Suprime-Cam Year 3 results: Cosmology from galaxy clustering and weak lensing with HSC and SDSS using the minimal bias model}},}\ }\href {\doibase 10.1103/PhysRevD.108.123521} {\bibfield  {journal} {\bibinfo  {journal} {Phys. Rev. D}\ }\textbf {\bibinfo {volume} {108}},\ \bibinfo {pages} {123521} (\bibinfo {year} {2023})},\ \Eprint {http://arxiv.org/abs/2304.00705} {arXiv:2304.00705 [astro-ph.CO]} \BibitemShut {NoStop}%
\bibitem [{\citenamefont {Farren}\ \emph {et~al.}(2025)\citenamefont {Farren} \emph {et~al.}}]{Farren:2024rla}%
  \BibitemOpen
  \bibfield  {author} {\bibinfo {author} {\bibfnamefont {Gerrit~S.}\ \bibnamefont {Farren}} \emph {et~al.},\ }\bibfield  {title} {\enquote {\bibinfo {title} {{Atacama Cosmology Telescope: Multiprobe cosmology with unWISE galaxies and ACT DR6 CMB lensing}},}\ }\href {\doibase 10.1103/PhysRevD.111.083516} {\bibfield  {journal} {\bibinfo  {journal} {Phys. Rev. D}\ }\textbf {\bibinfo {volume} {111}},\ \bibinfo {pages} {083516} (\bibinfo {year} {2025})},\ \Eprint {http://arxiv.org/abs/2409.02109} {arXiv:2409.02109 [astro-ph.CO]} \BibitemShut {NoStop}%
\bibitem [{\citenamefont {Nicola}\ \emph {et~al.}(2017)\citenamefont {Nicola}, \citenamefont {Refregier},\ and\ \citenamefont {Amara}}]{Nicola:2016qrc}%
  \BibitemOpen
  \bibfield  {author} {\bibinfo {author} {\bibfnamefont {Andrina}\ \bibnamefont {Nicola}}, \bibinfo {author} {\bibfnamefont {Alexandre}\ \bibnamefont {Refregier}}, \ and\ \bibinfo {author} {\bibfnamefont {Adam}\ \bibnamefont {Amara}},\ }\bibfield  {title} {\enquote {\bibinfo {title} {{Integrated Cosmological Probes: Extended Analysis}},}\ }\href {\doibase 10.1103/PhysRevD.95.083523} {\bibfield  {journal} {\bibinfo  {journal} {Phys. Rev. D}\ }\textbf {\bibinfo {volume} {95}},\ \bibinfo {pages} {083523} (\bibinfo {year} {2017})},\ \Eprint {http://arxiv.org/abs/1612.03121} {arXiv:1612.03121 [astro-ph.CO]} \BibitemShut {NoStop}%
\bibitem [{\citenamefont {Nicola}\ \emph {et~al.}(2016)\citenamefont {Nicola}, \citenamefont {Refregier},\ and\ \citenamefont {Amara}}]{Nicola:2016eua}%
  \BibitemOpen
  \bibfield  {author} {\bibinfo {author} {\bibfnamefont {Andrina}\ \bibnamefont {Nicola}}, \bibinfo {author} {\bibfnamefont {Alexandre}\ \bibnamefont {Refregier}}, \ and\ \bibinfo {author} {\bibfnamefont {Adam}\ \bibnamefont {Amara}},\ }\bibfield  {title} {\enquote {\bibinfo {title} {{Integrated approach to cosmology: Combining CMB, large-scale structure and weak lensing}},}\ }\href {\doibase 10.1103/PhysRevD.94.083517} {\bibfield  {journal} {\bibinfo  {journal} {Phys. Rev. D}\ }\textbf {\bibinfo {volume} {94}},\ \bibinfo {pages} {083517} (\bibinfo {year} {2016})},\ \Eprint {http://arxiv.org/abs/1607.01014} {arXiv:1607.01014 [astro-ph.CO]} \BibitemShut {NoStop}%
\bibitem [{\citenamefont {Reeves}\ \emph {et~al.}(2025{\natexlab{a}})\citenamefont {Reeves}, \citenamefont {Nicola},\ and\ \citenamefont {Refregier}}]{Reeves:2025axp}%
  \BibitemOpen
  \bibfield  {author} {\bibinfo {author} {\bibfnamefont {Alexander}\ \bibnamefont {Reeves}}, \bibinfo {author} {\bibfnamefont {Andrina}\ \bibnamefont {Nicola}}, \ and\ \bibinfo {author} {\bibfnamefont {Alexandre}\ \bibnamefont {Refregier}},\ }\bibfield  {title} {\enquote {\bibinfo {title} {{Tuning the cosmic instrument: robust cosmology through combined probes}},}\ }\href@noop {} {\  (\bibinfo {year} {2025}{\natexlab{a}})},\ \Eprint {http://arxiv.org/abs/2502.01722} {arXiv:2502.01722 [astro-ph.CO]} \BibitemShut {NoStop}%
\bibitem [{\citenamefont {Abbott}\ \emph {et~al.}(2023)\citenamefont {Abbott} \emph {et~al.}}]{DES:2022urg}%
  \BibitemOpen
  \bibfield  {author} {\bibinfo {author} {\bibfnamefont {T.~M.~C.}\ \bibnamefont {Abbott}} \emph {et~al.} (\bibinfo {collaboration} {DES, SPT}),\ }\bibfield  {title} {\enquote {\bibinfo {title} {{Joint analysis of Dark Energy Survey Year 3 data and CMB lensing from SPT and Planck. III. Combined cosmological constraints}},}\ }\href {\doibase 10.1103/PhysRevD.107.023531} {\bibfield  {journal} {\bibinfo  {journal} {Phys. Rev. D}\ }\textbf {\bibinfo {volume} {107}},\ \bibinfo {pages} {023531} (\bibinfo {year} {2023})},\ \Eprint {http://arxiv.org/abs/2206.10824} {arXiv:2206.10824 [astro-ph.CO]} \BibitemShut {NoStop}%
\bibitem [{\citenamefont {Xu}\ \emph {et~al.}(2024)\citenamefont {Xu}, \citenamefont {Eifler}, \citenamefont {Miranda}, \citenamefont {Fang}, \citenamefont {Saraivanov}, \citenamefont {Krause}, \citenamefont {Huang}, \citenamefont {Benabed},\ and\ \citenamefont {Zhong}}]{Xu:2023qmp}%
  \BibitemOpen
  \bibfield  {author} {\bibinfo {author} {\bibfnamefont {Jiachuan}\ \bibnamefont {Xu}}, \bibinfo {author} {\bibfnamefont {Tim}\ \bibnamefont {Eifler}}, \bibinfo {author} {\bibfnamefont {Vivian}\ \bibnamefont {Miranda}}, \bibinfo {author} {\bibfnamefont {Xiao}\ \bibnamefont {Fang}}, \bibinfo {author} {\bibfnamefont {Evan}\ \bibnamefont {Saraivanov}}, \bibinfo {author} {\bibfnamefont {Elisabeth}\ \bibnamefont {Krause}}, \bibinfo {author} {\bibfnamefont {Hung-Jin}\ \bibnamefont {Huang}}, \bibinfo {author} {\bibfnamefont {Karim}\ \bibnamefont {Benabed}}, \ and\ \bibinfo {author} {\bibfnamefont {Kunhao}\ \bibnamefont {Zhong}},\ }\bibfield  {title} {\enquote {\bibinfo {title} {{Constraining baryonic physics with DES Y1 and Planck data: Combining galaxy clustering, weak lensing, and CMB lensing}},}\ }\href {\doibase 10.1103/PhysRevD.110.063532} {\bibfield  {journal} {\bibinfo  {journal} {Phys. Rev. D}\ }\textbf {\bibinfo {volume} {110}},\ \bibinfo {pages} {063532} (\bibinfo {year} {2024})},\ \Eprint
  {http://arxiv.org/abs/2311.08047} {arXiv:2311.08047 [astro-ph.CO]} \BibitemShut {NoStop}%
\bibitem [{\citenamefont {{DESI Collaboration}}(2025)}]{DESI:2025dr2}%
  \BibitemOpen
  \bibfield  {author} {\bibinfo {author} {\bibnamefont {{DESI Collaboration}}},\ }\bibfield  {title} {\enquote {\bibinfo {title} {{DESI DR2 Results II: Measurements of Baryon Acoustic Oscillations}},}\ }\href@noop {} {\  (\bibinfo {year} {2025})},\ \Eprint {http://arxiv.org/abs/2503.14738} {arXiv:2503.14738 [astro-ph.CO]} \BibitemShut {NoStop}%
\bibitem [{\citenamefont {Lodha}\ \emph {et~al.}(2025)\citenamefont {Lodha} \emph {et~al.}}]{DESI:2025fii}%
  \BibitemOpen
  \bibfield  {author} {\bibinfo {author} {\bibfnamefont {K.}~\bibnamefont {Lodha}} \emph {et~al.} (\bibinfo {collaboration} {DESI}),\ }\bibfield  {title} {\enquote {\bibinfo {title} {{Extended Dark Energy analysis using DESI DR2 BAO measurements}},}\ }\href@noop {} {\  (\bibinfo {year} {2025})},\ \Eprint {http://arxiv.org/abs/2503.14743} {arXiv:2503.14743 [astro-ph.CO]} \BibitemShut {NoStop}%
\bibitem [{\citenamefont {Chaussidon}\ \emph {et~al.}(2025)\citenamefont {Chaussidon} \emph {et~al.}}]{Chaussidon:2025npr}%
  \BibitemOpen
  \bibfield  {author} {\bibinfo {author} {\bibfnamefont {E.}~\bibnamefont {Chaussidon}} \emph {et~al.},\ }\bibfield  {title} {\enquote {\bibinfo {title} {{Early time solution as an alternative to the late time evolving dark energy with DESI DR2 BAO}},}\ }\href@noop {} {\  (\bibinfo {year} {2025})},\ \Eprint {http://arxiv.org/abs/2503.24343} {arXiv:2503.24343 [astro-ph.CO]} \BibitemShut {NoStop}%
\bibitem [{\citenamefont {Poulin}\ \emph {et~al.}(2025)\citenamefont {Poulin}, \citenamefont {Smith}, \citenamefont {Calder{\'o}n},\ and\ \citenamefont {Simon}}]{Poulin:2025nfb}%
  \BibitemOpen
  \bibfield  {author} {\bibinfo {author} {\bibfnamefont {Vivian}\ \bibnamefont {Poulin}}, \bibinfo {author} {\bibfnamefont {Tristan~L.}\ \bibnamefont {Smith}}, \bibinfo {author} {\bibfnamefont {Rodrigo}\ \bibnamefont {Calder{\'o}n}}, \ and\ \bibinfo {author} {\bibfnamefont {Th{\'e}o}\ \bibnamefont {Simon}},\ }\bibfield  {title} {\enquote {\bibinfo {title} {{Impact of ACT DR6 and DESI DR2 for Early Dark Energy and the Hubble tension}},}\ }\href@noop {} {\  (\bibinfo {year} {2025})},\ \Eprint {http://arxiv.org/abs/2505.08051} {arXiv:2505.08051 [astro-ph.CO]} \BibitemShut {NoStop}%
\bibitem [{\citenamefont {Mirpoorian}\ \emph {et~al.}(2025)\citenamefont {Mirpoorian}, \citenamefont {Jedamzik},\ and\ \citenamefont {Pogosian}}]{Mirpoorian:2025rfp}%
  \BibitemOpen
  \bibfield  {author} {\bibinfo {author} {\bibfnamefont {Seyed~Hamidreza}\ \bibnamefont {Mirpoorian}}, \bibinfo {author} {\bibfnamefont {Karsten}\ \bibnamefont {Jedamzik}}, \ and\ \bibinfo {author} {\bibfnamefont {Levon}\ \bibnamefont {Pogosian}},\ }\bibfield  {title} {\enquote {\bibinfo {title} {{Is Dynamical Dark Energy Necessary? DESI BAO and Modified Recombination}},}\ }\href@noop {} {\  (\bibinfo {year} {2025})},\ \Eprint {http://arxiv.org/abs/2504.15274} {arXiv:2504.15274 [astro-ph.CO]} \BibitemShut {NoStop}%
\bibitem [{\citenamefont {Lynch}\ \emph {et~al.}(2024)\citenamefont {Lynch}, \citenamefont {Knox},\ and\ \citenamefont {Chluba}}]{Lynch:2024hzh}%
  \BibitemOpen
  \bibfield  {author} {\bibinfo {author} {\bibfnamefont {Gabriel~P.}\ \bibnamefont {Lynch}}, \bibinfo {author} {\bibfnamefont {Lloyd}\ \bibnamefont {Knox}}, \ and\ \bibinfo {author} {\bibfnamefont {Jens}\ \bibnamefont {Chluba}},\ }\bibfield  {title} {\enquote {\bibinfo {title} {{DESI observations and the Hubble tension in light of modified recombination}},}\ }\href {\doibase 10.1103/PhysRevD.110.083538} {\bibfield  {journal} {\bibinfo  {journal} {Phys. Rev. D}\ }\textbf {\bibinfo {volume} {110}},\ \bibinfo {pages} {083538} (\bibinfo {year} {2024})},\ \Eprint {http://arxiv.org/abs/2406.10202} {arXiv:2406.10202 [astro-ph.CO]} \BibitemShut {NoStop}%
\bibitem [{\citenamefont {Reeves}\ \emph {et~al.}(2024)\citenamefont {Reeves}, \citenamefont {Nicola}, \citenamefont {Refregier}, \citenamefont {Kacprzak},\ and\ \citenamefont {Valle}}]{Reeves:2023kjx}%
  \BibitemOpen
  \bibfield  {author} {\bibinfo {author} {\bibfnamefont {Alexander}\ \bibnamefont {Reeves}}, \bibinfo {author} {\bibfnamefont {Andrina}\ \bibnamefont {Nicola}}, \bibinfo {author} {\bibfnamefont {Alexandre}\ \bibnamefont {Refregier}}, \bibinfo {author} {\bibfnamefont {Tomasz}\ \bibnamefont {Kacprzak}}, \ and\ \bibinfo {author} {\bibfnamefont {Luis Fernando Machado~Poletti}\ \bibnamefont {Valle}},\ }\bibfield  {title} {\enquote {\bibinfo {title} {{12 \texttimes{} 2 pt combined probes: pipeline, neutrino mass, and data compression}},}\ }\href {\doibase 10.1088/1475-7516/2024/01/042} {\bibfield  {journal} {\bibinfo  {journal} {JCAP}\ }\textbf {\bibinfo {volume} {01}},\ \bibinfo {pages} {042} (\bibinfo {year} {2024})},\ \Eprint {http://arxiv.org/abs/2309.03258} {arXiv:2309.03258 [astro-ph.CO]} \BibitemShut {NoStop}%
\bibitem [{\citenamefont {Sailer}\ \emph {et~al.}(2024)\citenamefont {Sailer} \emph {et~al.}}]{Sailer:2024coh}%
  \BibitemOpen
  \bibfield  {author} {\bibinfo {author} {\bibfnamefont {Noah}\ \bibnamefont {Sailer}} \emph {et~al.},\ }\bibfield  {title} {\enquote {\bibinfo {title} {{Cosmological constraints from the cross-correlation of DESI Luminous Red Galaxies with CMB lensing from Planck PR4 and ACT DR6}},}\ }\href@noop {} {\  (\bibinfo {year} {2024})},\ \Eprint {http://arxiv.org/abs/2407.04607} {arXiv:2407.04607 [astro-ph.CO]} \BibitemShut {NoStop}%
\bibitem [{\citenamefont {Qu}\ \emph {et~al.}(2025{\natexlab{a}})\citenamefont {Qu} \emph {et~al.}}]{Qu:2024sfu}%
  \BibitemOpen
  \bibfield  {author} {\bibinfo {author} {\bibfnamefont {Frank~J.}\ \bibnamefont {Qu}} \emph {et~al.},\ }\bibfield  {title} {\enquote {\bibinfo {title} {{Atacama Cosmology Telescope DR6 and DESI: Structure growth measurements from the cross-correlation of DESI legacy imaging galaxies and CMB lensing from ACT DR6 and Planck PR4}},}\ }\href {\doibase 10.1103/PhysRevD.111.103503} {\bibfield  {journal} {\bibinfo  {journal} {Phys. Rev. D}\ }\textbf {\bibinfo {volume} {111}},\ \bibinfo {pages} {103503} (\bibinfo {year} {2025}{\natexlab{a}})},\ \Eprint {http://arxiv.org/abs/2410.10808} {arXiv:2410.10808 [astro-ph.CO]} \BibitemShut {NoStop}%
\bibitem [{\citenamefont {Kim}\ \emph {et~al.}(2024)\citenamefont {Kim} \emph {et~al.}}]{Kim:2024dmg}%
  \BibitemOpen
  \bibfield  {author} {\bibinfo {author} {\bibfnamefont {Joshua}\ \bibnamefont {Kim}} \emph {et~al.},\ }\bibfield  {title} {\enquote {\bibinfo {title} {{The Atacama Cosmology Telescope DR6 and DESI: structure formation over cosmic time with a measurement of the cross-correlation of CMB lensing and luminous red galaxies}},}\ }\href {\doibase 10.1088/1475-7516/2024/12/022} {\bibfield  {journal} {\bibinfo  {journal} {JCAP}\ }\textbf {\bibinfo {volume} {12}},\ \bibinfo {pages} {022} (\bibinfo {year} {2024})},\ \Eprint {http://arxiv.org/abs/2407.04606} {arXiv:2407.04606 [astro-ph.CO]} \BibitemShut {NoStop}%
\bibitem [{\citenamefont {Sabogal}\ and\ \citenamefont {Nunes}(2025)}]{Sabogal:2025jbo}%
  \BibitemOpen
  \bibfield  {author} {\bibinfo {author} {\bibfnamefont {Miguel~A.}\ \bibnamefont {Sabogal}}\ and\ \bibinfo {author} {\bibfnamefont {Rafael~C.}\ \bibnamefont {Nunes}},\ }\bibfield  {title} {\enquote {\bibinfo {title} {{Robust Evidence for Dynamical Dark Energy from DESI Galaxy-CMB Lensing Cross-Correlation and Geometric Probes}},}\ }\href@noop {} {\  (\bibinfo {year} {2025})},\ \Eprint {http://arxiv.org/abs/2505.24465} {arXiv:2505.24465 [astro-ph.CO]} \BibitemShut {NoStop}%
\bibitem [{\citenamefont {Brout}\ and\ \citenamefont {{et al.}}(2022)}]{Brout:2022vxf}%
  \BibitemOpen
  \bibfield  {author} {\bibinfo {author} {\bibfnamefont {D.}~\bibnamefont {Brout}}\ and\ \bibinfo {author} {\bibnamefont {{et al.}}},\ }\bibfield  {title} {\enquote {\bibinfo {title} {{The Pantheon\,+ Analysis: The Full SN Ia Data Set and Light‑curve Release}},}\ }\href@noop {} {\bibfield  {journal} {\bibinfo  {journal} {ApJ}\ }\textbf {\bibinfo {volume} {938}},\ \bibinfo {pages} {113} (\bibinfo {year} {2022})},\ \Eprint {http://arxiv.org/abs/2112.03863} {2112.03863} \BibitemShut {NoStop}%
\bibitem [{\citenamefont {Riess}\ \emph {et~al.}(2022)\citenamefont {Riess} \emph {et~al.}}]{Riess:2021jrx}%
  \BibitemOpen
  \bibfield  {author} {\bibinfo {author} {\bibfnamefont {Adam~G.}\ \bibnamefont {Riess}} \emph {et~al.},\ }\bibfield  {title} {\enquote {\bibinfo {title} {{A Comprehensive Measurement of the Local Value of the Hubble Constant with 1 km s$^{−1}$ Mpc$^{−1}$ Uncertainty from the Hubble Space Telescope and the SH0ES Team}},}\ }\href {\doibase 10.3847/2041-8213/ac5c5b} {\bibfield  {journal} {\bibinfo  {journal} {Astrophys. J. Lett.}\ }\textbf {\bibinfo {volume} {934}},\ \bibinfo {pages} {L7} (\bibinfo {year} {2022})},\ \Eprint {http://arxiv.org/abs/2112.04510} {arXiv:2112.04510 [astro-ph.CO]} \BibitemShut {NoStop}%
\bibitem [{\citenamefont {Zhou}\ and\ \citenamefont {{et al.}}(2023)}]{Zhou:2023gji}%
  \BibitemOpen
  \bibfield  {author} {\bibinfo {author} {\bibfnamefont {R.}~\bibnamefont {Zhou}}\ and\ \bibinfo {author} {\bibnamefont {{et al.}}},\ }\bibfield  {title} {\enquote {\bibinfo {title} {{DESI Luminous Red Galaxy Samples for Cross‐correlations}},}\ }\href@noop {} {\bibfield  {journal} {\bibinfo  {journal} {JCAP}\ }\textbf {\bibinfo {volume} {11}},\ \bibinfo {pages} {097} (\bibinfo {year} {2023})},\ \Eprint {http://arxiv.org/abs/2309.06443} {2309.06443} \BibitemShut {NoStop}%
\bibitem [{\citenamefont {Dey}\ and\ \citenamefont {{et al.}}(2019)}]{Dey:2019}%
  \BibitemOpen
  \bibfield  {author} {\bibinfo {author} {\bibfnamefont {A.}~\bibnamefont {Dey}}\ and\ \bibinfo {author} {\bibnamefont {{et al.}}},\ }\bibfield  {title} {\enquote {\bibinfo {title} {{Overview of the DESI Legacy Imaging Surveys}},}\ }\href@noop {} {\bibfield  {journal} {\bibinfo  {journal} {AJ}\ }\textbf {\bibinfo {volume} {157}},\ \bibinfo {pages} {168} (\bibinfo {year} {2019})},\ \Eprint {http://arxiv.org/abs/1804.08657} {1804.08657} \BibitemShut {NoStop}%
\bibitem [{\citenamefont {Adame}\ \emph {et~al.}(2024)\citenamefont {Adame} \emph {et~al.}}]{DESI:2023dwi}%
  \BibitemOpen
  \bibfield  {author} {\bibinfo {author} {\bibfnamefont {A.~G.}\ \bibnamefont {Adame}} \emph {et~al.} (\bibinfo {collaboration} {DESI}),\ }\bibfield  {title} {\enquote {\bibinfo {title} {{Validation of the Scientific Program for the Dark Energy Spectroscopic Instrument}},}\ }\href {\doibase 10.3847/1538-3881/ad0b08} {\bibfield  {journal} {\bibinfo  {journal} {Astron. J.}\ }\textbf {\bibinfo {volume} {167}},\ \bibinfo {pages} {62} (\bibinfo {year} {2024})},\ \Eprint {http://arxiv.org/abs/2306.06307} {arXiv:2306.06307 [astro-ph.CO]} \BibitemShut {NoStop}%
\bibitem [{\citenamefont {Louis}\ and\ \citenamefont {Collaboration)}(2025)}]{ACTDR6:PowerSpectra}%
  \BibitemOpen
  \bibfield  {author} {\bibinfo {author} {\bibfnamefont {T.}~\bibnamefont {Louis}}\ and\ \bibinfo {author} {\bibfnamefont {{et al.}~(ACT}\ \bibnamefont {Collaboration)}},\ }\bibfield  {title} {\enquote {\bibinfo {title} {{The Atacama Cosmology Telescope: DR6 Power Spectra, Likelihoods and \ensuremath{\Lambda}CDM Parameters}},}\ }\href@noop {} {\  (\bibinfo {year} {2025})},\ \Eprint {http://arxiv.org/abs/2503.14452} {arXiv:2503.14452 [astro-ph.CO]} \BibitemShut {NoStop}%
\bibitem [{\citenamefont {Pagano}\ \emph {et~al.}(2020)\citenamefont {Pagano}, \citenamefont {Delouis}, \citenamefont {Mottet}, \citenamefont {Puget},\ and\ \citenamefont {Vibert}}]{Pagano:2019tci}%
  \BibitemOpen
  \bibfield  {author} {\bibinfo {author} {\bibfnamefont {L.}~\bibnamefont {Pagano}}, \bibinfo {author} {\bibfnamefont {J.~M.}\ \bibnamefont {Delouis}}, \bibinfo {author} {\bibfnamefont {S.}~\bibnamefont {Mottet}}, \bibinfo {author} {\bibfnamefont {J.~L.}\ \bibnamefont {Puget}}, \ and\ \bibinfo {author} {\bibfnamefont {L.}~\bibnamefont {Vibert}},\ }\bibfield  {title} {\enquote {\bibinfo {title} {{Reionization optical depth determination from Planck HFI data with ten percent accuracy}},}\ }\href {\doibase 10.1051/0004-6361/201936630} {\bibfield  {journal} {\bibinfo  {journal} {Astron. Astrophys.}\ }\textbf {\bibinfo {volume} {635}},\ \bibinfo {pages} {A99} (\bibinfo {year} {2020})},\ \Eprint {http://arxiv.org/abs/1908.09856} {arXiv:1908.09856 [astro-ph.CO]} \BibitemShut {NoStop}%
\bibitem [{\citenamefont {Prince}\ and\ \citenamefont {Dunkley}(2022)}]{Prince:2021fdv}%
  \BibitemOpen
  \bibfield  {author} {\bibinfo {author} {\bibfnamefont {Heather}\ \bibnamefont {Prince}}\ and\ \bibinfo {author} {\bibfnamefont {Jo}~\bibnamefont {Dunkley}},\ }\bibfield  {title} {\enquote {\bibinfo {title} {{Compressed Python likelihood for large scale temperature and polarization from Planck}},}\ }\href {\doibase 10.1103/PhysRevD.105.023518} {\bibfield  {journal} {\bibinfo  {journal} {Phys. Rev. D}\ }\textbf {\bibinfo {volume} {105}},\ \bibinfo {pages} {023518} (\bibinfo {year} {2022})},\ \Eprint {http://arxiv.org/abs/2104.05715} {arXiv:2104.05715 [astro-ph.CO]} \BibitemShut {NoStop}%
\bibitem [{\citenamefont {Qu}\ \emph {et~al.}(2024)\citenamefont {Qu} \emph {et~al.}}]{ACT:2023dou}%
  \BibitemOpen
  \bibfield  {author} {\bibinfo {author} {\bibfnamefont {Frank~J.}\ \bibnamefont {Qu}} \emph {et~al.} (\bibinfo {collaboration} {ACT}),\ }\bibfield  {title} {\enquote {\bibinfo {title} {{The Atacama Cosmology Telescope: A Measurement of the DR6 CMB Lensing Power Spectrum and Its Implications for Structure Growth}},}\ }\href {\doibase 10.3847/1538-4357/acfe06} {\bibfield  {journal} {\bibinfo  {journal} {Astrophys. J.}\ }\textbf {\bibinfo {volume} {962}},\ \bibinfo {pages} {112} (\bibinfo {year} {2024})},\ \Eprint {http://arxiv.org/abs/2304.05202} {arXiv:2304.05202 [astro-ph.CO]} \BibitemShut {NoStop}%
\bibitem [{\citenamefont {Akrami}\ \emph {et~al.}(2020)\citenamefont {Akrami} \emph {et~al.}}]{Planck:2020olo}%
  \BibitemOpen
  \bibfield  {author} {\bibinfo {author} {\bibfnamefont {Y.}~\bibnamefont {Akrami}} \emph {et~al.} (\bibinfo {collaboration} {Planck}),\ }\bibfield  {title} {\enquote {\bibinfo {title} {{$Planck$ intermediate results. LVII. Joint Planck LFI and HFI data processing}},}\ }\href {\doibase 10.1051/0004-6361/202038073} {\bibfield  {journal} {\bibinfo  {journal} {Astron. Astrophys.}\ }\textbf {\bibinfo {volume} {643}},\ \bibinfo {pages} {A42} (\bibinfo {year} {2020})},\ \Eprint {http://arxiv.org/abs/2007.04997} {arXiv:2007.04997 [astro-ph.CO]} \BibitemShut {NoStop}%
\bibitem [{\citenamefont {Tristram}\ and\ \citenamefont {{et al.}}(2023)}]{Tristram:2023haj}%
  \BibitemOpen
  \bibfield  {author} {\bibinfo {author} {\bibfnamefont {M.}~\bibnamefont {Tristram}}\ and\ \bibinfo {author} {\bibnamefont {{et al.}}},\ }\bibfield  {title} {\enquote {\bibinfo {title} {{Cosmological Parameters from the Final Planck Data Release (PR4)}},}\ }\href@noop {} {\  (\bibinfo {year} {2023})},\ \Eprint {http://arxiv.org/abs/2309.10034} {arXiv:2309.10034 [astro-ph.CO]} \BibitemShut {NoStop}%
\bibitem [{\citenamefont {{DESI Collaboration}}(2016)}]{DESI:instrument}%
  \BibitemOpen
  \bibfield  {author} {\bibinfo {author} {\bibnamefont {{DESI Collaboration}}},\ }\bibfield  {title} {\enquote {\bibinfo {title} {{The DESI Experiment Part II: Instrument Design}},}\ }\href@noop {} {\  (\bibinfo {year} {2016})},\ \Eprint {http://arxiv.org/abs/1611.00037} {arXiv:1611.00037 [astro-ph.IM]} \BibitemShut {NoStop}%
\bibitem [{\citenamefont {Torrado}\ and\ \citenamefont {Lewis}(2021)}]{Torrado:2020dgo}%
  \BibitemOpen
  \bibfield  {author} {\bibinfo {author} {\bibfnamefont {Jesus}\ \bibnamefont {Torrado}}\ and\ \bibinfo {author} {\bibfnamefont {Antony}\ \bibnamefont {Lewis}},\ }\bibfield  {title} {\enquote {\bibinfo {title} {{Cobaya: Code for Bayesian Analysis of hierarchical physical models}},}\ }\href {\doibase 10.1088/1475-7516/2021/05/057} {\bibfield  {journal} {\bibinfo  {journal} {JCAP}\ }\textbf {\bibinfo {volume} {05}},\ \bibinfo {pages} {057} (\bibinfo {year} {2021})},\ \Eprint {http://arxiv.org/abs/2005.05290} {arXiv:2005.05290 [astro-ph.IM]} \BibitemShut {NoStop}%
\bibitem [{\citenamefont {Fang}\ \emph {et~al.}(2020)\citenamefont {Fang}, \citenamefont {Krause}, \citenamefont {Eifler},\ and\ \citenamefont {MacCrann}}]{Fang:2019xat}%
  \BibitemOpen
  \bibfield  {author} {\bibinfo {author} {\bibfnamefont {Xiao}\ \bibnamefont {Fang}}, \bibinfo {author} {\bibfnamefont {Elisabeth}\ \bibnamefont {Krause}}, \bibinfo {author} {\bibfnamefont {Tim}\ \bibnamefont {Eifler}}, \ and\ \bibinfo {author} {\bibfnamefont {Niall}\ \bibnamefont {MacCrann}},\ }\bibfield  {title} {\enquote {\bibinfo {title} {{Beyond Limber: Efficient computation of angular power spectra for galaxy clustering and weak lensing}},}\ }\href {\doibase 10.1088/1475-7516/2020/05/010} {\bibfield  {journal} {\bibinfo  {journal} {JCAP}\ }\textbf {\bibinfo {volume} {05}},\ \bibinfo {pages} {010} (\bibinfo {year} {2020})},\ \Eprint {http://arxiv.org/abs/1911.11947} {arXiv:1911.11947 [astro-ph.CO]} \BibitemShut {NoStop}%
\bibitem [{\citenamefont {Chisari}\ \emph {et~al.}(2019)\citenamefont {Chisari} \emph {et~al.}}]{LSSTDarkEnergyScience:2018yem}%
  \BibitemOpen
  \bibfield  {author} {\bibinfo {author} {\bibfnamefont {Nora~Elisa}\ \bibnamefont {Chisari}} \emph {et~al.} (\bibinfo {collaboration} {LSST Dark Energy Science}),\ }\bibfield  {title} {\enquote {\bibinfo {title} {{Core Cosmology Library: Precision Cosmological Predictions for LSST}},}\ }\href {\doibase 10.3847/1538-4365/ab1658} {\bibfield  {journal} {\bibinfo  {journal} {Astrophys. J. Suppl.}\ }\textbf {\bibinfo {volume} {242}},\ \bibinfo {pages} {2} (\bibinfo {year} {2019})},\ \Eprint {http://arxiv.org/abs/1812.05995} {arXiv:1812.05995 [astro-ph.CO]} \BibitemShut {NoStop}%
\bibitem [{\citenamefont {Reymond}\ \emph {et~al.}(2025)\citenamefont {Reymond}, \citenamefont {Reeves}, \citenamefont {Zhang},\ and\ \citenamefont {Refregier}}]{Reymond:2025ixl}%
  \BibitemOpen
  \bibfield  {author} {\bibinfo {author} {\bibfnamefont {Laura}\ \bibnamefont {Reymond}}, \bibinfo {author} {\bibfnamefont {Alexander}\ \bibnamefont {Reeves}}, \bibinfo {author} {\bibfnamefont {Pierre}\ \bibnamefont {Zhang}}, \ and\ \bibinfo {author} {\bibfnamefont {Alexandre}\ \bibnamefont {Refregier}},\ }\bibfield  {title} {\enquote {\bibinfo {title} {{$\texttt{SwiftC}_\ell$: fast differentiable angular power spectra beyond Limber}},}\ }\href@noop {} {\  (\bibinfo {year} {2025})},\ \Eprint {http://arxiv.org/abs/2505.22718} {arXiv:2505.22718 [astro-ph.CO]} \BibitemShut {NoStop}%
\bibitem [{\citenamefont {{Limber}}(1953)}]{limber_approx}%
  \BibitemOpen
  \bibfield  {author} {\bibinfo {author} {\bibfnamefont {D.~Nelson}\ \bibnamefont {{Limber}}},\ }\bibfield  {title} {\enquote {\bibinfo {title} {{The Analysis of Counts of the Extragalactic Nebulae in Terms of a Fluctuating Density Field.}}}\ }\href {\doibase 10.1086/145672} {\bibfield  {journal} {\bibinfo  {journal} {\apj}\ }\textbf {\bibinfo {volume} {117}},\ \bibinfo {pages} {134} (\bibinfo {year} {1953})}\BibitemShut {NoStop}%
\bibitem [{\citenamefont {Poulin}\ \emph {et~al.}(2018)\citenamefont {Poulin}, \citenamefont {Smith}, \citenamefont {Grin}, \citenamefont {Karwal},\ and\ \citenamefont {Kamionkowski}}]{Poulin:2018dzj}%
  \BibitemOpen
  \bibfield  {author} {\bibinfo {author} {\bibfnamefont {Vivian}\ \bibnamefont {Poulin}}, \bibinfo {author} {\bibfnamefont {Tristan~L.}\ \bibnamefont {Smith}}, \bibinfo {author} {\bibfnamefont {Daniel}\ \bibnamefont {Grin}}, \bibinfo {author} {\bibfnamefont {Tanvi}\ \bibnamefont {Karwal}}, \ and\ \bibinfo {author} {\bibfnamefont {Marc}\ \bibnamefont {Kamionkowski}},\ }\bibfield  {title} {\enquote {\bibinfo {title} {{Cosmological implications of ultralight axionlike fields}},}\ }\href {\doibase 10.1103/PhysRevD.98.083525} {\bibfield  {journal} {\bibinfo  {journal} {Phys. Rev. D}\ }\textbf {\bibinfo {volume} {98}},\ \bibinfo {pages} {083525} (\bibinfo {year} {2018})},\ \Eprint {http://arxiv.org/abs/1806.10608} {arXiv:1806.10608 [astro-ph.CO]} \BibitemShut {NoStop}%
\bibitem [{\citenamefont {Blas}\ \emph {et~al.}(2011)\citenamefont {Blas}, \citenamefont {Lesgourgues},\ and\ \citenamefont {Tram}}]{Blas:2011rf}%
  \BibitemOpen
  \bibfield  {author} {\bibinfo {author} {\bibfnamefont {Diego}\ \bibnamefont {Blas}}, \bibinfo {author} {\bibfnamefont {Julien}\ \bibnamefont {Lesgourgues}}, \ and\ \bibinfo {author} {\bibfnamefont {Thomas}\ \bibnamefont {Tram}},\ }\bibfield  {title} {\enquote {\bibinfo {title} {{The Cosmic Linear Anisotropy Solving System (CLASS) II: Approximation schemes}},}\ }\href {\doibase 10.1088/1475-7516/2011/07/034} {\bibfield  {journal} {\bibinfo  {journal} {JCAP}\ }\textbf {\bibinfo {volume} {07}},\ \bibinfo {pages} {034} (\bibinfo {year} {2011})},\ \Eprint {http://arxiv.org/abs/1104.2933} {arXiv:1104.2933 [astro-ph.CO]} \BibitemShut {NoStop}%
\bibitem [{\citenamefont {Calabrese}\ \emph {et~al.}(2025)\citenamefont {Calabrese} \emph {et~al.}}]{ACT:2025tim}%
  \BibitemOpen
  \bibfield  {author} {\bibinfo {author} {\bibfnamefont {Erminia}\ \bibnamefont {Calabrese}} \emph {et~al.} (\bibinfo {collaboration} {ACT}),\ }\bibfield  {title} {\enquote {\bibinfo {title} {{The Atacama Cosmology Telescope: DR6 Constraints on Extended Cosmological Models}},}\ }\href@noop {} {\  (\bibinfo {year} {2025})},\ \Eprint {http://arxiv.org/abs/2503.14454} {arXiv:2503.14454 [astro-ph.CO]} \BibitemShut {NoStop}%
\bibitem [{\citenamefont {Padmanabhan}\ \emph {et~al.}(2007)\citenamefont {Padmanabhan} \emph {et~al.}}]{SDSS:2006egz}%
  \BibitemOpen
  \bibfield  {author} {\bibinfo {author} {\bibfnamefont {Nikhil}\ \bibnamefont {Padmanabhan}} \emph {et~al.} (\bibinfo {collaboration} {SDSS}),\ }\bibfield  {title} {\enquote {\bibinfo {title} {{The Clustering of Luminous Red Galaxies in the Sloan Digital Sky Survey Imaging Data}},}\ }\href {\doibase 10.1111/j.1365-2966.2007.11593.x} {\bibfield  {journal} {\bibinfo  {journal} {Mon. Not. Roy. Astron. Soc.}\ }\textbf {\bibinfo {volume} {378}},\ \bibinfo {pages} {852--872} (\bibinfo {year} {2007})},\ \Eprint {http://arxiv.org/abs/astro-ph/0605302} {arXiv:astro-ph/0605302} \BibitemShut {NoStop}%
\bibitem [{\citenamefont {Madhavacheril}\ \emph {et~al.}(2024)\citenamefont {Madhavacheril} \emph {et~al.}}]{ACT:2023kun}%
  \BibitemOpen
  \bibfield  {author} {\bibinfo {author} {\bibfnamefont {Mathew~S.}\ \bibnamefont {Madhavacheril}} \emph {et~al.} (\bibinfo {collaboration} {ACT}),\ }\bibfield  {title} {\enquote {\bibinfo {title} {{The Atacama Cosmology Telescope: DR6 Gravitational Lensing Map and Cosmological Parameters}},}\ }\href {\doibase 10.3847/1538-4357/acff5f} {\bibfield  {journal} {\bibinfo  {journal} {Astrophys. J.}\ }\textbf {\bibinfo {volume} {962}},\ \bibinfo {pages} {113} (\bibinfo {year} {2024})},\ \Eprint {http://arxiv.org/abs/2304.05203} {arXiv:2304.05203 [astro-ph.CO]} \BibitemShut {NoStop}%
\bibitem [{\citenamefont {MacCrann}\ \emph {et~al.}(2024)\citenamefont {MacCrann} \emph {et~al.}}]{ACT:2023ubw}%
  \BibitemOpen
  \bibfield  {author} {\bibinfo {author} {\bibfnamefont {Niall}\ \bibnamefont {MacCrann}} \emph {et~al.} (\bibinfo {collaboration} {ACT}),\ }\bibfield  {title} {\enquote {\bibinfo {title} {{The Atacama Cosmology Telescope: Mitigating the Impact of Extragalactic Foregrounds for the DR6 Cosmic Microwave Background Lensing Analysis}},}\ }\href {\doibase 10.3847/1538-4357/ad2610} {\bibfield  {journal} {\bibinfo  {journal} {Astrophys. J.}\ }\textbf {\bibinfo {volume} {966}},\ \bibinfo {pages} {138} (\bibinfo {year} {2024})},\ \Eprint {http://arxiv.org/abs/2304.05196} {arXiv:2304.05196 [astro-ph.CO]} \BibitemShut {NoStop}%
\bibitem [{\citenamefont {Spurio~Mancini}\ \emph {et~al.}(2022)\citenamefont {Spurio~Mancini}, \citenamefont {Piras}, \citenamefont {Alsing}, \citenamefont {Joachimi},\ and\ \citenamefont {Hobson}}]{SpurioMancini:2021ppk}%
  \BibitemOpen
  \bibfield  {author} {\bibinfo {author} {\bibfnamefont {Alessio}\ \bibnamefont {Spurio~Mancini}}, \bibinfo {author} {\bibfnamefont {Davide}\ \bibnamefont {Piras}}, \bibinfo {author} {\bibfnamefont {Justin}\ \bibnamefont {Alsing}}, \bibinfo {author} {\bibfnamefont {Benjamin}\ \bibnamefont {Joachimi}}, \ and\ \bibinfo {author} {\bibfnamefont {Michael~P.}\ \bibnamefont {Hobson}},\ }\bibfield  {title} {\enquote {\bibinfo {title} {{CosmoPower: emulating cosmological power spectra for accelerated Bayesian inference from next-generation surveys}},}\ }\href {\doibase 10.1093/mnras/stac064} {\bibfield  {journal} {\bibinfo  {journal} {Mon. Not. Roy. Astron. Soc.}\ }\textbf {\bibinfo {volume} {511}},\ \bibinfo {pages} {1771--1788} (\bibinfo {year} {2022})},\ \Eprint {http://arxiv.org/abs/2106.03846} {arXiv:2106.03846 [astro-ph.CO]} \BibitemShut {NoStop}%
\bibitem [{\citenamefont {Abadi}\ \emph {et~al.}(2015)\citenamefont {Abadi}, \citenamefont {Agarwal}, \citenamefont {Barham}, \citenamefont {Brevdo}, \citenamefont {Chen}, \citenamefont {Citro}, \citenamefont {Corrado}, \citenamefont {Davis}, \citenamefont {Dean}, \citenamefont {Devin}, \citenamefont {Ghemawat}, \citenamefont {Goodfellow}, \citenamefont {Harp}, \citenamefont {Irving}, \citenamefont {Isard}, \citenamefont {Jia}, \citenamefont {Jozefowicz}, \citenamefont {Kaiser}, \citenamefont {Kudlur}, \citenamefont {Levenberg}, \citenamefont {Man\'{e}}, \citenamefont {Monga}, \citenamefont {Moore}, \citenamefont {Murray}, \citenamefont {Olah}, \citenamefont {Schuster}, \citenamefont {Shlens}, \citenamefont {Steiner}, \citenamefont {Sutskever}, \citenamefont {Talwar}, \citenamefont {Tucker}, \citenamefont {Vanhoucke}, \citenamefont {Vasudevan}, \citenamefont {Vi\'{e}gas}, \citenamefont {Vinyals}, \citenamefont {Warden}, \citenamefont {Wattenberg}, \citenamefont {Wicke}, \citenamefont {Yu},\ and\ \citenamefont
  {Zheng}}]{tensorflow2015-whitepaper}%
  \BibitemOpen
  \bibfield  {author} {\bibinfo {author} {\bibfnamefont {Mart\'{i}n}\ \bibnamefont {Abadi}}, \bibinfo {author} {\bibfnamefont {Ashish}\ \bibnamefont {Agarwal}}, \bibinfo {author} {\bibfnamefont {Paul}\ \bibnamefont {Barham}}, \bibinfo {author} {\bibfnamefont {Eugene}\ \bibnamefont {Brevdo}}, \bibinfo {author} {\bibfnamefont {Zhifeng}\ \bibnamefont {Chen}}, \bibinfo {author} {\bibfnamefont {Craig}\ \bibnamefont {Citro}}, \bibinfo {author} {\bibfnamefont {Greg~S.}\ \bibnamefont {Corrado}}, \bibinfo {author} {\bibfnamefont {Andy}\ \bibnamefont {Davis}}, \bibinfo {author} {\bibfnamefont {Jeffrey}\ \bibnamefont {Dean}}, \bibinfo {author} {\bibfnamefont {Matthieu}\ \bibnamefont {Devin}}, \bibinfo {author} {\bibfnamefont {Sanjay}\ \bibnamefont {Ghemawat}}, \bibinfo {author} {\bibfnamefont {Ian}\ \bibnamefont {Goodfellow}}, \bibinfo {author} {\bibfnamefont {Andrew}\ \bibnamefont {Harp}}, \bibinfo {author} {\bibfnamefont {Geoffrey}\ \bibnamefont {Irving}}, \bibinfo {author} {\bibfnamefont {Michael}\ \bibnamefont
  {Isard}}, \bibinfo {author} {\bibfnamefont {Yangqing}\ \bibnamefont {Jia}}, \bibinfo {author} {\bibfnamefont {Rafal}\ \bibnamefont {Jozefowicz}}, \bibinfo {author} {\bibfnamefont {Lukasz}\ \bibnamefont {Kaiser}}, \bibinfo {author} {\bibfnamefont {Manjunath}\ \bibnamefont {Kudlur}}, \bibinfo {author} {\bibfnamefont {Josh}\ \bibnamefont {Levenberg}}, \bibinfo {author} {\bibfnamefont {Dandelion}\ \bibnamefont {Man\'{e}}}, \bibinfo {author} {\bibfnamefont {Rajat}\ \bibnamefont {Monga}}, \bibinfo {author} {\bibfnamefont {Sherry}\ \bibnamefont {Moore}}, \bibinfo {author} {\bibfnamefont {Derek}\ \bibnamefont {Murray}}, \bibinfo {author} {\bibfnamefont {Chris}\ \bibnamefont {Olah}}, \bibinfo {author} {\bibfnamefont {Mike}\ \bibnamefont {Schuster}}, \bibinfo {author} {\bibfnamefont {Jonathon}\ \bibnamefont {Shlens}}, \bibinfo {author} {\bibfnamefont {Benoit}\ \bibnamefont {Steiner}}, \bibinfo {author} {\bibfnamefont {Ilya}\ \bibnamefont {Sutskever}}, \bibinfo {author} {\bibfnamefont {Kunal}\ \bibnamefont {Talwar}},
  \bibinfo {author} {\bibfnamefont {Paul}\ \bibnamefont {Tucker}}, \bibinfo {author} {\bibfnamefont {Vincent}\ \bibnamefont {Vanhoucke}}, \bibinfo {author} {\bibfnamefont {Vijay}\ \bibnamefont {Vasudevan}}, \bibinfo {author} {\bibfnamefont {Fernanda}\ \bibnamefont {Vi\'{e}gas}}, \bibinfo {author} {\bibfnamefont {Oriol}\ \bibnamefont {Vinyals}}, \bibinfo {author} {\bibfnamefont {Pete}\ \bibnamefont {Warden}}, \bibinfo {author} {\bibfnamefont {Martin}\ \bibnamefont {Wattenberg}}, \bibinfo {author} {\bibfnamefont {Martin}\ \bibnamefont {Wicke}}, \bibinfo {author} {\bibfnamefont {Yuan}\ \bibnamefont {Yu}}, \ and\ \bibinfo {author} {\bibfnamefont {Xiaoqiang}\ \bibnamefont {Zheng}},\ }\href {https://www.tensorflow.org/} {\enquote {\bibinfo {title} {{TensorFlow}: Large-scale machine learning on heterogeneous systems},}\ } (\bibinfo {year} {2015}),\ \bibinfo {note} {software available from tensorflow.org}\BibitemShut {NoStop}%
\bibitem [{\citenamefont {Heek}\ \emph {et~al.}(2024)\citenamefont {Heek}, \citenamefont {Levskaya}, \citenamefont {Oliver}, \citenamefont {Ritter}, \citenamefont {Rondepierre}, \citenamefont {Steiner},\ and\ \citenamefont {van {Z}ee}}]{flax2020github}%
  \BibitemOpen
  \bibfield  {author} {\bibinfo {author} {\bibfnamefont {Jonathan}\ \bibnamefont {Heek}}, \bibinfo {author} {\bibfnamefont {Anselm}\ \bibnamefont {Levskaya}}, \bibinfo {author} {\bibfnamefont {Avital}\ \bibnamefont {Oliver}}, \bibinfo {author} {\bibfnamefont {Marvin}\ \bibnamefont {Ritter}}, \bibinfo {author} {\bibfnamefont {Bertrand}\ \bibnamefont {Rondepierre}}, \bibinfo {author} {\bibfnamefont {Andreas}\ \bibnamefont {Steiner}}, \ and\ \bibinfo {author} {\bibfnamefont {Marc}\ \bibnamefont {van {Z}ee}},\ }\href {http://github.com/google/flax} {\enquote {\bibinfo {title} {{F}lax: A neural network library and ecosystem for {JAX}},}\ } (\bibinfo {year} {2024})\BibitemShut {NoStop}%
\bibitem [{\citenamefont {Hivon}\ \emph {et~al.}(2002)\citenamefont {Hivon}, \citenamefont {Gorski}, \citenamefont {Netterfield}, \citenamefont {Crill}, \citenamefont {Prunet},\ and\ \citenamefont {Hansen}}]{Hivon:2001jp}%
  \BibitemOpen
  \bibfield  {author} {\bibinfo {author} {\bibfnamefont {E.}~\bibnamefont {Hivon}}, \bibinfo {author} {\bibfnamefont {K.~M.}\ \bibnamefont {Gorski}}, \bibinfo {author} {\bibfnamefont {C.~B.}\ \bibnamefont {Netterfield}}, \bibinfo {author} {\bibfnamefont {B.~P.}\ \bibnamefont {Crill}}, \bibinfo {author} {\bibfnamefont {S.}~\bibnamefont {Prunet}}, \ and\ \bibinfo {author} {\bibfnamefont {F.}~\bibnamefont {Hansen}},\ }\bibfield  {title} {\enquote {\bibinfo {title} {{Master of the cosmic microwave background anisotropy power spectrum: a fast method for statistical analysis of large and complex cosmic microwave background data sets}},}\ }\href {\doibase 10.1086/338126} {\bibfield  {journal} {\bibinfo  {journal} {Astrophys. J.}\ }\textbf {\bibinfo {volume} {567}},\ \bibinfo {pages} {2} (\bibinfo {year} {2002})},\ \Eprint {http://arxiv.org/abs/astro-ph/0105302} {arXiv:astro-ph/0105302} \BibitemShut {NoStop}%
\bibitem [{\citenamefont {Alonso}\ \emph {et~al.}(2019)\citenamefont {Alonso}, \citenamefont {Sanchez},\ and\ \citenamefont {Slosar}}]{Alonso:2018jzx}%
  \BibitemOpen
  \bibfield  {author} {\bibinfo {author} {\bibfnamefont {David}\ \bibnamefont {Alonso}}, \bibinfo {author} {\bibfnamefont {Javier}\ \bibnamefont {Sanchez}}, \ and\ \bibinfo {author} {\bibfnamefont {An\v{z}e}\ \bibnamefont {Slosar}} (\bibinfo {collaboration} {LSST Dark Energy Science}),\ }\bibfield  {title} {\enquote {\bibinfo {title} {{A unified pseudo-$C_\ell$ framework}},}\ }\href {\doibase 10.1093/mnras/stz093} {\bibfield  {journal} {\bibinfo  {journal} {Mon. Not. Roy. Astron. Soc.}\ }\textbf {\bibinfo {volume} {484}},\ \bibinfo {pages} {4127--4151} (\bibinfo {year} {2019})},\ \Eprint {http://arxiv.org/abs/1809.09603} {arXiv:1809.09603 [astro-ph.CO]} \BibitemShut {NoStop}%
\bibitem [{\citenamefont {Nicola}\ \emph {et~al.}(2021)\citenamefont {Nicola}, \citenamefont {Garc\'\i{}a-Garc\'\i{}a}, \citenamefont {Alonso}, \citenamefont {Dunkley}, \citenamefont {Ferreira}, \citenamefont {Slosar},\ and\ \citenamefont {Spergel}}]{Nicola:2020lhi}%
  \BibitemOpen
  \bibfield  {author} {\bibinfo {author} {\bibfnamefont {Andrina}\ \bibnamefont {Nicola}}, \bibinfo {author} {\bibfnamefont {Carlos}\ \bibnamefont {Garc\'\i{}a-Garc\'\i{}a}}, \bibinfo {author} {\bibfnamefont {David}\ \bibnamefont {Alonso}}, \bibinfo {author} {\bibfnamefont {Jo}~\bibnamefont {Dunkley}}, \bibinfo {author} {\bibfnamefont {Pedro~G.}\ \bibnamefont {Ferreira}}, \bibinfo {author} {\bibfnamefont {An\v{z}e}\ \bibnamefont {Slosar}}, \ and\ \bibinfo {author} {\bibfnamefont {David~N.}\ \bibnamefont {Spergel}},\ }\bibfield  {title} {\enquote {\bibinfo {title} {{Cosmic shear power spectra in practice}},}\ }\href {\doibase 10.1088/1475-7516/2021/03/067} {\bibfield  {journal} {\bibinfo  {journal} {JCAP}\ }\textbf {\bibinfo {volume} {03}},\ \bibinfo {pages} {067} (\bibinfo {year} {2021})},\ \Eprint {http://arxiv.org/abs/2010.09717} {arXiv:2010.09717 [astro-ph.CO]} \BibitemShut {NoStop}%
\bibitem [{\citenamefont {Wolz}\ \emph {et~al.}(2025)\citenamefont {Wolz}, \citenamefont {Alonso},\ and\ \citenamefont {Nicola}}]{Wolz:2024dro}%
  \BibitemOpen
  \bibfield  {author} {\bibinfo {author} {\bibfnamefont {Kevin}\ \bibnamefont {Wolz}}, \bibinfo {author} {\bibfnamefont {David}\ \bibnamefont {Alonso}}, \ and\ \bibinfo {author} {\bibfnamefont {Andrina}\ \bibnamefont {Nicola}},\ }\bibfield  {title} {\enquote {\bibinfo {title} {{Catalog-based pseudo-C\ensuremath{\ell}s}},}\ }\href {\doibase 10.1088/1475-7516/2025/01/028} {\bibfield  {journal} {\bibinfo  {journal} {JCAP}\ }\textbf {\bibinfo {volume} {01}},\ \bibinfo {pages} {028} (\bibinfo {year} {2025})},\ \Eprint {http://arxiv.org/abs/2407.21013} {arXiv:2407.21013 [astro-ph.CO]} \BibitemShut {NoStop}%
\bibitem [{\citenamefont {Baleato~Lizancos}\ and\ \citenamefont {White}(2024)}]{BaleatoLizancos:2023jbr}%
  \BibitemOpen
  \bibfield  {author} {\bibinfo {author} {\bibfnamefont {Ant\'on}\ \bibnamefont {Baleato~Lizancos}}\ and\ \bibinfo {author} {\bibfnamefont {Martin}\ \bibnamefont {White}},\ }\bibfield  {title} {\enquote {\bibinfo {title} {{Harmonic analysis of discrete tracers of large-scale structure}},}\ }\href {\doibase 10.1088/1475-7516/2024/05/010} {\bibfield  {journal} {\bibinfo  {journal} {JCAP}\ }\textbf {\bibinfo {volume} {05}},\ \bibinfo {pages} {010} (\bibinfo {year} {2024})},\ \Eprint {http://arxiv.org/abs/2312.12285} {arXiv:2312.12285 [astro-ph.CO]} \BibitemShut {NoStop}%
\bibitem [{\citenamefont {Aghanim}\ \emph {et~al.}(2020)\citenamefont {Aghanim} \emph {et~al.}}]{Planck:2018vyg}%
  \BibitemOpen
  \bibfield  {author} {\bibinfo {author} {\bibfnamefont {N.}~\bibnamefont {Aghanim}} \emph {et~al.} (\bibinfo {collaboration} {Planck}),\ }\bibfield  {title} {\enquote {\bibinfo {title} {{Planck 2018 results. VI. Cosmological parameters}},}\ }\href {\doibase 10.1051/0004-6361/201833910} {\bibfield  {journal} {\bibinfo  {journal} {Astron. Astrophys.}\ }\textbf {\bibinfo {volume} {641}},\ \bibinfo {pages} {A6} (\bibinfo {year} {2020})},\ \bibinfo {note} {[Erratum: Astron.Astrophys. 652, C4 (2021)]},\ \Eprint {http://arxiv.org/abs/1807.06209} {arXiv:1807.06209 [astro-ph.CO]} \BibitemShut {NoStop}%
\bibitem [{\citenamefont {Potter}\ \emph {et~al.}(2016)\citenamefont {Potter}, \citenamefont {Stadel},\ and\ \citenamefont {Teyssier}}]{Potter:2016ttn}%
  \BibitemOpen
  \bibfield  {author} {\bibinfo {author} {\bibfnamefont {Douglas}\ \bibnamefont {Potter}}, \bibinfo {author} {\bibfnamefont {Joachim}\ \bibnamefont {Stadel}}, \ and\ \bibinfo {author} {\bibfnamefont {Romain}\ \bibnamefont {Teyssier}},\ }\bibfield  {title} {\enquote {\bibinfo {title} {{PKDGRAV3: Beyond Trillion Particle Cosmological Simulations for the Next Era of Galaxy Surveys}},}\ }\href@noop {} {\  (\bibinfo {year} {2016})},\ \Eprint {http://arxiv.org/abs/1609.08621} {arXiv:1609.08621 [astro-ph.IM]} \BibitemShut {NoStop}%
\bibitem [{\citenamefont {Sgier}\ \emph {et~al.}(2021)\citenamefont {Sgier}, \citenamefont {Fluri}, \citenamefont {Herbel}, \citenamefont {R\'efr\'egier}, \citenamefont {Amara}, \citenamefont {Kacprzak},\ and\ \citenamefont {Nicola}}]{Sgier:2020das}%
  \BibitemOpen
  \bibfield  {author} {\bibinfo {author} {\bibfnamefont {Raphael}\ \bibnamefont {Sgier}}, \bibinfo {author} {\bibfnamefont {Janis}\ \bibnamefont {Fluri}}, \bibinfo {author} {\bibfnamefont {J\"org}\ \bibnamefont {Herbel}}, \bibinfo {author} {\bibfnamefont {Alexandre}\ \bibnamefont {R\'efr\'egier}}, \bibinfo {author} {\bibfnamefont {Adam}\ \bibnamefont {Amara}}, \bibinfo {author} {\bibfnamefont {Tomasz}\ \bibnamefont {Kacprzak}}, \ and\ \bibinfo {author} {\bibfnamefont {Andrina}\ \bibnamefont {Nicola}},\ }\bibfield  {title} {\enquote {\bibinfo {title} {{Fast Lightcones for Combined Cosmological Probes}},}\ }\href {\doibase 10.1088/1475-7516/2021/02/047} {\bibfield  {journal} {\bibinfo  {journal} {JCAP}\ }\textbf {\bibinfo {volume} {02}},\ \bibinfo {pages} {047} (\bibinfo {year} {2021})},\ \Eprint {http://arxiv.org/abs/2007.05735} {arXiv:2007.05735 [astro-ph.CO]} \BibitemShut {NoStop}%
\bibitem [{\citenamefont {Hartlap}\ \emph {et~al.}(2007)\citenamefont {Hartlap}, \citenamefont {Simon},\ and\ \citenamefont {Schneider}}]{Hartlap:2006kj}%
  \BibitemOpen
  \bibfield  {author} {\bibinfo {author} {\bibfnamefont {J.}~\bibnamefont {Hartlap}}, \bibinfo {author} {\bibfnamefont {Patrick}\ \bibnamefont {Simon}}, \ and\ \bibinfo {author} {\bibfnamefont {P.}~\bibnamefont {Schneider}},\ }\bibfield  {title} {\enquote {\bibinfo {title} {{Why your model parameter confidences might be too optimistic: Unbiased estimation of the inverse covariance matrix}},}\ }\href {\doibase 10.1051/0004-6361:20066170} {\bibfield  {journal} {\bibinfo  {journal} {Astron. Astrophys.}\ }\textbf {\bibinfo {volume} {464}},\ \bibinfo {pages} {399} (\bibinfo {year} {2007})},\ \Eprint {http://arxiv.org/abs/astro-ph/0608064} {arXiv:astro-ph/0608064} \BibitemShut {NoStop}%
\bibitem [{\citenamefont {Schmittfull}\ \emph {et~al.}(2013)\citenamefont {Schmittfull}, \citenamefont {Challinor}, \citenamefont {Hanson},\ and\ \citenamefont {Lewis}}]{Schmittfull:2013uea}%
  \BibitemOpen
  \bibfield  {author} {\bibinfo {author} {\bibfnamefont {Marcel~M.}\ \bibnamefont {Schmittfull}}, \bibinfo {author} {\bibfnamefont {Anthony}\ \bibnamefont {Challinor}}, \bibinfo {author} {\bibfnamefont {Duncan}\ \bibnamefont {Hanson}}, \ and\ \bibinfo {author} {\bibfnamefont {Antony}\ \bibnamefont {Lewis}},\ }\bibfield  {title} {\enquote {\bibinfo {title} {{Joint analysis of CMB temperature and lensing-reconstruction power spectra}},}\ }\href {\doibase 10.1103/PhysRevD.88.063012} {\bibfield  {journal} {\bibinfo  {journal} {Phys. Rev. D}\ }\textbf {\bibinfo {volume} {88}},\ \bibinfo {pages} {063012} (\bibinfo {year} {2013})},\ \Eprint {http://arxiv.org/abs/1308.0286} {arXiv:1308.0286 [astro-ph.CO]} \BibitemShut {NoStop}%
\bibitem [{\citenamefont {Peloton}\ \emph {et~al.}(2017)\citenamefont {Peloton}, \citenamefont {Schmittfull}, \citenamefont {Lewis}, \citenamefont {Carron},\ and\ \citenamefont {Zahn}}]{Peloton:2016kbw}%
  \BibitemOpen
  \bibfield  {author} {\bibinfo {author} {\bibfnamefont {Julien}\ \bibnamefont {Peloton}}, \bibinfo {author} {\bibfnamefont {Marcel}\ \bibnamefont {Schmittfull}}, \bibinfo {author} {\bibfnamefont {Antony}\ \bibnamefont {Lewis}}, \bibinfo {author} {\bibfnamefont {Julien}\ \bibnamefont {Carron}}, \ and\ \bibinfo {author} {\bibfnamefont {Oliver}\ \bibnamefont {Zahn}},\ }\bibfield  {title} {\enquote {\bibinfo {title} {{Full covariance of CMB and lensing reconstruction power spectra}},}\ }\href {\doibase 10.1103/PhysRevD.95.043508} {\bibfield  {journal} {\bibinfo  {journal} {Phys. Rev. D}\ }\textbf {\bibinfo {volume} {95}},\ \bibinfo {pages} {043508} (\bibinfo {year} {2017})},\ \Eprint {http://arxiv.org/abs/1611.01446} {arXiv:1611.01446 [astro-ph.CO]} \BibitemShut {NoStop}%
\bibitem [{\citenamefont {Kou}\ and\ \citenamefont {Lewis}(2025)}]{Kou:2025hvg}%
  \BibitemOpen
  \bibfield  {author} {\bibinfo {author} {\bibfnamefont {Rapha{\"e}l}\ \bibnamefont {Kou}}\ and\ \bibinfo {author} {\bibfnamefont {Antony}\ \bibnamefont {Lewis}},\ }\bibfield  {title} {\enquote {\bibinfo {title} {{Improving CMB constraints on early Universe physics with LSS: a multi-probe forecast including cross-covariance}},}\ }\href {\doibase 10.1088/1475-7516/2025/08/031} {\bibfield  {journal} {\bibinfo  {journal} {JCAP}\ }\textbf {\bibinfo {volume} {08}},\ \bibinfo {pages} {031} (\bibinfo {year} {2025})},\ \Eprint {http://arxiv.org/abs/2504.13913} {arXiv:2504.13913 [astro-ph.CO]} \BibitemShut {NoStop}%
\bibitem [{\citenamefont {D'Amico}\ \emph {et~al.}(2021)\citenamefont {D'Amico}, \citenamefont {Senatore}, \citenamefont {Zhang},\ and\ \citenamefont {Zheng}}]{DAmico:2020ods}%
  \BibitemOpen
  \bibfield  {author} {\bibinfo {author} {\bibfnamefont {Guido}\ \bibnamefont {D'Amico}}, \bibinfo {author} {\bibfnamefont {Leonardo}\ \bibnamefont {Senatore}}, \bibinfo {author} {\bibfnamefont {Pierre}\ \bibnamefont {Zhang}}, \ and\ \bibinfo {author} {\bibfnamefont {Henry}\ \bibnamefont {Zheng}},\ }\bibfield  {title} {\enquote {\bibinfo {title} {{The Hubble Tension in Light of the Full-Shape Analysis of Large-Scale Structure Data}},}\ }\href {\doibase 10.1088/1475-7516/2021/05/072} {\bibfield  {journal} {\bibinfo  {journal} {JCAP}\ }\textbf {\bibinfo {volume} {05}},\ \bibinfo {pages} {072} (\bibinfo {year} {2021})},\ \Eprint {http://arxiv.org/abs/2006.12420} {arXiv:2006.12420 [astro-ph.CO]} \BibitemShut {NoStop}%
\bibitem [{\citenamefont {Smith}\ \emph {et~al.}(2021)\citenamefont {Smith}, \citenamefont {Poulin}, \citenamefont {Bernal}, \citenamefont {Boddy}, \citenamefont {Kamionkowski},\ and\ \citenamefont {Murgia}}]{Smith:2020rxx}%
  \BibitemOpen
  \bibfield  {author} {\bibinfo {author} {\bibfnamefont {Tristan~L.}\ \bibnamefont {Smith}}, \bibinfo {author} {\bibfnamefont {Vivian}\ \bibnamefont {Poulin}}, \bibinfo {author} {\bibfnamefont {Jos{\'e}~Luis}\ \bibnamefont {Bernal}}, \bibinfo {author} {\bibfnamefont {Kimberly~K.}\ \bibnamefont {Boddy}}, \bibinfo {author} {\bibfnamefont {Marc}\ \bibnamefont {Kamionkowski}}, \ and\ \bibinfo {author} {\bibfnamefont {Riccardo}\ \bibnamefont {Murgia}},\ }\bibfield  {title} {\enquote {\bibinfo {title} {{Early dark energy is not excluded by current large-scale structure data}},}\ }\href {\doibase 10.1103/PhysRevD.103.123542} {\bibfield  {journal} {\bibinfo  {journal} {Phys. Rev. D}\ }\textbf {\bibinfo {volume} {103}},\ \bibinfo {pages} {123542} (\bibinfo {year} {2021})},\ \Eprint {http://arxiv.org/abs/2009.10740} {arXiv:2009.10740 [astro-ph.CO]} \BibitemShut {NoStop}%
\bibitem [{\citenamefont {Foreman-Mackey}\ \emph {et~al.}(2013)\citenamefont {Foreman-Mackey}, \citenamefont {Hogg}, \citenamefont {Lang},\ and\ \citenamefont {Goodman}}]{Foreman-Mackey:2012any}%
  \BibitemOpen
  \bibfield  {author} {\bibinfo {author} {\bibfnamefont {Daniel}\ \bibnamefont {Foreman-Mackey}}, \bibinfo {author} {\bibfnamefont {David~W.}\ \bibnamefont {Hogg}}, \bibinfo {author} {\bibfnamefont {Dustin}\ \bibnamefont {Lang}}, \ and\ \bibinfo {author} {\bibfnamefont {Jonathan}\ \bibnamefont {Goodman}},\ }\bibfield  {title} {\enquote {\bibinfo {title} {{emcee: The MCMC Hammer}},}\ }\href {\doibase 10.1086/670067} {\bibfield  {journal} {\bibinfo  {journal} {Publ. Astron. Soc. Pac.}\ }\textbf {\bibinfo {volume} {125}},\ \bibinfo {pages} {306--312} (\bibinfo {year} {2013})},\ \Eprint {http://arxiv.org/abs/1202.3665} {arXiv:1202.3665 [astro-ph.IM]} \BibitemShut {NoStop}%
\bibitem [{\citenamefont {Reeves}\ \emph {et~al.}(2025{\natexlab{b}})\citenamefont {Reeves}, \citenamefont {Zhang},\ and\ \citenamefont {Zheng}}]{Reeves:2025bxn}%
  \BibitemOpen
  \bibfield  {author} {\bibinfo {author} {\bibfnamefont {Alexander}\ \bibnamefont {Reeves}}, \bibinfo {author} {\bibfnamefont {Pierre}\ \bibnamefont {Zhang}}, \ and\ \bibinfo {author} {\bibfnamefont {Henry}\ \bibnamefont {Zheng}},\ }\bibfield  {title} {\enquote {\bibinfo {title} {{PyBird-JAX: Accelerated inference in large-scale structure with model-independent emulation of one-loop galaxy power spectra}},}\ }\href@noop {} {\  (\bibinfo {year} {2025}{\natexlab{b}})},\ \Eprint {http://arxiv.org/abs/2507.20990} {arXiv:2507.20990 [astro-ph.CO]} \BibitemShut {NoStop}%
\bibitem [{\citenamefont {Herold}\ \emph {et~al.}(2025)\citenamefont {Herold}, \citenamefont {Ferreira},\ and\ \citenamefont {Heinrich}}]{Herold:2024enb}%
  \BibitemOpen
  \bibfield  {author} {\bibinfo {author} {\bibfnamefont {Laura}\ \bibnamefont {Herold}}, \bibinfo {author} {\bibfnamefont {Elisa G.~M.}\ \bibnamefont {Ferreira}}, \ and\ \bibinfo {author} {\bibfnamefont {Lukas}\ \bibnamefont {Heinrich}},\ }\bibfield  {title} {\enquote {\bibinfo {title} {{Profile likelihoods in cosmology: When, why, and how illustrated with {\ensuremath{\Lambda}}CDM, massive neutrinos, and dark energy}},}\ }\href {\doibase 10.1103/PhysRevD.111.083504} {\bibfield  {journal} {\bibinfo  {journal} {Phys. Rev. D}\ }\textbf {\bibinfo {volume} {111}},\ \bibinfo {pages} {083504} (\bibinfo {year} {2025})},\ \Eprint {http://arxiv.org/abs/2408.07700} {arXiv:2408.07700 [astro-ph.CO]} \BibitemShut {NoStop}%
\bibitem [{\citenamefont {Sch{\"o}neberg}\ \emph {et~al.}(2022)\citenamefont {Sch{\"o}neberg}, \citenamefont {Franco~Abell{\'a}n}, \citenamefont {P{\'e}rez~S{\'a}nchez}, \citenamefont {Witte}, \citenamefont {Poulin},\ and\ \citenamefont {Lesgourgues}}]{Schoneberg:2021qvd}%
  \BibitemOpen
  \bibfield  {author} {\bibinfo {author} {\bibfnamefont {Nils}\ \bibnamefont {Sch{\"o}neberg}}, \bibinfo {author} {\bibfnamefont {Guillermo}\ \bibnamefont {Franco~Abell{\'a}n}}, \bibinfo {author} {\bibfnamefont {Andrea}\ \bibnamefont {P{\'e}rez~S{\'a}nchez}}, \bibinfo {author} {\bibfnamefont {Samuel~J.}\ \bibnamefont {Witte}}, \bibinfo {author} {\bibfnamefont {Vivian}\ \bibnamefont {Poulin}}, \ and\ \bibinfo {author} {\bibfnamefont {Julien}\ \bibnamefont {Lesgourgues}},\ }\bibfield  {title} {\enquote {\bibinfo {title} {{The H0 Olympics: A fair ranking of proposed models}},}\ }\href {\doibase 10.1016/j.physrep.2022.07.001} {\bibfield  {journal} {\bibinfo  {journal} {Phys. Rept.}\ }\textbf {\bibinfo {volume} {984}},\ \bibinfo {pages} {1--55} (\bibinfo {year} {2022})},\ \Eprint {http://arxiv.org/abs/2107.10291} {arXiv:2107.10291 [astro-ph.CO]} \BibitemShut {NoStop}%
\bibitem [{\citenamefont {Hannestad}(2000)}]{Hannestad:2000wx}%
  \BibitemOpen
  \bibfield  {author} {\bibinfo {author} {\bibfnamefont {Steen}\ \bibnamefont {Hannestad}},\ }\bibfield  {title} {\enquote {\bibinfo {title} {{Stochastic optimization methods for extracting cosmological parameters from cosmic microwave background radiation power spectra}},}\ }\href {\doibase 10.1103/PhysRevD.61.023002} {\bibfield  {journal} {\bibinfo  {journal} {Phys. Rev. D}\ }\textbf {\bibinfo {volume} {61}},\ \bibinfo {pages} {023002} (\bibinfo {year} {2000})},\ \Eprint {http://arxiv.org/abs/astro-ph/9911330} {arXiv:astro-ph/9911330} \BibitemShut {NoStop}%
\bibitem [{\citenamefont {{Kingma}}\ and\ \citenamefont {{Ba}}(2014)}]{adamxyz}%
  \BibitemOpen
  \bibfield  {author} {\bibinfo {author} {\bibfnamefont {Diederik~P.}\ \bibnamefont {{Kingma}}}\ and\ \bibinfo {author} {\bibfnamefont {Jimmy}\ \bibnamefont {{Ba}}},\ }\bibfield  {title} {\enquote {\bibinfo {title} {{Adam: A Method for Stochastic Optimization}},}\ }\href {\doibase 10.48550/arXiv.1412.6980} {\bibfield  {journal} {\bibinfo  {journal} {arXiv e-prints}\ ,\ \bibinfo {eid} {arXiv:1412.6980}} (\bibinfo {year} {2014})},\ \Eprint {http://arxiv.org/abs/1412.6980} {arXiv:1412.6980 [cs.LG]} \BibitemShut {NoStop}%
\bibitem [{\citenamefont {DeepMind}\ \emph {et~al.}(2020)\citenamefont {DeepMind}, \citenamefont {Babuschkin}, \citenamefont {Baumli}, \citenamefont {Bell}, \citenamefont {Bhupatiraju}, \citenamefont {Bruce}, \citenamefont {Buchlovsky}, \citenamefont {Budden}, \citenamefont {Cai}, \citenamefont {Clark}, \citenamefont {Danihelka}, \citenamefont {Dedieu}, \citenamefont {Fantacci}, \citenamefont {Godwin}, \citenamefont {Jones}, \citenamefont {Hemsley}, \citenamefont {Hennigan}, \citenamefont {Hessel}, \citenamefont {Hou}, \citenamefont {Kapturowski}, \citenamefont {Keck}, \citenamefont {Kemaev}, \citenamefont {King}, \citenamefont {Kunesch}, \citenamefont {Martens}, \citenamefont {Merzic}, \citenamefont {Mikulik}, \citenamefont {Norman}, \citenamefont {Papamakarios}, \citenamefont {Quan}, \citenamefont {Ring}, \citenamefont {Ruiz}, \citenamefont {Sanchez}, \citenamefont {Sartran}, \citenamefont {Schneider}, \citenamefont {Sezener}, \citenamefont {Spencer}, \citenamefont {Srinivasan}, \citenamefont
  {Stanojevi\'{c}}, \citenamefont {Stokowiec}, \citenamefont {Wang}, \citenamefont {Zhou},\ and\ \citenamefont {Viola}}]{deepmind2020jax}%
  \BibitemOpen
  \bibfield  {author} {\bibinfo {author} {\bibnamefont {DeepMind}}, \bibinfo {author} {\bibfnamefont {Igor}\ \bibnamefont {Babuschkin}}, \bibinfo {author} {\bibfnamefont {Kate}\ \bibnamefont {Baumli}}, \bibinfo {author} {\bibfnamefont {Alison}\ \bibnamefont {Bell}}, \bibinfo {author} {\bibfnamefont {Surya}\ \bibnamefont {Bhupatiraju}}, \bibinfo {author} {\bibfnamefont {Jake}\ \bibnamefont {Bruce}}, \bibinfo {author} {\bibfnamefont {Peter}\ \bibnamefont {Buchlovsky}}, \bibinfo {author} {\bibfnamefont {David}\ \bibnamefont {Budden}}, \bibinfo {author} {\bibfnamefont {Trevor}\ \bibnamefont {Cai}}, \bibinfo {author} {\bibfnamefont {Aidan}\ \bibnamefont {Clark}}, \bibinfo {author} {\bibfnamefont {Ivo}\ \bibnamefont {Danihelka}}, \bibinfo {author} {\bibfnamefont {Antoine}\ \bibnamefont {Dedieu}}, \bibinfo {author} {\bibfnamefont {Claudio}\ \bibnamefont {Fantacci}}, \bibinfo {author} {\bibfnamefont {Jonathan}\ \bibnamefont {Godwin}}, \bibinfo {author} {\bibfnamefont {Chris}\ \bibnamefont {Jones}}, \bibinfo {author}
  {\bibfnamefont {Ross}\ \bibnamefont {Hemsley}}, \bibinfo {author} {\bibfnamefont {Tom}\ \bibnamefont {Hennigan}}, \bibinfo {author} {\bibfnamefont {Matteo}\ \bibnamefont {Hessel}}, \bibinfo {author} {\bibfnamefont {Shaobo}\ \bibnamefont {Hou}}, \bibinfo {author} {\bibfnamefont {Steven}\ \bibnamefont {Kapturowski}}, \bibinfo {author} {\bibfnamefont {Thomas}\ \bibnamefont {Keck}}, \bibinfo {author} {\bibfnamefont {Iurii}\ \bibnamefont {Kemaev}}, \bibinfo {author} {\bibfnamefont {Michael}\ \bibnamefont {King}}, \bibinfo {author} {\bibfnamefont {Markus}\ \bibnamefont {Kunesch}}, \bibinfo {author} {\bibfnamefont {Lena}\ \bibnamefont {Martens}}, \bibinfo {author} {\bibfnamefont {Hamza}\ \bibnamefont {Merzic}}, \bibinfo {author} {\bibfnamefont {Vladimir}\ \bibnamefont {Mikulik}}, \bibinfo {author} {\bibfnamefont {Tamara}\ \bibnamefont {Norman}}, \bibinfo {author} {\bibfnamefont {George}\ \bibnamefont {Papamakarios}}, \bibinfo {author} {\bibfnamefont {John}\ \bibnamefont {Quan}}, \bibinfo {author} {\bibfnamefont
  {Roman}\ \bibnamefont {Ring}}, \bibinfo {author} {\bibfnamefont {Francisco}\ \bibnamefont {Ruiz}}, \bibinfo {author} {\bibfnamefont {Alvaro}\ \bibnamefont {Sanchez}}, \bibinfo {author} {\bibfnamefont {Laurent}\ \bibnamefont {Sartran}}, \bibinfo {author} {\bibfnamefont {Rosalia}\ \bibnamefont {Schneider}}, \bibinfo {author} {\bibfnamefont {Eren}\ \bibnamefont {Sezener}}, \bibinfo {author} {\bibfnamefont {Stephen}\ \bibnamefont {Spencer}}, \bibinfo {author} {\bibfnamefont {Srivatsan}\ \bibnamefont {Srinivasan}}, \bibinfo {author} {\bibfnamefont {Milo\v{s}}\ \bibnamefont {Stanojevi\'{c}}}, \bibinfo {author} {\bibfnamefont {Wojciech}\ \bibnamefont {Stokowiec}}, \bibinfo {author} {\bibfnamefont {Luyu}\ \bibnamefont {Wang}}, \bibinfo {author} {\bibfnamefont {Guangyao}\ \bibnamefont {Zhou}}, \ and\ \bibinfo {author} {\bibfnamefont {Fabio}\ \bibnamefont {Viola}},\ }\href {http://github.com/google-deepmind} {\enquote {\bibinfo {title} {The {D}eep{M}ind {JAX} {E}cosystem},}\ } (\bibinfo {year} {2020})\BibitemShut
  {NoStop}%
\bibitem [{\citenamefont {Wilks}(1938)}]{wilks_theorem}%
  \BibitemOpen
  \bibfield  {author} {\bibinfo {author} {\bibfnamefont {S.~S.}\ \bibnamefont {Wilks}},\ }\bibfield  {title} {\enquote {\bibinfo {title} {{The Large-Sample Distribution of the Likelihood Ratio for Testing Composite Hypotheses}},}\ }\href {\doibase 10.1214/aoms/1177732360} {\bibfield  {journal} {\bibinfo  {journal} {The Annals of Mathematical Statistics}\ }\textbf {\bibinfo {volume} {9}},\ \bibinfo {pages} {60 -- 62} (\bibinfo {year} {1938})}\BibitemShut {NoStop}%
\bibitem [{\citenamefont {McDonough}\ \emph {et~al.}(2022)\citenamefont {McDonough}, \citenamefont {Lin}, \citenamefont {Hill}, \citenamefont {Hu},\ and\ \citenamefont {Zhou}}]{McDonough:2021pdg}%
  \BibitemOpen
  \bibfield  {author} {\bibinfo {author} {\bibfnamefont {Evan}\ \bibnamefont {McDonough}}, \bibinfo {author} {\bibfnamefont {Meng-Xiang}\ \bibnamefont {Lin}}, \bibinfo {author} {\bibfnamefont {J.~Colin}\ \bibnamefont {Hill}}, \bibinfo {author} {\bibfnamefont {Wayne}\ \bibnamefont {Hu}}, \ and\ \bibinfo {author} {\bibfnamefont {Shengjia}\ \bibnamefont {Zhou}},\ }\bibfield  {title} {\enquote {\bibinfo {title} {{Early dark sector, the Hubble tension, and the swampland}},}\ }\href {\doibase 10.1103/PhysRevD.106.043525} {\bibfield  {journal} {\bibinfo  {journal} {Phys. Rev. D}\ }\textbf {\bibinfo {volume} {106}},\ \bibinfo {pages} {043525} (\bibinfo {year} {2022})},\ \Eprint {http://arxiv.org/abs/2112.09128} {arXiv:2112.09128 [astro-ph.CO]} \BibitemShut {NoStop}%
\bibitem [{\citenamefont {Niedermann}\ and\ \citenamefont {Sloth}(2022)}]{Niedermann:2021ijp}%
  \BibitemOpen
  \bibfield  {author} {\bibinfo {author} {\bibfnamefont {Florian}\ \bibnamefont {Niedermann}}\ and\ \bibinfo {author} {\bibfnamefont {Martin~S.}\ \bibnamefont {Sloth}},\ }\bibfield  {title} {\enquote {\bibinfo {title} {{Hot new early dark energy: Towards a unified dark sector of neutrinos, dark energy and dark matter}},}\ }\href {\doibase 10.1016/j.physletb.2022.137555} {\bibfield  {journal} {\bibinfo  {journal} {Phys. Lett. B}\ }\textbf {\bibinfo {volume} {835}},\ \bibinfo {pages} {137555} (\bibinfo {year} {2022})},\ \Eprint {http://arxiv.org/abs/2112.00759} {arXiv:2112.00759 [hep-ph]} \BibitemShut {NoStop}%
\bibitem [{\citenamefont {Zumalacarregui}(2020)}]{Zumalacarregui:2020cjh}%
  \BibitemOpen
  \bibfield  {author} {\bibinfo {author} {\bibfnamefont {Miguel}\ \bibnamefont {Zumalacarregui}},\ }\bibfield  {title} {\enquote {\bibinfo {title} {{Gravity in the Era of Equality: Towards solutions to the Hubble problem without fine-tuned initial conditions}},}\ }\href {\doibase 10.1103/PhysRevD.102.023523} {\bibfield  {journal} {\bibinfo  {journal} {Phys. Rev. D}\ }\textbf {\bibinfo {volume} {102}},\ \bibinfo {pages} {023523} (\bibinfo {year} {2020})},\ \Eprint {http://arxiv.org/abs/2003.06396} {arXiv:2003.06396 [astro-ph.CO]} \BibitemShut {NoStop}%
\bibitem [{\citenamefont {Oikonomou}(2021)}]{Oikonomou:2020qah}%
  \BibitemOpen
  \bibfield  {author} {\bibinfo {author} {\bibfnamefont {V.~K.}\ \bibnamefont {Oikonomou}},\ }\bibfield  {title} {\enquote {\bibinfo {title} {{Unifying inflation with early and late dark energy epochs in axion $F(R)$ gravity}},}\ }\href {\doibase 10.1103/PhysRevD.103.044036} {\bibfield  {journal} {\bibinfo  {journal} {Phys. Rev. D}\ }\textbf {\bibinfo {volume} {103}},\ \bibinfo {pages} {044036} (\bibinfo {year} {2021})},\ \Eprint {http://arxiv.org/abs/2012.00586} {arXiv:2012.00586 [astro-ph.CO]} \BibitemShut {NoStop}%
\bibitem [{\citenamefont {Ye}\ \emph {et~al.}(2023)\citenamefont {Ye}, \citenamefont {Zhang},\ and\ \citenamefont {Piao}}]{Ye:2021iwa}%
  \BibitemOpen
  \bibfield  {author} {\bibinfo {author} {\bibfnamefont {Gen}\ \bibnamefont {Ye}}, \bibinfo {author} {\bibfnamefont {Jun}\ \bibnamefont {Zhang}}, \ and\ \bibinfo {author} {\bibfnamefont {Yun-Song}\ \bibnamefont {Piao}},\ }\bibfield  {title} {\enquote {\bibinfo {title} {{Alleviating both H0 and S8 tensions: Early dark energy lifts the CMB-lockdown on ultralight axion}},}\ }\href {\doibase 10.1016/j.physletb.2023.137770} {\bibfield  {journal} {\bibinfo  {journal} {Phys. Lett. B}\ }\textbf {\bibinfo {volume} {839}},\ \bibinfo {pages} {137770} (\bibinfo {year} {2023})},\ \Eprint {http://arxiv.org/abs/2107.13391} {arXiv:2107.13391 [astro-ph.CO]} \BibitemShut {NoStop}%
\bibitem [{\citenamefont {Smith}\ \emph {et~al.}(2020)\citenamefont {Smith}, \citenamefont {Poulin},\ and\ \citenamefont {Amin}}]{Smith:2019ihp}%
  \BibitemOpen
  \bibfield  {author} {\bibinfo {author} {\bibfnamefont {Tristan~L.}\ \bibnamefont {Smith}}, \bibinfo {author} {\bibfnamefont {Vivian}\ \bibnamefont {Poulin}}, \ and\ \bibinfo {author} {\bibfnamefont {Mustafa~A.}\ \bibnamefont {Amin}},\ }\bibfield  {title} {\enquote {\bibinfo {title} {{Oscillating scalar fields and the Hubble tension: a resolution with novel signatures}},}\ }\href {\doibase 10.1103/PhysRevD.101.063523} {\bibfield  {journal} {\bibinfo  {journal} {Phys. Rev. D}\ }\textbf {\bibinfo {volume} {101}},\ \bibinfo {pages} {063523} (\bibinfo {year} {2020})},\ \Eprint {http://arxiv.org/abs/1908.06995} {arXiv:1908.06995 [astro-ph.CO]} \BibitemShut {NoStop}%
\bibitem [{\citenamefont {Chevallier}\ and\ \citenamefont {Polarski}(2001)}]{Chevallier:2000qy}%
  \BibitemOpen
  \bibfield  {author} {\bibinfo {author} {\bibfnamefont {Michel}\ \bibnamefont {Chevallier}}\ and\ \bibinfo {author} {\bibfnamefont {David}\ \bibnamefont {Polarski}},\ }\bibfield  {title} {\enquote {\bibinfo {title} {{Accelerating universes with scaling dark matter}},}\ }\href {\doibase 10.1142/S0218271801000822} {\bibfield  {journal} {\bibinfo  {journal} {Int. J. Mod. Phys. D}\ }\textbf {\bibinfo {volume} {10}},\ \bibinfo {pages} {213--224} (\bibinfo {year} {2001})},\ \Eprint {http://arxiv.org/abs/gr-qc/0009008} {arXiv:gr-qc/0009008} \BibitemShut {NoStop}%
\bibitem [{\citenamefont {Linder}(2003)}]{Linder:2002et}%
  \BibitemOpen
  \bibfield  {author} {\bibinfo {author} {\bibfnamefont {Eric~V.}\ \bibnamefont {Linder}},\ }\bibfield  {title} {\enquote {\bibinfo {title} {{Exploring the expansion history of the universe}},}\ }\href {\doibase 10.1103/PhysRevLett.90.091301} {\bibfield  {journal} {\bibinfo  {journal} {Phys. Rev. Lett.}\ }\textbf {\bibinfo {volume} {90}},\ \bibinfo {pages} {091301} (\bibinfo {year} {2003})},\ \Eprint {http://arxiv.org/abs/astro-ph/0208512} {arXiv:astro-ph/0208512} \BibitemShut {NoStop}%
\bibitem [{\citenamefont {Wright}\ \emph {et~al.}(2025)\citenamefont {Wright} \emph {et~al.}}]{Wright:2025xka}%
  \BibitemOpen
  \bibfield  {author} {\bibinfo {author} {\bibfnamefont {Angus~H.}\ \bibnamefont {Wright}} \emph {et~al.},\ }\bibfield  {title} {\enquote {\bibinfo {title} {{KiDS-Legacy: Cosmological constraints from cosmic shear with the complete Kilo-Degree Survey}},}\ }\href@noop {} {\  (\bibinfo {year} {2025})},\ \Eprint {http://arxiv.org/abs/2503.19441} {arXiv:2503.19441 [astro-ph.CO]} \BibitemShut {NoStop}%
\bibitem [{\citenamefont {Maus}\ \emph {et~al.}(2025)\citenamefont {Maus} \emph {et~al.}}]{Maus:2025rvz}%
  \BibitemOpen
  \bibfield  {author} {\bibinfo {author} {\bibfnamefont {M.}~\bibnamefont {Maus}} \emph {et~al.},\ }\bibfield  {title} {\enquote {\bibinfo {title} {{A joint analysis of 3D clustering and galaxy x CMB-lensing cross-correlations with DESI DR1 galaxies}},}\ }\href@noop {} {\  (\bibinfo {year} {2025})},\ \Eprint {http://arxiv.org/abs/2505.20656} {arXiv:2505.20656 [astro-ph.CO]} \BibitemShut {NoStop}%
\bibitem [{\citenamefont {Hang}\ \emph {et~al.}(2021)\citenamefont {Hang}, \citenamefont {Alam}, \citenamefont {Peacock},\ and\ \citenamefont {Cai}}]{Hang:2020gwn}%
  \BibitemOpen
  \bibfield  {author} {\bibinfo {author} {\bibfnamefont {Qianjun}\ \bibnamefont {Hang}}, \bibinfo {author} {\bibfnamefont {Shadab}\ \bibnamefont {Alam}}, \bibinfo {author} {\bibfnamefont {John~A.}\ \bibnamefont {Peacock}}, \ and\ \bibinfo {author} {\bibfnamefont {Yan-Chuan}\ \bibnamefont {Cai}},\ }\bibfield  {title} {\enquote {\bibinfo {title} {{Galaxy clustering in the DESI Legacy Survey and its imprint on the CMB}},}\ }\href {\doibase 10.1093/mnras/staa3738} {\bibfield  {journal} {\bibinfo  {journal} {Mon. Not. Roy. Astron. Soc.}\ }\textbf {\bibinfo {volume} {501}},\ \bibinfo {pages} {1481--1498} (\bibinfo {year} {2021})},\ \Eprint {http://arxiv.org/abs/2010.00466} {arXiv:2010.00466 [astro-ph.CO]} \BibitemShut {NoStop}%
\bibitem [{\citenamefont {Madhavacheril}\ and\ \citenamefont {{et al.}}(2023)}]{Madhavacheril:2023}%
  \BibitemOpen
  \bibfield  {author} {\bibinfo {author} {\bibfnamefont {M. S.}\ \bibnamefont {Madhavacheril}}\ and\ \bibinfo {author} {\bibnamefont {{et al.}}},\ }\bibfield  {title} {\enquote {\bibinfo {title} {{The Atacama Cosmology Telescope: DR6 Gravitational Lensing Map}},}\ }\href@noop {} {\bibfield  {journal} {\bibinfo  {journal} {ApJ}\ }\textbf {\bibinfo {volume} {945}},\ \bibinfo {pages} {123} (\bibinfo {year} {2023})},\ \Eprint {http://arxiv.org/abs/2304.05203} {2304.05203} \BibitemShut {NoStop}%
\bibitem [{\citenamefont {Leauthaud}\ \emph {et~al.}(2017)\citenamefont {Leauthaud} \emph {et~al.}}]{Leauthaud:2016jdb}%
  \BibitemOpen
  \bibfield  {author} {\bibinfo {author} {\bibfnamefont {Alexie}\ \bibnamefont {Leauthaud}} \emph {et~al.},\ }\bibfield  {title} {\enquote {\bibinfo {title} {{Lensing is Low: Cosmology, Galaxy Formation, or New Physics?}}}\ }\href {\doibase 10.1093/mnras/stx258} {\bibfield  {journal} {\bibinfo  {journal} {Mon. Not. Roy. Astron. Soc.}\ }\textbf {\bibinfo {volume} {467}},\ \bibinfo {pages} {3024--3047} (\bibinfo {year} {2017})},\ \Eprint {http://arxiv.org/abs/1611.08606} {arXiv:1611.08606 [astro-ph.CO]} \BibitemShut {NoStop}%
\bibitem [{\citenamefont {Zaborowski}\ \emph {et~al.}(2025)\citenamefont {Zaborowski} \emph {et~al.}}]{Zaborowski:2024car}%
  \BibitemOpen
  \bibfield  {author} {\bibinfo {author} {\bibfnamefont {E.~A.}\ \bibnamefont {Zaborowski}} \emph {et~al.},\ }\bibfield  {title} {\enquote {\bibinfo {title} {{A Sound Horizon-Free Measurement of $H_0$ in DESI 2024}},}\ }\href {\doibase 10.1088/1475-7516/2025/06/020} {\bibfield  {journal} {\bibinfo  {journal} {JCAP}\ }\textbf {\bibinfo {volume} {06}},\ \bibinfo {pages} {020} (\bibinfo {year} {2025})},\ \Eprint {http://arxiv.org/abs/2411.16677} {arXiv:2411.16677 [astro-ph.CO]} \BibitemShut {NoStop}%
\bibitem [{\citenamefont {Farren}\ \emph {et~al.}(2022)\citenamefont {Farren}, \citenamefont {Philcox},\ and\ \citenamefont {Sherwin}}]{Farren:2021grl}%
  \BibitemOpen
  \bibfield  {author} {\bibinfo {author} {\bibfnamefont {Gerrit~S.}\ \bibnamefont {Farren}}, \bibinfo {author} {\bibfnamefont {Oliver H.~E.}\ \bibnamefont {Philcox}}, \ and\ \bibinfo {author} {\bibfnamefont {Blake~D.}\ \bibnamefont {Sherwin}},\ }\bibfield  {title} {\enquote {\bibinfo {title} {{Determining the Hubble constant without the sound horizon: Perspectives with future galaxy surveys}},}\ }\href {\doibase 10.1103/PhysRevD.105.063503} {\bibfield  {journal} {\bibinfo  {journal} {Phys. Rev. D}\ }\textbf {\bibinfo {volume} {105}},\ \bibinfo {pages} {063503} (\bibinfo {year} {2022})},\ \Eprint {http://arxiv.org/abs/2112.10749} {arXiv:2112.10749 [astro-ph.CO]} \BibitemShut {NoStop}%
\bibitem [{\citenamefont {Philcox}\ \emph {et~al.}(2022)\citenamefont {Philcox}, \citenamefont {Farren}, \citenamefont {Sherwin}, \citenamefont {Baxter},\ and\ \citenamefont {Brout}}]{Philcox:2022sgj}%
  \BibitemOpen
  \bibfield  {author} {\bibinfo {author} {\bibfnamefont {Oliver H.~E.}\ \bibnamefont {Philcox}}, \bibinfo {author} {\bibfnamefont {Gerrit~S.}\ \bibnamefont {Farren}}, \bibinfo {author} {\bibfnamefont {Blake~D.}\ \bibnamefont {Sherwin}}, \bibinfo {author} {\bibfnamefont {Eric~J.}\ \bibnamefont {Baxter}}, \ and\ \bibinfo {author} {\bibfnamefont {Dillon~J.}\ \bibnamefont {Brout}},\ }\bibfield  {title} {\enquote {\bibinfo {title} {{Determining the Hubble constant without the sound horizon: A 3.6{\%} constraint on H0 from galaxy surveys, CMB lensing, and supernovae}},}\ }\href {\doibase 10.1103/PhysRevD.106.063530} {\bibfield  {journal} {\bibinfo  {journal} {Phys. Rev. D}\ }\textbf {\bibinfo {volume} {106}},\ \bibinfo {pages} {063530} (\bibinfo {year} {2022})},\ \Eprint {http://arxiv.org/abs/2204.02984} {arXiv:2204.02984 [astro-ph.CO]} \BibitemShut {NoStop}%
\bibitem [{\citenamefont {Garc{\'\i}a~Escudero}\ \emph {et~al.}(2025)\citenamefont {Garc{\'\i}a~Escudero}, \citenamefont {Mirpoorian},\ and\ \citenamefont {Pogosian}}]{GarciaEscudero:2025lef}%
  \BibitemOpen
  \bibfield  {author} {\bibinfo {author} {\bibfnamefont {Helena}\ \bibnamefont {Garc{\'\i}a~Escudero}}, \bibinfo {author} {\bibfnamefont {Seyed~Hamidreza}\ \bibnamefont {Mirpoorian}}, \ and\ \bibinfo {author} {\bibfnamefont {Levon}\ \bibnamefont {Pogosian}},\ }\bibfield  {title} {\enquote {\bibinfo {title} {{Sound-Horizon-Agnostic Inference of the Hubble Constant and Neutrino Mass from BAO, CMB Lensing, and Galaxy Weak Lensing and Clustering}},}\ }\href@noop {} {\  (\bibinfo {year} {2025})},\ \Eprint {http://arxiv.org/abs/2509.16202} {arXiv:2509.16202 [astro-ph.CO]} \BibitemShut {NoStop}%
\bibitem [{\citenamefont {Gelman}(1996)}]{Gelman1996}%
  \BibitemOpen
  \bibfield  {author} {\bibinfo {author} {\bibfnamefont {Andrew}\ \bibnamefont {Gelman}},\ }\bibfield  {title} {\enquote {\bibinfo {title} {Posterior predictive assessment of model fitness via realized discrepancies},}\ }\href@noop {} {\bibfield  {journal} {\bibinfo  {journal} {Statistica Sinica}\ }\textbf {\bibinfo {volume} {6}},\ \bibinfo {pages} {733--760} (\bibinfo {year} {1996})}\BibitemShut {NoStop}%
\bibitem [{\citenamefont {Meng}(1994)}]{Meng1994}%
  \BibitemOpen
  \bibfield  {author} {\bibinfo {author} {\bibfnamefont {Xiao-Li}\ \bibnamefont {Meng}},\ }\bibfield  {title} {\enquote {\bibinfo {title} {Posterior predictive $p$-values},}\ }\href@noop {} {\bibfield  {journal} {\bibinfo  {journal} {The Annals of Statistics}\ }\textbf {\bibinfo {volume} {22}},\ \bibinfo {pages} {1142--1160} (\bibinfo {year} {1994})}\BibitemShut {NoStop}%
\bibitem [{\citenamefont {Freedman}\ \emph {et~al.}(2025)\citenamefont {Freedman}, \citenamefont {Madore}, \citenamefont {Hoyt}, \citenamefont {Jang}, \citenamefont {Lee},\ and\ \citenamefont {Owens}}]{Freedman:2024eph}%
  \BibitemOpen
  \bibfield  {author} {\bibinfo {author} {\bibfnamefont {Wendy~L.}\ \bibnamefont {Freedman}}, \bibinfo {author} {\bibfnamefont {Barry~F.}\ \bibnamefont {Madore}}, \bibinfo {author} {\bibfnamefont {Taylor~J.}\ \bibnamefont {Hoyt}}, \bibinfo {author} {\bibfnamefont {In~Sung}\ \bibnamefont {Jang}}, \bibinfo {author} {\bibfnamefont {Abigail~J.}\ \bibnamefont {Lee}}, \ and\ \bibinfo {author} {\bibfnamefont {Kayla~A.}\ \bibnamefont {Owens}},\ }\bibfield  {title} {\enquote {\bibinfo {title} {{Status Report on the Chicago-Carnegie Hubble Program (CCHP): Measurement of the Hubble Constant Using the Hubble and James Webb Space Telescopes}},}\ }\href {\doibase 10.3847/1538-4357/adce78} {\bibfield  {journal} {\bibinfo  {journal} {Astrophys. J.}\ }\textbf {\bibinfo {volume} {985}},\ \bibinfo {pages} {203} (\bibinfo {year} {2025})},\ \Eprint {http://arxiv.org/abs/2408.06153} {arXiv:2408.06153 [astro-ph.CO]} \BibitemShut {NoStop}%
\bibitem [{\citenamefont {Camphuis}\ \emph {et~al.}(2025)\citenamefont {Camphuis} \emph {et~al.}}]{SPT-3G:2025bzu}%
  \BibitemOpen
  \bibfield  {author} {\bibinfo {author} {\bibfnamefont {E.}~\bibnamefont {Camphuis}} \emph {et~al.} (\bibinfo {collaboration} {SPT-3G}),\ }\bibfield  {title} {\enquote {\bibinfo {title} {{SPT-3G D1: CMB temperature and polarization power spectra and cosmology from 2019 and 2020 observations of the SPT-3G Main field}},}\ }\href@noop {} {\  (\bibinfo {year} {2025})},\ \Eprint {http://arxiv.org/abs/2506.20707} {arXiv:2506.20707 [astro-ph.CO]} \BibitemShut {NoStop}%
\bibitem [{\citenamefont {Hill}\ \emph {et~al.}(2022)\citenamefont {Hill} \emph {et~al.}}]{Hill:2021yec}%
  \BibitemOpen
  \bibfield  {author} {\bibinfo {author} {\bibfnamefont {J.~Colin}\ \bibnamefont {Hill}} \emph {et~al.},\ }\bibfield  {title} {\enquote {\bibinfo {title} {{Atacama Cosmology Telescope: Constraints on prerecombination early dark energy}},}\ }\href {\doibase 10.1103/PhysRevD.105.123536} {\bibfield  {journal} {\bibinfo  {journal} {Phys. Rev. D}\ }\textbf {\bibinfo {volume} {105}},\ \bibinfo {pages} {123536} (\bibinfo {year} {2022})},\ \Eprint {http://arxiv.org/abs/2109.04451} {arXiv:2109.04451 [astro-ph.CO]} \BibitemShut {NoStop}%
\bibitem [{\citenamefont {La~Posta}\ \emph {et~al.}(2022)\citenamefont {La~Posta}, \citenamefont {Louis}, \citenamefont {Garrido},\ and\ \citenamefont {Hill}}]{LaPosta:2021pgm}%
  \BibitemOpen
  \bibfield  {author} {\bibinfo {author} {\bibfnamefont {Adrien}\ \bibnamefont {La~Posta}}, \bibinfo {author} {\bibfnamefont {Thibaut}\ \bibnamefont {Louis}}, \bibinfo {author} {\bibfnamefont {Xavier}\ \bibnamefont {Garrido}}, \ and\ \bibinfo {author} {\bibfnamefont {J.~Colin}\ \bibnamefont {Hill}},\ }\bibfield  {title} {\enquote {\bibinfo {title} {{Constraints on prerecombination early dark energy from SPT-3G public data}},}\ }\href {\doibase 10.1103/PhysRevD.105.083519} {\bibfield  {journal} {\bibinfo  {journal} {Phys. Rev. D}\ }\textbf {\bibinfo {volume} {105}},\ \bibinfo {pages} {083519} (\bibinfo {year} {2022})},\ \Eprint {http://arxiv.org/abs/2112.10754} {arXiv:2112.10754 [astro-ph.CO]} \BibitemShut {NoStop}%
\bibitem [{\citenamefont {Khalife}\ \emph {et~al.}(2025)\citenamefont {Khalife} \emph {et~al.}}]{SPT-3G:2025vyw}%
  \BibitemOpen
  \bibfield  {author} {\bibinfo {author} {\bibfnamefont {A.~R.}\ \bibnamefont {Khalife}} \emph {et~al.} (\bibinfo {collaboration} {SPT-3G}),\ }\bibfield  {title} {\enquote {\bibinfo {title} {{SPT-3G D1: Axion Early Dark Energy with CMB experiments and DESI}},}\ }\href@noop {} {\  (\bibinfo {year} {2025})},\ \Eprint {http://arxiv.org/abs/2507.23355} {arXiv:2507.23355 [astro-ph.CO]} \BibitemShut {NoStop}%
\bibitem [{\citenamefont {Rosenberg}\ \emph {et~al.}(2022)\citenamefont {Rosenberg}, \citenamefont {Gratton},\ and\ \citenamefont {Efstathiou}}]{Rosenberg:2022sdy}%
  \BibitemOpen
  \bibfield  {author} {\bibinfo {author} {\bibfnamefont {Erik}\ \bibnamefont {Rosenberg}}, \bibinfo {author} {\bibfnamefont {Steven}\ \bibnamefont {Gratton}}, \ and\ \bibinfo {author} {\bibfnamefont {George}\ \bibnamefont {Efstathiou}},\ }\bibfield  {title} {\enquote {\bibinfo {title} {{CMB power spectra and cosmological parameters from Planck PR4 with CamSpec}},}\ }\href {\doibase 10.1093/mnras/stac2744} {\bibfield  {journal} {\bibinfo  {journal} {Mon. Not. Roy. Astron. Soc.}\ }\textbf {\bibinfo {volume} {517}},\ \bibinfo {pages} {4620--4636} (\bibinfo {year} {2022})},\ \Eprint {http://arxiv.org/abs/2205.10869} {arXiv:2205.10869 [astro-ph.CO]} \BibitemShut {NoStop}%
\bibitem [{\citenamefont {McDonough}\ \emph {et~al.}(2024)\citenamefont {McDonough}, \citenamefont {Hill}, \citenamefont {Ivanov}, \citenamefont {La~Posta},\ and\ \citenamefont {Toomey}}]{McDonough:2023qcu}%
  \BibitemOpen
  \bibfield  {author} {\bibinfo {author} {\bibfnamefont {Evan}\ \bibnamefont {McDonough}}, \bibinfo {author} {\bibfnamefont {J.~Colin}\ \bibnamefont {Hill}}, \bibinfo {author} {\bibfnamefont {Mikhail~M.}\ \bibnamefont {Ivanov}}, \bibinfo {author} {\bibfnamefont {Adrien}\ \bibnamefont {La~Posta}}, \ and\ \bibinfo {author} {\bibfnamefont {Michael~W.}\ \bibnamefont {Toomey}},\ }\bibfield  {title} {\enquote {\bibinfo {title} {{Observational constraints on early dark energy}},}\ }\href {\doibase 10.1142/S0218271824300039} {\bibfield  {journal} {\bibinfo  {journal} {Int. J. Mod. Phys. D}\ }\textbf {\bibinfo {volume} {33}},\ \bibinfo {pages} {2430003} (\bibinfo {year} {2024})},\ \Eprint {http://arxiv.org/abs/2310.19899} {arXiv:2310.19899 [astro-ph.CO]} \BibitemShut {NoStop}%
\bibitem [{\citenamefont {Lesgourgues}\ and\ \citenamefont {Pastor}(2006)}]{Lesgourgues:2006nd}%
  \BibitemOpen
  \bibfield  {author} {\bibinfo {author} {\bibfnamefont {Julien}\ \bibnamefont {Lesgourgues}}\ and\ \bibinfo {author} {\bibfnamefont {Sergio}\ \bibnamefont {Pastor}},\ }\bibfield  {title} {\enquote {\bibinfo {title} {{Massive neutrinos and cosmology}},}\ }\href {\doibase 10.1016/j.physrep.2006.04.001} {\bibfield  {journal} {\bibinfo  {journal} {Phys. Rept.}\ }\textbf {\bibinfo {volume} {429}},\ \bibinfo {pages} {307--379} (\bibinfo {year} {2006})},\ \Eprint {http://arxiv.org/abs/astro-ph/0603494} {arXiv:astro-ph/0603494} \BibitemShut {NoStop}%
\bibitem [{\citenamefont {Wong}(2011)}]{Wong:2011ip}%
  \BibitemOpen
  \bibfield  {author} {\bibinfo {author} {\bibfnamefont {Yvonne Y.~Y.}\ \bibnamefont {Wong}},\ }\bibfield  {title} {\enquote {\bibinfo {title} {{Neutrino mass in cosmology: status and prospects}},}\ }\href {\doibase 10.1146/annurev-nucl-102010-130252} {\bibfield  {journal} {\bibinfo  {journal} {Ann. Rev. Nucl. Part. Sci.}\ }\textbf {\bibinfo {volume} {61}},\ \bibinfo {pages} {69--98} (\bibinfo {year} {2011})},\ \Eprint {http://arxiv.org/abs/1111.1436} {arXiv:1111.1436 [astro-ph.CO]} \BibitemShut {NoStop}%
\bibitem [{\citenamefont {Dvorkin}\ \emph {et~al.}(2019)\citenamefont {Dvorkin} \emph {et~al.}}]{Dvorkin:2019jgs}%
  \BibitemOpen
  \bibfield  {author} {\bibinfo {author} {\bibfnamefont {Cora}\ \bibnamefont {Dvorkin}} \emph {et~al.},\ }\bibfield  {title} {\enquote {\bibinfo {title} {{Neutrino Mass from Cosmology: Probing Physics Beyond the Standard Model}},}\ }\href@noop {} {\  (\bibinfo {year} {2019})},\ \Eprint {http://arxiv.org/abs/1903.03689} {arXiv:1903.03689 [astro-ph.CO]} \BibitemShut {NoStop}%
\bibitem [{\citenamefont {Craig}\ \emph {et~al.}(2024)\citenamefont {Craig}, \citenamefont {Green}, \citenamefont {Meyers},\ and\ \citenamefont {Rajendran}}]{Craig:2024tky}%
  \BibitemOpen
  \bibfield  {author} {\bibinfo {author} {\bibfnamefont {Nathaniel}\ \bibnamefont {Craig}}, \bibinfo {author} {\bibfnamefont {Daniel}\ \bibnamefont {Green}}, \bibinfo {author} {\bibfnamefont {Joel}\ \bibnamefont {Meyers}}, \ and\ \bibinfo {author} {\bibfnamefont {Surjeet}\ \bibnamefont {Rajendran}},\ }\bibfield  {title} {\enquote {\bibinfo {title} {{No $\nu$s is Good News}},}\ }\href@noop {} {\  (\bibinfo {year} {2024})},\ \Eprint {http://arxiv.org/abs/2405.00836} {arXiv:2405.00836 [astro-ph.CO]} \BibitemShut {NoStop}%
\bibitem [{\citenamefont {Green}\ and\ \citenamefont {Meyers}(2024)}]{Green:2024xbb}%
  \BibitemOpen
  \bibfield  {author} {\bibinfo {author} {\bibfnamefont {Daniel}\ \bibnamefont {Green}}\ and\ \bibinfo {author} {\bibfnamefont {Joel}\ \bibnamefont {Meyers}},\ }\bibfield  {title} {\enquote {\bibinfo {title} {{The Cosmological Preference for Negative Neutrino Mass}},}\ }\href@noop {} {\  (\bibinfo {year} {2024})},\ \Eprint {http://arxiv.org/abs/2407.07878} {arXiv:2407.07878 [astro-ph.CO]} \BibitemShut {NoStop}%
\bibitem [{\citenamefont {Elbers}\ \emph {et~al.}(2025)\citenamefont {Elbers} \emph {et~al.}}]{DESI:2025ejh}%
  \BibitemOpen
  \bibfield  {author} {\bibinfo {author} {\bibfnamefont {W.}~\bibnamefont {Elbers}} \emph {et~al.} (\bibinfo {collaboration} {DESI}),\ }\bibfield  {title} {\enquote {\bibinfo {title} {{Constraints on Neutrino Physics from DESI DR2 BAO and DR1 Full Shape}},}\ }\href@noop {} {\  (\bibinfo {year} {2025})},\ \Eprint {http://arxiv.org/abs/2503.14744} {arXiv:2503.14744 [astro-ph.CO]} \BibitemShut {NoStop}%
\bibitem [{\citenamefont {Sailer}\ \emph {et~al.}(2025)\citenamefont {Sailer}, \citenamefont {Farren}, \citenamefont {Ferraro},\ and\ \citenamefont {White}}]{Sailer:2025lxj}%
  \BibitemOpen
  \bibfield  {author} {\bibinfo {author} {\bibfnamefont {Noah}\ \bibnamefont {Sailer}}, \bibinfo {author} {\bibfnamefont {Gerrit~S.}\ \bibnamefont {Farren}}, \bibinfo {author} {\bibfnamefont {Simone}\ \bibnamefont {Ferraro}}, \ and\ \bibinfo {author} {\bibfnamefont {Martin}\ \bibnamefont {White}},\ }\bibfield  {title} {\enquote {\bibinfo {title} {{Dispu$\tau$able: the high cost of a low optical depth}},}\ }\href@noop {} {\  (\bibinfo {year} {2025})},\ \Eprint {http://arxiv.org/abs/2504.16932} {arXiv:2504.16932 [astro-ph.CO]} \BibitemShut {NoStop}%
\bibitem [{\citenamefont {Jhaveri}\ \emph {et~al.}(2025)\citenamefont {Jhaveri}, \citenamefont {Karwal},\ and\ \citenamefont {Hu}}]{Jhaveri:2025neg}%
  \BibitemOpen
  \bibfield  {author} {\bibinfo {author} {\bibfnamefont {Tanisha}\ \bibnamefont {Jhaveri}}, \bibinfo {author} {\bibfnamefont {Tanvi}\ \bibnamefont {Karwal}}, \ and\ \bibinfo {author} {\bibfnamefont {Wayne}\ \bibnamefont {Hu}},\ }\bibfield  {title} {\enquote {\bibinfo {title} {{Turning a negative neutrino mass into a positive optical depth}},}\ }\href@noop {} {\  (\bibinfo {year} {2025})},\ \Eprint {http://arxiv.org/abs/2504.21813} {arXiv:2504.21813 [astro-ph.CO]} \BibitemShut {NoStop}%
\bibitem [{\citenamefont {Elbers}\ \emph {et~al.}(2024)\citenamefont {Elbers}, \citenamefont {Frenk}, \citenamefont {Jenkins}, \citenamefont {Li},\ and\ \citenamefont {Pascoli}}]{Elbers:2024sha}%
  \BibitemOpen
  \bibfield  {author} {\bibinfo {author} {\bibfnamefont {Willem}\ \bibnamefont {Elbers}}, \bibinfo {author} {\bibfnamefont {Carlos~S.}\ \bibnamefont {Frenk}}, \bibinfo {author} {\bibfnamefont {Adrian}\ \bibnamefont {Jenkins}}, \bibinfo {author} {\bibfnamefont {Baojiu}\ \bibnamefont {Li}}, \ and\ \bibinfo {author} {\bibfnamefont {Silvia}\ \bibnamefont {Pascoli}},\ }\bibfield  {title} {\enquote {\bibinfo {title} {{Negative neutrino masses as a mirage of dark energy}},}\ }\href@noop {} {\  (\bibinfo {year} {2024})},\ \Eprint {http://arxiv.org/abs/2407.10965} {arXiv:2407.10965 [astro-ph.CO]} \BibitemShut {NoStop}%
\bibitem [{\citenamefont {Ahlen}\ \emph {et~al.}(2025)\citenamefont {Ahlen} \emph {et~al.}}]{DESI:2025ffm}%
  \BibitemOpen
  \bibfield  {author} {\bibinfo {author} {\bibfnamefont {S.}~\bibnamefont {Ahlen}} \emph {et~al.} (\bibinfo {collaboration} {DESI}),\ }\bibfield  {title} {\enquote {\bibinfo {title} {{Positive neutrino masses with DESI DR2 via matter conversion to dark energy}},}\ }\href@noop {} {\  (\bibinfo {year} {2025})},\ \Eprint {http://arxiv.org/abs/2504.20338} {arXiv:2504.20338 [astro-ph.CO]} \BibitemShut {NoStop}%
\bibitem [{\citenamefont {Shao}\ \emph {et~al.}(2025)\citenamefont {Shao}, \citenamefont {Givans}, \citenamefont {Dunkley}, \citenamefont {Madhavacheril}, \citenamefont {Qu}, \citenamefont {Farren},\ and\ \citenamefont {Sherwin}}]{Shao:2024mag}%
  \BibitemOpen
  \bibfield  {author} {\bibinfo {author} {\bibfnamefont {Helen}\ \bibnamefont {Shao}}, \bibinfo {author} {\bibfnamefont {Jahmour~J.}\ \bibnamefont {Givans}}, \bibinfo {author} {\bibfnamefont {Jo}~\bibnamefont {Dunkley}}, \bibinfo {author} {\bibfnamefont {Mathew}\ \bibnamefont {Madhavacheril}}, \bibinfo {author} {\bibfnamefont {Frank~J.}\ \bibnamefont {Qu}}, \bibinfo {author} {\bibfnamefont {Gerrit}\ \bibnamefont {Farren}}, \ and\ \bibinfo {author} {\bibfnamefont {Blake}\ \bibnamefont {Sherwin}},\ }\bibfield  {title} {\enquote {\bibinfo {title} {{Cosmological limits on the neutrino mass sum for beyond-{\ensuremath{\Lambda}}CDM models}},}\ }\href {\doibase 10.1103/PhysRevD.111.083535} {\bibfield  {journal} {\bibinfo  {journal} {Phys. Rev. D}\ }\textbf {\bibinfo {volume} {111}},\ \bibinfo {pages} {083535} (\bibinfo {year} {2025})},\ \Eprint {http://arxiv.org/abs/2409.02295} {arXiv:2409.02295 [astro-ph.CO]} \BibitemShut {NoStop}%
\bibitem [{\citenamefont {Roy~Choudhury}\ and\ \citenamefont {Okumura}(2024)}]{RoyChoudhury:2024wri}%
  \BibitemOpen
  \bibfield  {author} {\bibinfo {author} {\bibfnamefont {Shouvik}\ \bibnamefont {Roy~Choudhury}}\ and\ \bibinfo {author} {\bibfnamefont {Teppei}\ \bibnamefont {Okumura}},\ }\bibfield  {title} {\enquote {\bibinfo {title} {{Updated Cosmological Constraints in Extended Parameter Space with Planck PR4, DESI Baryon Acoustic Oscillations, and Supernovae: Dynamical Dark Energy, Neutrino Masses, Lensing Anomaly, and the Hubble Tension}},}\ }\href {\doibase 10.3847/2041-8213/ad8c26} {\bibfield  {journal} {\bibinfo  {journal} {Astrophys. J. Lett.}\ }\textbf {\bibinfo {volume} {976}},\ \bibinfo {pages} {L11} (\bibinfo {year} {2024})},\ \Eprint {http://arxiv.org/abs/2409.13022} {arXiv:2409.13022 [astro-ph.CO]} \BibitemShut {NoStop}%
\bibitem [{\citenamefont {Qu}\ \emph {et~al.}(2025{\natexlab{b}})\citenamefont {Qu}, \citenamefont {Surrao}, \citenamefont {Bolliet}, \citenamefont {Hill}, \citenamefont {Sherwin}, \citenamefont {Jense},\ and\ \citenamefont {La~Posta}}]{Qu:2024lpx}%
  \BibitemOpen
  \bibfield  {author} {\bibinfo {author} {\bibfnamefont {Frank~J.}\ \bibnamefont {Qu}}, \bibinfo {author} {\bibfnamefont {Kristen~M.}\ \bibnamefont {Surrao}}, \bibinfo {author} {\bibfnamefont {Boris}\ \bibnamefont {Bolliet}}, \bibinfo {author} {\bibfnamefont {J.~Colin}\ \bibnamefont {Hill}}, \bibinfo {author} {\bibfnamefont {Blake~D.}\ \bibnamefont {Sherwin}}, \bibinfo {author} {\bibfnamefont {Hidde~T.}\ \bibnamefont {Jense}}, \ and\ \bibinfo {author} {\bibfnamefont {Adrien}\ \bibnamefont {La~Posta}},\ }\bibfield  {title} {\enquote {\bibinfo {title} {{Accelerated inference on accelerated cosmic expansion: New constraints on axionlike early dark energy with DESI BAO and ACT DR6 CMB lensing}},}\ }\href {\doibase 10.1103/xhh6-9v62} {\bibfield  {journal} {\bibinfo  {journal} {Phys. Rev. D}\ }\textbf {\bibinfo {volume} {111}},\ \bibinfo {pages} {123507} (\bibinfo {year} {2025}{\natexlab{b}})},\ \Eprint {http://arxiv.org/abs/2404.16805} {arXiv:2404.16805 [astro-ph.CO]} \BibitemShut {NoStop}%
\bibitem [{\citenamefont {Reeves}\ \emph {et~al.}(2023)\citenamefont {Reeves}, \citenamefont {Herold}, \citenamefont {Vagnozzi}, \citenamefont {Sherwin},\ and\ \citenamefont {Ferreira}}]{Reeves:2022aoi}%
  \BibitemOpen
  \bibfield  {author} {\bibinfo {author} {\bibfnamefont {Alexander}\ \bibnamefont {Reeves}}, \bibinfo {author} {\bibfnamefont {Laura}\ \bibnamefont {Herold}}, \bibinfo {author} {\bibfnamefont {Sunny}\ \bibnamefont {Vagnozzi}}, \bibinfo {author} {\bibfnamefont {Blake~D.}\ \bibnamefont {Sherwin}}, \ and\ \bibinfo {author} {\bibfnamefont {Elisa G.~M.}\ \bibnamefont {Ferreira}},\ }\bibfield  {title} {\enquote {\bibinfo {title} {{Restoring cosmological concordance with early dark energy and massive neutrinos?}}}\ }\href {\doibase 10.1093/mnras/stad317} {\bibfield  {journal} {\bibinfo  {journal} {Mon. Not. Roy. Astron. Soc.}\ }\textbf {\bibinfo {volume} {520}},\ \bibinfo {pages} {3688--3695} (\bibinfo {year} {2023})},\ \Eprint {http://arxiv.org/abs/2207.01501} {arXiv:2207.01501 [astro-ph.CO]} \BibitemShut {NoStop}%
\bibitem [{\citenamefont {Rubin}\ \emph {et~al.}(2023)\citenamefont {Rubin} \emph {et~al.}}]{Rubin:2023jdq}%
  \BibitemOpen
  \bibfield  {author} {\bibinfo {author} {\bibfnamefont {David}\ \bibnamefont {Rubin}} \emph {et~al.},\ }\bibfield  {title} {\enquote {\bibinfo {title} {{Union Through UNITY: Cosmology with 2,000 SNe Using a Unified Bayesian Framework}},}\ }\href@noop {} {\  (\bibinfo {year} {2023})},\ \Eprint {http://arxiv.org/abs/2311.12098} {arXiv:2311.12098 [astro-ph.CO]} \BibitemShut {NoStop}%
\bibitem [{\citenamefont {Abbott}\ \emph {et~al.}(2024)\citenamefont {Abbott} \emph {et~al.}}]{DES:2024jxu}%
  \BibitemOpen
  \bibfield  {author} {\bibinfo {author} {\bibfnamefont {T.~M.~C.}\ \bibnamefont {Abbott}} \emph {et~al.} (\bibinfo {collaboration} {DES}),\ }\bibfield  {title} {\enquote {\bibinfo {title} {{The Dark Energy Survey: Cosmology Results with {\ensuremath{\sim}}1500 New High-redshift Type Ia Supernovae Using the Full 5 yr Data Set}},}\ }\href {\doibase 10.3847/2041-8213/ad6f9f} {\bibfield  {journal} {\bibinfo  {journal} {Astrophys. J. Lett.}\ }\textbf {\bibinfo {volume} {973}},\ \bibinfo {pages} {L14} (\bibinfo {year} {2024})},\ \Eprint {http://arxiv.org/abs/2401.02929} {arXiv:2401.02929 [astro-ph.CO]} \BibitemShut {NoStop}%
\bibitem [{\citenamefont {Hill}\ \emph {et~al.}(2020)\citenamefont {Hill}, \citenamefont {McDonough}, \citenamefont {Toomey},\ and\ \citenamefont {Alexander}}]{Hill:2020osr}%
  \BibitemOpen
  \bibfield  {author} {\bibinfo {author} {\bibfnamefont {J.~Colin}\ \bibnamefont {Hill}}, \bibinfo {author} {\bibfnamefont {Evan}\ \bibnamefont {McDonough}}, \bibinfo {author} {\bibfnamefont {Michael~W.}\ \bibnamefont {Toomey}}, \ and\ \bibinfo {author} {\bibfnamefont {Stephon}\ \bibnamefont {Alexander}},\ }\bibfield  {title} {\enquote {\bibinfo {title} {{Early dark energy does not restore cosmological concordance}},}\ }\href {\doibase 10.1103/PhysRevD.102.043507} {\bibfield  {journal} {\bibinfo  {journal} {Phys. Rev. D}\ }\textbf {\bibinfo {volume} {102}},\ \bibinfo {pages} {043507} (\bibinfo {year} {2020})},\ \Eprint {http://arxiv.org/abs/2003.07355} {arXiv:2003.07355 [astro-ph.CO]} \BibitemShut {NoStop}%
\bibitem [{\citenamefont {Vagnozzi}(2023)}]{Vagnozzi:2023nrq}%
  \BibitemOpen
  \bibfield  {author} {\bibinfo {author} {\bibfnamefont {Sunny}\ \bibnamefont {Vagnozzi}},\ }\bibfield  {title} {\enquote {\bibinfo {title} {{Seven Hints That Early-Time New Physics Alone Is Not Sufficient to Solve the Hubble Tension}},}\ }\href {\doibase 10.3390/universe9090393} {\bibfield  {journal} {\bibinfo  {journal} {Universe}\ }\textbf {\bibinfo {volume} {9}},\ \bibinfo {pages} {393} (\bibinfo {year} {2023})},\ \Eprint {http://arxiv.org/abs/2308.16628} {arXiv:2308.16628 [astro-ph.CO]} \BibitemShut {NoStop}%
\bibitem [{\citenamefont {Asgari}\ \emph {et~al.}(2021)\citenamefont {Asgari} \emph {et~al.}}]{KiDS:2020suj}%
  \BibitemOpen
  \bibfield  {author} {\bibinfo {author} {\bibfnamefont {Marika}\ \bibnamefont {Asgari}} \emph {et~al.} (\bibinfo {collaboration} {KiDS}),\ }\bibfield  {title} {\enquote {\bibinfo {title} {{KiDS-1000 Cosmology: Cosmic shear constraints and comparison between two point statistics}},}\ }\href {\doibase 10.1051/0004-6361/202039070} {\bibfield  {journal} {\bibinfo  {journal} {Astron. Astrophys.}\ }\textbf {\bibinfo {volume} {645}},\ \bibinfo {pages} {A104} (\bibinfo {year} {2021})},\ \Eprint {http://arxiv.org/abs/2007.15633} {arXiv:2007.15633 [astro-ph.CO]} \BibitemShut {NoStop}%
\bibitem [{\citenamefont {Di~Valentino}\ \emph {et~al.}(2025)\citenamefont {Di~Valentino} \emph {et~al.}}]{CosmoVerseNetwork:2025alb}%
  \BibitemOpen
  \bibfield  {author} {\bibinfo {author} {\bibfnamefont {Eleonora}\ \bibnamefont {Di~Valentino}} \emph {et~al.} (\bibinfo {collaboration} {CosmoVerse Network}),\ }\bibfield  {title} {\enquote {\bibinfo {title} {{The CosmoVerse White Paper: Addressing observational tensions in cosmology with systematics and fundamental physics}},}\ }\href {\doibase 10.1016/j.dark.2025.101965} {\bibfield  {journal} {\bibinfo  {journal} {Phys. Dark Univ.}\ }\textbf {\bibinfo {volume} {49}},\ \bibinfo {pages} {101965} (\bibinfo {year} {2025})},\ \Eprint {http://arxiv.org/abs/2504.01669} {arXiv:2504.01669 [astro-ph.CO]} \BibitemShut {NoStop}%
\bibitem [{\citenamefont {Laureijs}\ \emph {et~al.}(2011)\citenamefont {Laureijs} \emph {et~al.}}]{EUCLID:2011zbd}%
  \BibitemOpen
  \bibfield  {author} {\bibinfo {author} {\bibfnamefont {R.}~\bibnamefont {Laureijs}} \emph {et~al.} (\bibinfo {collaboration} {EUCLID}),\ }\bibfield  {title} {\enquote {\bibinfo {title} {{Euclid Definition Study Report}},}\ }\href@noop {} {\  (\bibinfo {year} {2011})},\ \Eprint {http://arxiv.org/abs/1110.3193} {arXiv:1110.3193 [astro-ph.CO]} \BibitemShut {NoStop}%
\bibitem [{\citenamefont {Abell}\ \emph {et~al.}(2009)\citenamefont {Abell} \emph {et~al.}}]{LSSTScience:2009jmu}%
  \BibitemOpen
  \bibfield  {author} {\bibinfo {author} {\bibfnamefont {Paul~A.}\ \bibnamefont {Abell}} \emph {et~al.} (\bibinfo {collaboration} {LSST Science, LSST Project}),\ }\bibfield  {title} {\enquote {\bibinfo {title} {{LSST Science Book, Version 2.0}},}\ }\href@noop {} {\  (\bibinfo {year} {2009})},\ \Eprint {http://arxiv.org/abs/0912.0201} {arXiv:0912.0201 [astro-ph.IM]} \BibitemShut {NoStop}%
\bibitem [{\citenamefont {Philcox}\ \emph {et~al.}(2020)\citenamefont {Philcox}, \citenamefont {Ivanov}, \citenamefont {Simonovi\'c},\ and\ \citenamefont {Zaldarriaga}}]{Philcox:2020vvt}%
  \BibitemOpen
  \bibfield  {author} {\bibinfo {author} {\bibfnamefont {Oliver H.~E.}\ \bibnamefont {Philcox}}, \bibinfo {author} {\bibfnamefont {Mikhail~M.}\ \bibnamefont {Ivanov}}, \bibinfo {author} {\bibfnamefont {Marko}\ \bibnamefont {Simonovi\'c}}, \ and\ \bibinfo {author} {\bibfnamefont {Matias}\ \bibnamefont {Zaldarriaga}},\ }\bibfield  {title} {\enquote {\bibinfo {title} {{Combining Full-Shape and BAO Analyses of Galaxy Power Spectra: A 1.6\textbackslash{}\% CMB-independent constraint on H$_0$}},}\ }\href {\doibase 10.1088/1475-7516/2020/05/032} {\bibfield  {journal} {\bibinfo  {journal} {JCAP}\ }\textbf {\bibinfo {volume} {05}},\ \bibinfo {pages} {032} (\bibinfo {year} {2020})},\ \Eprint {http://arxiv.org/abs/2002.04035} {arXiv:2002.04035 [astro-ph.CO]} \BibitemShut {NoStop}%
\bibitem [{\citenamefont {Hamann}\ \emph {et~al.}(2010)\citenamefont {Hamann}, \citenamefont {Hannestad}, \citenamefont {Lesgourgues}, \citenamefont {Rampf},\ and\ \citenamefont {Wong}}]{Hamann:2010pw}%
  \BibitemOpen
  \bibfield  {author} {\bibinfo {author} {\bibfnamefont {Jan}\ \bibnamefont {Hamann}}, \bibinfo {author} {\bibfnamefont {Steen}\ \bibnamefont {Hannestad}}, \bibinfo {author} {\bibfnamefont {Julien}\ \bibnamefont {Lesgourgues}}, \bibinfo {author} {\bibfnamefont {Cornelius}\ \bibnamefont {Rampf}}, \ and\ \bibinfo {author} {\bibfnamefont {Yvonne Y.~Y.}\ \bibnamefont {Wong}},\ }\bibfield  {title} {\enquote {\bibinfo {title} {{Cosmological parameters from large scale structure - geometric versus shape information}},}\ }\href {\doibase 10.1088/1475-7516/2010/07/022} {\bibfield  {journal} {\bibinfo  {journal} {JCAP}\ }\textbf {\bibinfo {volume} {07}},\ \bibinfo {pages} {022} (\bibinfo {year} {2010})},\ \Eprint {http://arxiv.org/abs/1003.3999} {arXiv:1003.3999 [astro-ph.CO]} \BibitemShut {NoStop}%
\bibitem [{\citenamefont {Chudaykin}\ \emph {et~al.}(2020)\citenamefont {Chudaykin}, \citenamefont {Ivanov}, \citenamefont {Philcox},\ and\ \citenamefont {Simonovi\'c}}]{Chudaykin:2020aoj}%
  \BibitemOpen
  \bibfield  {author} {\bibinfo {author} {\bibfnamefont {Anton}\ \bibnamefont {Chudaykin}}, \bibinfo {author} {\bibfnamefont {Mikhail~M.}\ \bibnamefont {Ivanov}}, \bibinfo {author} {\bibfnamefont {Oliver H.~E.}\ \bibnamefont {Philcox}}, \ and\ \bibinfo {author} {\bibfnamefont {Marko}\ \bibnamefont {Simonovi\'c}},\ }\bibfield  {title} {\enquote {\bibinfo {title} {{Nonlinear perturbation theory extension of the Boltzmann code CLASS}},}\ }\href {\doibase 10.1103/PhysRevD.102.063533} {\bibfield  {journal} {\bibinfo  {journal} {Phys. Rev. D}\ }\textbf {\bibinfo {volume} {102}},\ \bibinfo {pages} {063533} (\bibinfo {year} {2020})},\ \Eprint {http://arxiv.org/abs/2004.10607} {arXiv:2004.10607 [astro-ph.CO]} \BibitemShut {NoStop}%
\bibitem [{\citenamefont {Herold}\ \emph {et~al.}(2022)\citenamefont {Herold}, \citenamefont {Ferreira},\ and\ \citenamefont {Komatsu}}]{Herold:2021ksg}%
  \BibitemOpen
  \bibfield  {author} {\bibinfo {author} {\bibfnamefont {Laura}\ \bibnamefont {Herold}}, \bibinfo {author} {\bibfnamefont {Elisa G.~M.}\ \bibnamefont {Ferreira}}, \ and\ \bibinfo {author} {\bibfnamefont {Eiichiro}\ \bibnamefont {Komatsu}},\ }\bibfield  {title} {\enquote {\bibinfo {title} {{New Constraint on Early Dark Energy from Planck and BOSS Data Using the Profile Likelihood}},}\ }\href {\doibase 10.3847/2041-8213/ac63a3} {\bibfield  {journal} {\bibinfo  {journal} {Astrophys. J. Lett.}\ }\textbf {\bibinfo {volume} {929}},\ \bibinfo {pages} {L16} (\bibinfo {year} {2022})},\ \Eprint {http://arxiv.org/abs/2112.12140} {arXiv:2112.12140 [astro-ph.CO]} \BibitemShut {NoStop}%
\bibitem [{\citenamefont {Herold}\ and\ \citenamefont {Ferreira}(2023)}]{Herold:2022iib}%
  \BibitemOpen
  \bibfield  {author} {\bibinfo {author} {\bibfnamefont {Laura}\ \bibnamefont {Herold}}\ and\ \bibinfo {author} {\bibfnamefont {Elisa G.~M.}\ \bibnamefont {Ferreira}},\ }\bibfield  {title} {\enquote {\bibinfo {title} {{Resolving the Hubble tension with early dark energy}},}\ }\href {\doibase 10.1103/PhysRevD.108.043513} {\bibfield  {journal} {\bibinfo  {journal} {Phys. Rev. D}\ }\textbf {\bibinfo {volume} {108}},\ \bibinfo {pages} {043513} (\bibinfo {year} {2023})},\ \Eprint {http://arxiv.org/abs/2210.16296} {arXiv:2210.16296 [astro-ph.CO]} \BibitemShut {NoStop}%
\bibitem [{\citenamefont {Farren}\ \emph {et~al.}(2024)\citenamefont {Farren} \emph {et~al.}}]{ACT:2023oei}%
  \BibitemOpen
  \bibfield  {author} {\bibinfo {author} {\bibfnamefont {Gerrit~S.}\ \bibnamefont {Farren}} \emph {et~al.} (\bibinfo {collaboration} {ACT}),\ }\bibfield  {title} {\enquote {\bibinfo {title} {{The Atacama Cosmology Telescope: Cosmology from Cross-correlations of unWISE Galaxies and ACT DR6 CMB Lensing}},}\ }\href {\doibase 10.3847/1538-4357/ad31a5} {\bibfield  {journal} {\bibinfo  {journal} {Astrophys. J.}\ }\textbf {\bibinfo {volume} {966}},\ \bibinfo {pages} {157} (\bibinfo {year} {2024})},\ \Eprint {http://arxiv.org/abs/2309.05659} {arXiv:2309.05659 [astro-ph.CO]} \BibitemShut {NoStop}%
\end{thebibliography}%

\appendix
\section{Compressed \hillipop\ likelihood \label{appendix:compressed_hillipop}}
\begin{figure}
    \centering
    \includegraphics[width=\linewidth]{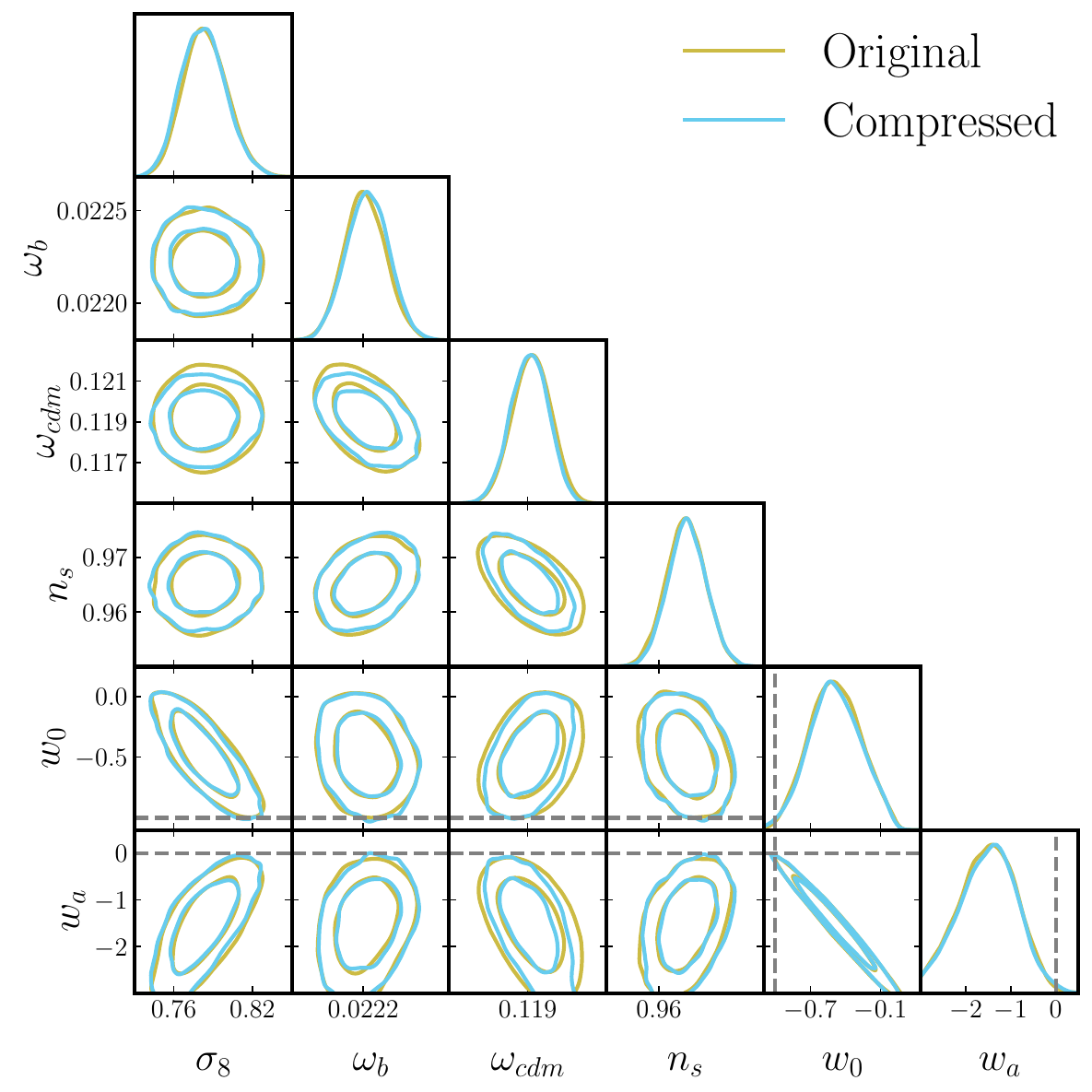}
    \caption{\textbf{Compressed \hillipop\ vs original for combination of \textit{Planck} PR4 + BAO data in the DDE model.}}
    \label{fig:compressed_vs_uncompressed}
\end{figure}

To compress the \hillipop\ likelihood, we bin the full high-$\ell$ likelihood (which has $N_d=29758$ data points corresponding to a concatenation of unbinned $TTTEEE$ spectra for several frequency channels) into $N=1573$ bandpowers. Specifically, we adopt a simple-minded binning scheme of
\begin{equation} 
\Delta\ell = 
\begin{cases}
5, & 30 \le \ell < 100,\\ 
10, & 100 \le \ell < 500,\\ 
20, & 500 \le \ell < 1000,\\ 
50, & 1000 \le \ell \le 2500~,
\end{cases}
\end{equation}
we additionally weight each mode entering the bins with a cosmic variance weighting of $w_\ell=\ell(\ell+1)$. This reduces the data-vector size by a factor of $\sim 19$ (from 29,758 to 1,573 points) and we checked that the posteriors obtained in the $\Lambda\mathrm{CDM}$, $EDE$ and DDE model are identical to the uncompressed posteriors (see Fig.~\ref{fig:compressed_vs_uncompressed} for the DDE comparison) as might be expected as cosmological parameters typically induce smooth changes to the associated CMB power spectra. This compression is useful as it speeds up the analysis by a factor of $\sim1.8$ by reducing the memory overhead of evaluating one likelihood and thereby reducing memory latency when running inference on GPUs. We make the \texttt{JAX}-version of the \hillipop\ likelihood and the corresponding binning matrix for this compression publicly available \url{https://github.com/alexander-reeves/jax-loglike}. 

\section{To what extent are we double-counting BAO information?} \label{appendix:bao_information}

\begin{figure*}
    \centering
    \includegraphics[width=\linewidth]{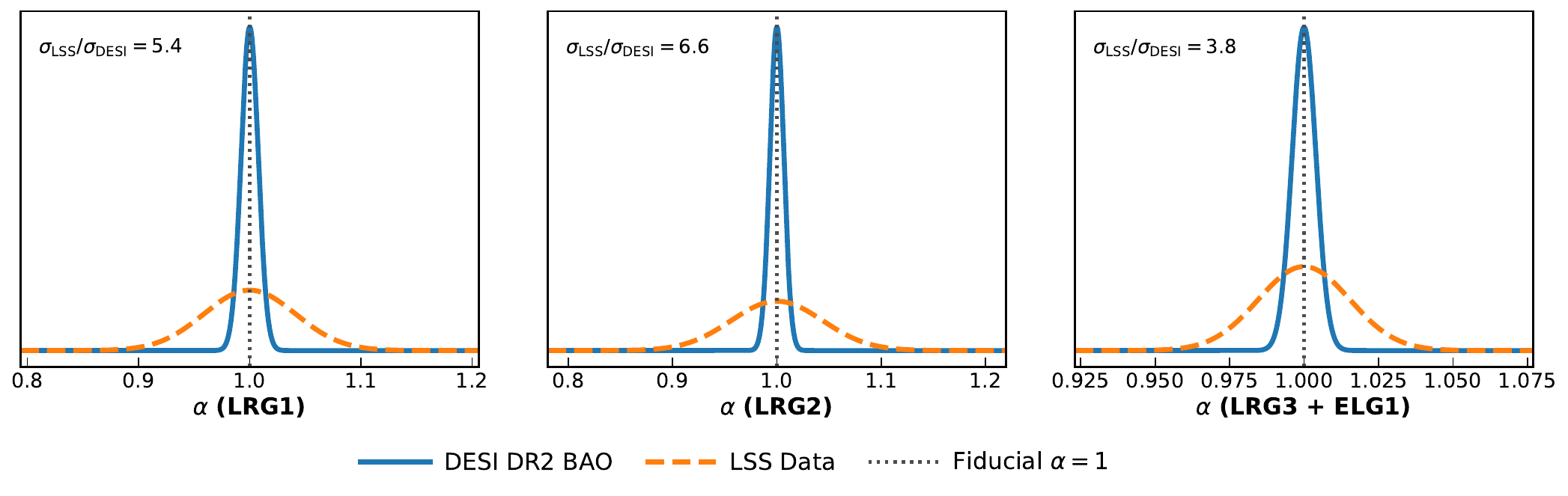}
    \caption{\textbf{Theoretical BAO information content in LSS data.} The BAO information content is compared for the three redshift bins that are most tightly constrained by the LSS data. For all other DESI DR2 BAO redshift bins, the LSS data provides essentially no BAO information.}
    \label{fig:lss_bao_information}
\end{figure*}

We perform the same analysis presented in Appendix~A of Ref.~\cite{Reeves:2025axp} using these updated data to assess the validity of the assumption that we can combine the LSS and BAO information at the likelihood level while ignoring their cross-correlation. We begin by defining a wiggle--no-wiggle split of the linear matter power spectrum following Refs.~\cite{Philcox:2020vvt, Hamann:2010pw, Chudaykin:2020aoj}, using an approximate analytic expression for the ``wiggly'' component:
\begin{equation}\label{eqt:wiggly_bao}
\begin{split}
P_{w}(P_{nw}, k, \alpha) \approx A\,P_{nw}(k)\,
\sin\!\left(\frac{k\,\ell_{\mathrm{BAO}}}{\alpha}\right)\, \\
\qquad \times \exp\!\bigl[-k^2\bigl(\Sigma_{\mathrm{NL}}^2 + \Sigma_{\mathrm{Silk}}^2\bigr)\bigr]\,.
\end{split}
\end{equation}
where $P_{nw}$ is the smooth (no-wiggle) power spectrum, $A\approx0.05$ is an overall amplitude rescaling, $\ell_{\mathrm{BAO}}\approx 105\,h^{-1}\mathrm{Mpc}$, $\Sigma_{\mathrm{NL}}\approx 3\,h^{-1}\mathrm{Mpc}$, $\Sigma_{\mathrm{Silk}}\approx 5\,h^{-1}\mathrm{Mpc}$, and $\alpha$ is the isotropic Alcock--Paczynski parameter. We obtain $P_{nw}$ by performing a ``wiggly--non-wiggly split’’ via the discrete spectral analysis method of Refs.~\cite{Hamann:2010pw, Chudaykin:2020aoj}. This method identifies and removes BAO-induced features in the Fourier transform of $\ln\!\bigl(k\,P(k)\bigr)$, replacing them with a smooth spline before inverse-transforming back to $k$-space. This functional form accurately (to within $5\%$) reproduces the BAO residual in our fiducial (\textit{Planck}~2018) cosmology, providing a suitable order-of-magnitude model of the BAO signal.

Next, to examine the sensitivity of our framework to the BAO scale, we construct a \texttt{JAX}-based theoretical calculation that combines the smooth component $P_{nw}(k,z)$ with its ``wiggly’’ counterpart $P_w$ in each DESI DR2 BAO redshift bin. Assuming linear theory (which is accurate for an order-of-magnitude assessment of the cross-covariance), we write 
\begin{equation}
\begin{split}
P_\mathrm{tot}(k,z) \;=\; P_{nw}(k,0)\,D^2(z) \\
\qquad +\; P_w\!\Bigl[P_{nw}(k,0)\,D^2(z),\,k,\,\alpha(z)\Bigr]\,.
\end{split}
\end{equation}
where $D(z)$ is the linear growth factor. We then project $P_\mathrm{tot}(k,z)$ into angular power spectra $C_\ell$ via a Limber integral, using each probe’s window function. In this way, we can systematically vary $\alpha(z)$ and assess the extent to which BAO wiggles affect our $5\times2$pt observables. Since we vary $\alpha(z)$ uniformly across each DESI redshift bin without including detailed selection effects, our Fisher estimates should be regarded as an upper limit on the BAO information available to the $5\times2$pt observables. In practice, survey window functions and non-uniform redshift weighting would further dilute this sensitivity.

We can derive the Fisher information matrix following
\begin{equation}
    \mathcal{F} \;=\; 
    \Biggl(\frac{\partial \vec{T}}{\partial \alpha_i}\Biggr)^T
    C^{-1}_{ij}
    \Biggl(\frac{\partial \vec{T}}{\partial \alpha_j}\Biggr)\,,
\end{equation}
where $\vec{T}(\vec{\alpha})$ is the theory prediction for the $5\times2$pt data vector, which depends on the $\alpha^{\mathrm{iso}}$ parameters. We make use of \texttt{JAX}'s automatic differentiation feature to accurately compute these derivatives. The inverse of $\mathcal{F}$ (the precision matrix) provides an estimate of how sensitively our $5\times2$pt data constrain each $\alpha$ parameter. We present the one-dimensional Gaussian distributions representing this constraining power in Fig.~\ref{fig:lss_bao_information}. Comparing these constraints with the DESI DR2 BAO uncertainties from Table~IV of Ref.~\cite{DESI:2025dr2}, we find that the $5\times2$pt data vector is at least four times weaker (i.e., it yields parameter uncertainties about four times larger) than the DESI DR2 BAO results in constraining the $\alpha$ parameters. The BAO-scale parameter for the LRG1 sample is the most tightly constrained by the $5\times2$pt data for the LRG3 + ELG1 DESI bin, which overlaps with the fourth and third tomographic bins of the LS LRGs. We do not display the other BAO bins, as the constraining power from the LSS data vector is found to be negligible due to the limited overlap in redshift with the LS LRGS. 

Overall, this back-of-the-envelope calculation, which likely overestimates the sensitivity of projected LSS probes due to the simplification of uniformly varying $\alpha(z)$ across the redshift bins associated with the DR2 BAO data, demonstrates that the BAO information content in our joint analysis comes almost exclusively from the DESI DR2 BAO sample. This justifies our assumption of ignoring the cross-covariance between the BAO and $5\times2$pt data vectors.

\section{Comparison with Sailer et. al}\label{appendix:comapre_to_noah}
\begin{figure*}
    \centering
    \includegraphics[width=0.6\linewidth]{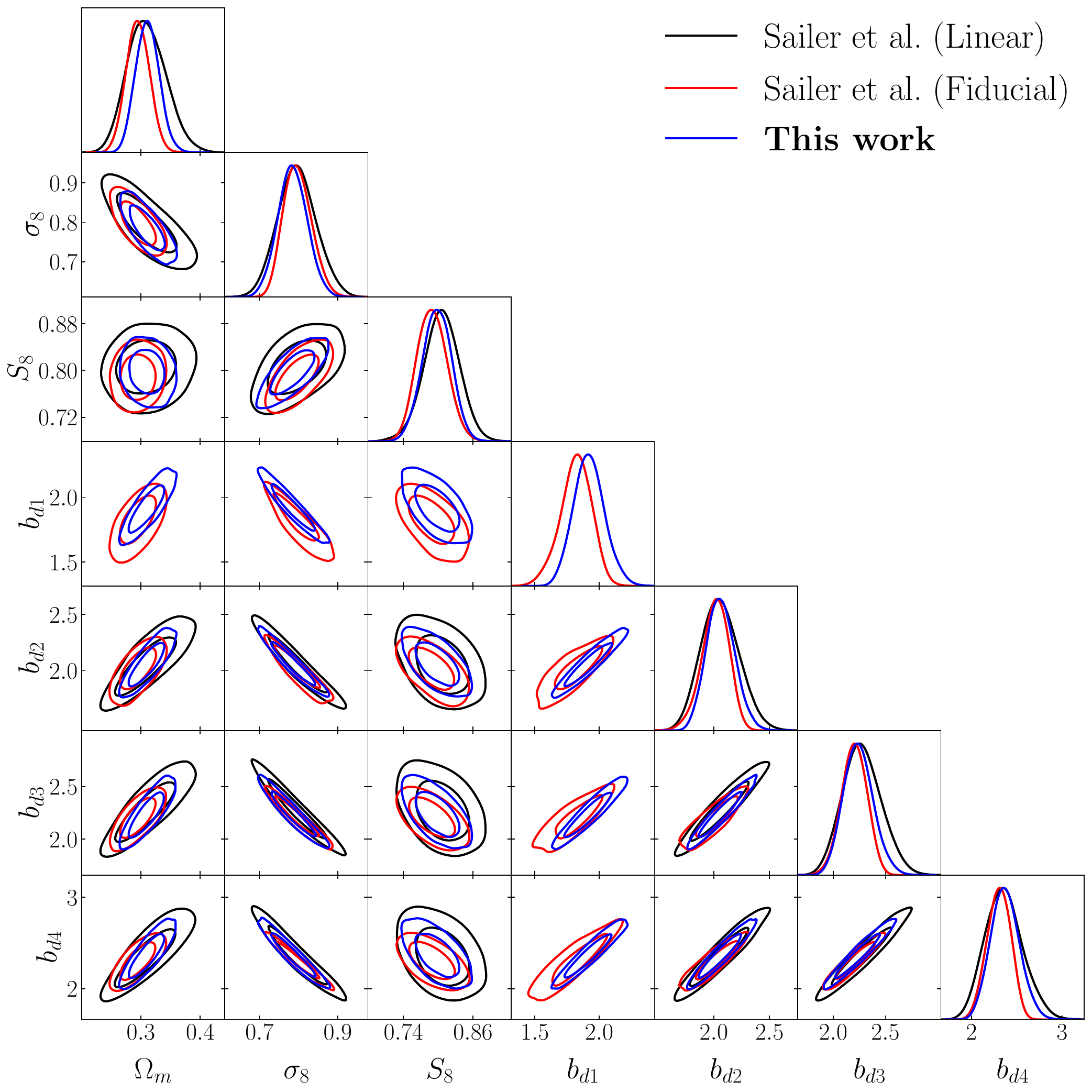}
    \caption{\textbf{Comparison of $\Lambda$CDM results to Ref.~\cite{Sailer:2024coh}}.}
    \label{fig:comapre_to_noah}
\end{figure*}

In Fig.~\ref{fig:comapre_to_noah} we compare our $\Lambda$CDM $2\times2$pt (i.e. LS LRGs auto-correlation and the cross-correlation with ACT DR6 lensing) constraints to those of Ref.~\cite{Sailer:2024coh}, where we adopt the same set-up fixing $\omega_b= 0.02236$, $n_s = 0.9649$, $\Omega_mh^3 = 0.09633$ as Ref.~\cite{Sailer:2024coh} (this differs from the set-up used in the main text and is employed here only for this test). The two analyses differ in several substantive ways. First, our covariance is simulation-based, constructed from forward-modeled mock maps, whereas Ref.~\cite{Sailer:2024coh} uses an analytic approach to estimate the covariance. Second, we include the lowest-redshift LRG bin in our linear bias analysis, while Ref.~\cite{Sailer:2024coh} excludes it. Third, for theory modeling, we use linear galaxy bias and use \textsc{HMcode-2020} for the nonlinear matter power spectrum, whereas the fiducial constraints of Ref.~\cite{Sailer:2024coh} employ a hybrid effective field theory (HEFT) approach. Finally, our beyond-Limber modeling allows us to access larger angular scales (lower $\ell$) with slightly different scale cuts, while Ref.~\cite{Sailer:2024coh} adopts more conservative large-scale cuts, though on the other hand in their HEFT setup they extend to smaller angular scales than we do. Despite these differences in approach, the posteriors closely agree across parameters comparing the fiducial set-up of Ref.~\cite{Sailer:2024coh} and the constraints from this pipeline (Fig.~\ref{fig:lss_bao_information}), demonstrating that the LRG$\times\kappa_{\rm CMB}$ $\Lambda$CDM constraints are robust to these analysis choices. The constraints are tighter than those for the linear theory case of Ref.~\cite{Sailer:2024coh}, mostly due to our inclusion of the lowest redshift bin in the analysis.

\section{EDE parameter priors}~\label{appendix:ede_prior}
\begin{figure}
    \centering
    \includegraphics[width=\linewidth]{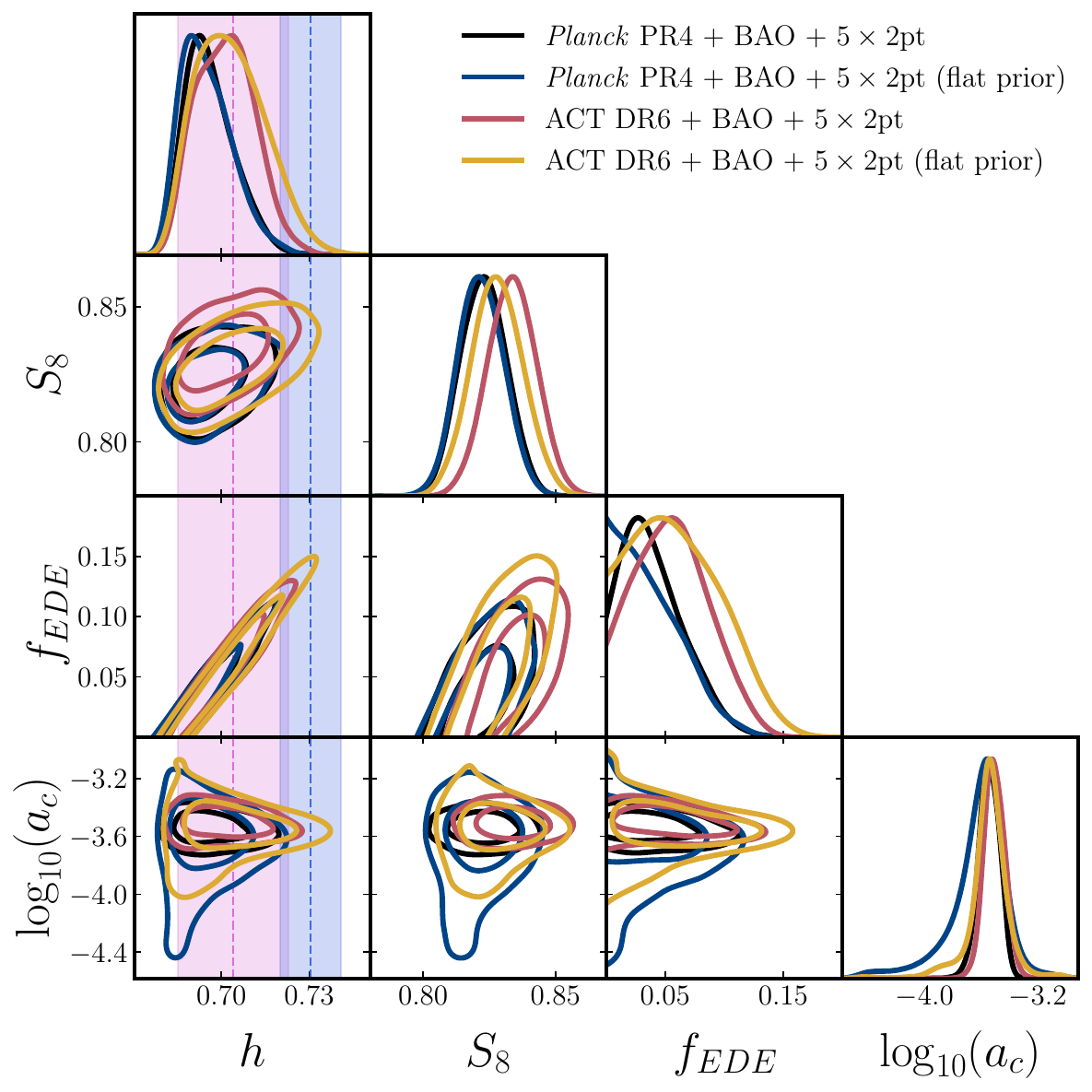}
    \caption{\textit{Comparison of baseline constraints for the EDE model with a flat prior on $\log_{10}a_c$.\label{fig:ede_flatprior}}}
\end{figure}
In this work, we adopted a Gaussian prior $\log_{10} a_c \sim \mathcal{N}(-3.531,\,0.10)$ following Ref.~\cite{Chaussidon:2025npr} to reduce prior volume effects when $f_{\mathrm{EDE}}\!\to\!0$, where $\log_{10} a_c$ becomes effectively unconstrained (see e.g. Refs.~\cite{Herold:2021ksg, Herold:2022iib} for a discussion of prior volume effects in EDE model constraints). As a robustness check (Fig.~\ref{fig:ede_flatprior}), we repeat the analysis with a flat prior on $\log_{10} a_c$ and observe a mild shift of the marginalized contours toward $f_{\mathrm{EDE}}=0$, consistent with what was found for a similar test in Ref.~\cite{DAmico:2020ods}. The \emph{upper limits} on $f_{\mathrm{EDE}}$ are essentially unchanged: for \textit{Planck} PR4\,+\,DESI/ACT BAO we find $f_{\mathrm{EDE}}<0.089$ (95\% CL), and with a flat prior $f_{\mathrm{EDE}}<0.090$; for ACT DR6\,+\,DESI/ACT BAO we obtain $f_{\mathrm{EDE}}<0.112$ (95\% CL), and with a flat prior $f_{\mathrm{EDE}}<0.126$. The use of the Gaussian prior also does not affect our model comparison results, which are based on best-fit $\chi^2$ values. We conclude that adopting this prior has a negligible impact on our results and leaves our overall conclusions regarding EDE unchanged.

\section{CMB lensing cross-correlation normalization correction}~\label{appendix:mc_norm}
When CMB lensing maps are reconstructed from masked and anisotropically filtered CMB data, the resulting quadratic estimator acquires a spurious normalization that depends on the geometry of the mask. This effect is important for lensing cross-correlation analysis, where the normalization differs from that of the auto-spectrum and, if uncorrected, can bias cosmological parameter inference~\cite{ACT:2023oei}. To correct for this, we follow Ref.~\cite{Sailer:2024coh} and compute a Monte Carlo (MC) normalization factor, defined as
\begin{equation}
\label{eq:MCcorr_final}
(\mathrm{MC})_{L} \;=\;
\frac{
  \sum_{\ell,i,m} W_{L\ell}\,
      \{ M^{\kappa_{\rm CMB}} \hat{\kappa_{\rm CMB}}^{\,i}\}_{\ell m}
      \{ M^g \kappa_{\rm CMB}^{\,i}\}^{*}_{\ell m}
}{
  \sum_{\ell,i,m} W_{L\ell}\,
      \{ M^{\kappa_{\rm CMB}} \kappa_{\rm CMB}^{\,i}\}_{\ell m}
      \{ M^g \kappa_{\rm CMB}^{\,i}\}^{*}_{\ell m}
}\,,
\end{equation}
where $M^{\kappa_{\rm CMB}}$ and $M^g$ are the CMB lensing and galaxy masks, $\hat{\kappa_{\rm CMB}}^{\,i}$ is the reconstructed convergence from the $i^\text{th}$ simulation, $\kappa_{\rm CMB}^{\,i}$ is the corresponding input convergence, and $W_{L\ell}$ denotes the bandpower window function. The operator 
$\{AB\}_{\ell m} \;\equiv\; \int d^2\hat{n}\,Y_{\ell m}^*(\hat{n})\,A(\hat{n})\,B(\hat{n})$
represents the masked spherical harmonic transform. In practice, we evaluate Eq.~\eqref{eq:MCcorr_final} using the $N_{\mathrm{sim}}=480$ official ACT DR6 simulations, applying the same ACT lensing mask $M^{\kappa_{\rm CMB}}$ and DESI LRG mask $M^g$ as used in the data. Averaging across simulations yields a multiplicative correction factor that we apply to the DESI LRG–CMB lensing cross-spectra. We do not compute this correction for the $\kappa_{\rm CMB} T$ ISW cross-correlation, since the sky overlap of the \textit{Planck} PR3 and ACT DR6 lensing masks is nearly identical, and the resulting correction is expected to be negligible.

\onecolumngrid

\acknowledgments
This work was supported by an ETH Zurich Doc.Mobility Fellowship. AR thanks Martin White, Noah Sailer, Rongpu Zhou and Gerrit Farren for useful correspondence regarding CMB lensing and DESI LS data. We also thank Colin Hill and Martin White for useful discussions related to early dark energy. We acknowledge the support of Euler Cluster by High Performance Computing Group from
ETHZ Scientific IT Services that we used for most of our computations. SF is supported by Lawrence Berkeley National Laboratory and the Director, Office of Science, Office of High Energy Physics of the U.S. Department of Energy under Contract No.\ DE-AC02-05CH11231. AN acknowledges support from the European Research Council (ERC) under the European Union’s Horizon 2020 research and innovation program with Grant agreement No. 101163128. 

\end{document}